\documentclass{aa}
\usepackage{natbib,psfig,graphicx,ulem}
\usepackage{txfonts}
\usepackage{longtable}
\usepackage{soul,color}

\def\mathnew{\mathsurround=0pt}
\def\simov#1#2{\lower .5pt\vbox{\baselineskip0pt \lineskip-.5pt
\ialign{$\mathnew#1\hfil##\hfil$\crcr#2\crcr\sim\crcr}}}

\def\MeV{Me\kern-0.11em V}
\def\keV{ke\kern-0.11em V}

\begin{document}

\title{Galaxy cluster searches by photometric redshifts in the CFHTLS
\thanks{Based on observations obtained with MegaPrime/MegaCam, a joint
project of CFHT and CEA/DAPNIA, at the Canada-France-Hawaii Telescope
(CFHT) which is operated by the National Research Council (NRC) of
Canada, the Institut National des Sciences de l'Univers of the Centre
National de la Recherche Scientifique (CNRS) of France, and the
University of Hawaii. This work is based in part on data products
produced at TERAPIX and the Canadian Astronomy Data Centre as part of
the Canada-France-Hawaii Telescope Legacy Survey, a collaborative
project of NRC and CNRS.}}

\offprints{C. Adami \email{christophe.adami@oamp.fr}}

\author{ C. Adami\inst{1}
\and
F.~Durret\inst{2,3}
\and
C.~Benoist\inst{4}
\and
J.~Coupon\inst{2,3}
\and
A.~Mazure\inst{1}
\and
B.~Meneux\inst{5}
\and
O.~Ilbert\inst{1}
\and
J.~Blaizot\inst{6}
\and
S.~Arnouts\inst{7}
\and
A.~Cappi\inst{8}
\and
B.~Garilli\inst{9}
\and
L.~Guennou\inst{1}
\and
V.~LeBrun\inst{1}
\and
O.~LeF\`evre\inst{1}
\and
S.~Maurogordato\inst{4}
\and
H.J.~McCracken\inst{2,3}
\and
Y.~Mellier\inst{2,3}
\and
E.~Slezak\inst{4}
\and
L.~Tresse\inst{1}
\and
M.P.~Ulmer\inst{10}
}

\institute{
LAM, OAMP, P\^ole de l'Etoile Site Ch\^ateau-Gombert 38, Rue Fr\'ed\'eric Juliot-Curie,  13388 Marseille, Cedex 13, France
\and
UPMC Universit\'e Paris 06, UMR~7095, Institut d'Astrophysique de Paris,
F-75014, Paris, France
\and
CNRS, UMR~7095, Institut d'Astrophysique de Paris, F-75014, Paris, France
\and
OCA, Cassiop\'ee, Boulevard de l'Observatoire, B.P. 4229, F-06304 NICE Cedex 4
\and
MPE, Giessenbachstrasse, 85748 Garching, Germany
\and
CRAL (UMR 5574), Universit\'e Claude Bernard Lyon 1 (UCBL),
 Ecole Normale Sup\'erieure de Lyon (ENS-L), and Centre National de la Recherche
 Scientifique (CNRS)
\and
Canada-France-Hawaii Telescope Corporation, Kamuela, HI-96743, USA
\and
INAF - Osservatorio Astronomico di Bologna, via Ranzani 1, 40127 Bologna, Italy
\and
INAF IASF - Milano, via Bassini 15, 20133 Milano, Italy
\and
Department Physics $\&$ Astronomy, Northwestern University, Evanston, IL 60208-2900, USA
}

\date{Accepted . Received ; Draft printed: \today}

\authorrunning{Adami et al}

\titlerunning{Galaxy cluster searches by photometric redshifts in the CFHTLS}

\abstract
{Counting clusters is one of the methods to constrain
  cosmological parameters, but has been up to now limited both by the
  redshift range and by the relatively small sizes of
  the homogeneously surveyed areas.}
{In order to enlarge publicly available optical cluster catalogs, in particular at high redshift, we
have performed a systematic search for clusters of galaxies in the
Canada France Hawaii Telescope Legacy Survey (CFHTLS).}
{We considered the deep 2, 3 and 4 CFHTLS Deep fields (each
  1$\times$1~deg$^2$), as well as the wide 1, 3 and 4 CFHTLS Wide fields. We used the Le
  Phare photometric redshifts for the galaxies detected in
  these fields with magnitude limits of i'=25 and 23 for the Deep and
  Wide fields respectively. We then constructed galaxy density maps in
  photometric redshift bins of 0.1 based on an adaptive kernel
  technique and detected structures with SExtractor at various
  detection levels. In order to assess the validity of our cluster
  detection rates, we applied a similar procedure to galaxies in
  Millennium simulations. We measured the correlation function of our cluster candidates.
  We analyzed large scale properties and substructures, including filaments,
  by applying a minimal spanning tree algorithm
  both to our data and to the Millennium simulations. }
{We have detected 1200 candidate clusters with various masses (minimal
  masses between 1.0 10$^{13}$ and 5.5 10$^{13}$ and mean masses
  between 1.3 10$^{14}$ and 12.6 10$^{14}$ M$_\odot$) in the CFHTLS
  Deep and Wide fields, thus notably increasing the number of known
  high redshift cluster candidates. We found a correlation function
  for these objects comparable to that obtained for high redshift
  cluster surveys.  We also show that the CFHTLS deep survey is able to
  trace the large scale structure of the universe up to z$\geq$1.  Our
  detections are fully consistent with those made in various CFHTLS
  analyses with other methods. We now need accurate mass
  determinations of these structures to constrain cosmological parameters.}
{We have shown that a search for galaxy clusters based on density maps
  built from galaxy catalogs in photometric redshift bins is
  successful and gives results comparable to or better than those
  obtained with other methods. By applying this technique to the
  CFHTLS survey we have increased the number of known optical high
  redshift cluster candidates by a large factor, an important step
  towards using cluster counts to measure cosmological parameters.}


\maketitle

\section{Introduction}\label{sec:intro}

The beginning of the 21st century is an exciting period for
cosmological studies. Several methods now allow to put strong
constraints on cosmological parameters. We can for example reconstruct
Hubble diagrams (supernovae or tomography) or use directly the
primordial fluctuation spectrum.  In addition, the cluster count
technique is probably the oldest one (see e.g. Gioia et al., 1990). Up
to now this technique was penalized by the redshift range of detected
clusters, which was too low to make the difference between flat and
open universes. Distant cluster surveys have also been mainly
conducted in areas too small or with inhomogeneous selection
functions.  Besides cluster mass knowledge, this technique requires
indeed large fields of view of several dozen square degrees to provide
large numbers of cluster detections at z$\geq$1 (e.g. Romer et
al. 2001).  Recent X-ray cluster surveys are beginning to produce
cluster catalogs at high z (e.g. the XMM-LSS survey, Pierre et al.,
2007) and it is the goal of the present paper to contribute to the
production of similar large cluster catalogs based on optical Canada
France Hawaii Telescope Legacy Survey data.

The Canada-France-Hawaii Telescope Legacy Deep and Wide Surveys
(CFHTLS-D and CFHTLS-W) respectively explore solid angles of 4 deg$^2$
and 171 deg$^2$ of the deep Universe, each in 4 independent patches
(http://www.cfht.hawaii.edu/Science/CFHLS/). For both surveys,
observations are carried out in five filters ($u*,g',r',i'$ and $z'$)
providing catalogs of sources that are 80\% complete up to
$i_{AB}$=26.0 (CFHTLS-D) and $i_{AB}$=24.0 (CFHTLS-W) (Mellier et al
2008,
http://terapix.iap.fr/cplt/oldSite/Descart/CFHTLS-T0005-Release.pdf).
The CFHTLS-W, in particular, encloses a sample of about 20 10$^6$
galaxies inside a volume size of $\sim 1$ Gpc$^3$, with a median
redshift of z$ \sim 0.92$ (Coupon et al 2009). According to the
standard cosmological model, the CFHTLS-W (W1, W2, W3, and W4
herafter) is then expected to contain 1000 to 5000 clusters of
galaxies with accurate photometric redshifts, most of them in the
$0.6<z<1.5$ range. Likewise, the CFHTLS-D (D1, D2, D3, and D4
hereafter) should contain 50 to 200 clusters, with a
significant fraction at higher redshift than the CFHTLS-W.  The two
surveys are therefore complementary data sets. They can produce
together two homogeneous optically selected samples of clusters of
galaxies that can be reliably compared and used jointly to explore the
mass function, the abundance of clusters of galaxies and the evolution
of cluster galaxy populations as a function of lookback time.

The construction of homogeneous catalogs of optically selected
clusters of galaxies is not, a simple task. Early searches for
clusters of galaxies in the CFHTLS were performed by Olsen et
al. (2007) based on a matched filter detection algorithm applied to
the Deep fields (see also the recent paper by Grove et al. 2009).
Galaxy density maps combined with photometric redshift catalogs were
considered by Mazure et al. (2007) in the D1 field.  Lensing techniques were
also employed to detect massive structures in the CFHTLS (e.g. Cabanac
et al. 2007, Gavazzi $\&$ Soucail 2007, Berg\'e et al. 2008). Other
cluster studies based on the CFHTLS data (e.g. the CFHTLS-CARS survey:
Erben et al. 2009, Hildebrandt et al. 2009) and based for part of them
on the red sequence in the color magnitude diagram are also in
progress.

We present here a systematic search for galaxy clusters in the D2, D3
and D4 Deep fields (the D1 field was already analyzed by Mazure et
al. 2007), as well as in the regions of the W1, W3 and W4 Wide fields
available in the T0004 release. Our approach is based on photometric
redshifts computed for all the galaxies extracted in each field
(Coupon et al. 2009). In this way, we take into account the full color
information and not only two bands as for example in red sequence
searches.  We divided the galaxy catalogs in slices of 0.1 in
redshift, each slice overlapping the previous one by 0.05, and built
density maps for each redshift slice. Structures in these density maps
were then detected with the SExtractor software in the different
redshift bins. We applied the same method to similar size mock samples
built from the Millennium simulation, in order to estimate the
reliability of our detections. We measured the clustering properties
of our catalog. We then analyzed substructuring and filamentary large
scale properties by applying a minimum spanning tree algorithm both to
our data and to the Millennium simulation.

In this paper we assume H$_0$ = 70 km s$^{-1}$ Mpc$^{-1}$, $\Omega
_m$=0.3, $\Omega _{\Lambda}$=0.7. All coordinates are given at the J2000
equinox and magnitudes are in the AB system.

\section{Searching for clusters in the CFHTLS}

A full description of the method applied to the D1 field is given in
Mazure et al. (2007) and we adopt here the same method, which we briefly
summarize below. We will not redo the D1 analysis because the
D1 photometric redshifts in Mazure et al. (2007) have
very similar quality compared to the present data. D1 cluster
detections will only be considered for an internal crosscheck with the W1 detections
of the present paper.

\subsection{Photometric redshifts}

We used the public photometric redshift catalogs from the CFHTLS data release
T0004 (available at http://terapix.iap.fr/) in the D2, D3, D4, W1, W3 and W4
fields. The regions we selected inside the wide
fields and covered by T0004 are mosaics of 19, 4 and 11 Megacam fields for W1, W3
and W4 respectively. In order
to avoid incompleteness effects and strong systematic biases in
photometric redshift computations, the catalogs were limited to i'=25
and 23 for the Deep and Wide fields respectively. This is slightly deeper than
the recommended cuts of Coupon et al. (2009) but proved to not be a problem in
our analysis.

Our approach is based on photometric redshifts, which can be estimated
with good precision up to z$\sim$1.5 (Coupon et al. 2009) thanks to
the optimal wavelength coverage achieved by the u*g'r'i'z' CFHTLS
data.  Photometric redshifts were computed for all the objects in the
CFHTLS galaxy catalogs with the Le Phare software developed by
S.~Arnouts and O.~Ilbert (Ilbert et al. 2006; also see
http://www.ifa.hawaii.edu/\~ ilbert/these.pdf.gz, pages 50 and
142). Details of this computation are given in Mellier et al. (2008).
Briefly, these photometric redshifts were computed with a large set of
templates, covering a broad domain in parameter space (see Coupon et
al. 2009 for a full description of the method and
sample). Spectroscopic redshifts from the VVDS (e.g. LeF\`evre et
al. 2004) were used to optimize the photometric redshift
estimates. This step is extensively described by Coupon et
  al. (2009) and the process consists in shifting the magnitude zero
  points until the difference between photometric and
  spectroscopic redshifts is minimized. These shifts were all lower than 0.1
  magnitude, see Table 2 of Coupon et al. (2009). The resulting
  statistical errors (including the $1+z$ dependence) on the
  photometric redshifts are also given in Coupon et al. (2009).  For
  example in the W1 field, they continuously increase (between i'=20.5
  and i'=24) from 0.025 to 0.053. At our limiting magnitude of i'=23,
  the redshift statistical error is 0.043. Deep fields have nearly
  constant redshift statistical errors of the order of 0.026  (maximum is 
  0.028) for i' magnitudes between 20.5 and 24.

We selected galaxies with photometric redshifts included in the range $0
\le z \le 1.5$ for the Deep fields and $0 \le z \le 1.2$ for the Wide
fields.

For each CFHTLS field and subfield, we give in Table~\ref{tab:data} the
numbers of galaxies taken into account. This table will be useful for
comparisons with future data releases.

\begin{table}
\caption{Number of galaxies in each CFHTLS field and subfield we studied. The considered
magnitude limit is i'=23 for the wide fields and i'=25 for the deep fields.}
\begin{center}
\begin{tabular}{rrrr}
\hline
Field & Subfield & Coordinates & Nb of galaxies \\
\hline
D2 &    & 100000+021220   & 376224 \\
D3 &    & 141754+523031   & 500307 \\
D4 &    & 221531-174405	  & 458296 \\
W1 &  1 & 021410-041200 & 221291 \\
W1 &  2 & 021410-050800 & 216063 \\
W1 &  3 & 021800-041200	& 218959 \\
W1 &  4 & 021800-050800	& 237444 \\
W1 &  5 & 021800-060400	& 221110 \\
W1 &  6 & 022150-041200	& 224493 \\
W1 &  7 & 022150-050800	& 205990 \\
W1 &  8 & 022150-060400	& 218958 \\
W1 &  9 & 022539-041200	& 228258 \\
W1 & 10 & 022539-050800	& 195797 \\
W1 & 11 & 022539-060400	& 177252 \\
W1 & 12 & 022929-041200	& 181313 \\
W1 & 13 & 022929-050800	& 201348 \\
W1 & 14 & 022929-060400	& 183899 \\
W1 & 15 & 022929-070000	& 195285 \\
W1 & 16 & 023319-041200	& 220113 \\
W1 & 17 & 023319-050800	& 198918 \\
W1 & 18 & 023319-060400	& 200125 \\
W1 & 19 & 023319-070000 & 190473 \\
W3 &  1 & 135955+523831 & 241036 \\
W3 &  2 & 140555+523831 & 226221 \\
W3 &  3 & 141154+523831 & 208163 \\
W3 &  4 & 141201+514231 & 204033 \\
W4 &  1 & 220930+002300 & 231078 \\
W4 &  2 & 220930-003100 & 204257 \\
W4 &  3 & 221318+002300 & 235100 \\
W4 &  4 & 221318-003100 & 230770 \\
W4 &  5 & 221318+011900 & 210846 \\
W4 &  6 & 221706+002300 & 207861 \\
W4 &  7 & 221706-003100 & 171636 \\
W4 &  8 & 221706+011900 & 213383 \\
W4 &  9 & 222054+002300 & 197228 \\
W4 & 10 & 222054-003100 & 215437 \\
W4 & 11 & 222054+011900 & 221290 \\
\hline
\end{tabular}
\end{center}
\label{tab:data}
\end{table}

\subsection{Density maps}

In order to obtain results directly comparable with those previously
obtained by Mazure et al. (2007), we applied the same procedure.  For
each field, galaxy catalogs were built in running slices of 0.1 in
redshift (see also Mazure et al. 2007), displaced by 0.05
(i.e. the first slice covers redshifts 0.0 to 0.1, the second 0.05 to
0.15 etc.). We assumed the most likely photometric redshift for
  each object in order to assign it to a redshift slice.  Density
maps were then computed for each redshift slice, based on an
adaptative kernel technique described in Mazure et al. (2007). The
highest redshift slices were 1.30-1.40 and 1.35-1.50 for the Deep
fields, and 1.05-1.15 for the Wide fields. An example of density map
obtained is displayed in Fig.~\ref{fig:densmaps}.

\begin{figure}[hbt]
\caption[]{Density maps for the D2 field for the z=0.65-0.75 redshift
bin. Two clusters are detected at S/N$\geq$6. } \label{fig:densmaps}
\end{figure}

The SExtractor software (Bertin \& Arnouts 1996) was then applied to the
galaxy density maps to detect structures at pre-defined significance levels
(called hereafter S/N)
of $2\sigma_S$, $3\sigma_S$, $4\sigma_S$, $5\sigma_S$ and $6\sigma_S$
(where $\sigma_S$ is the SExtractor detection threshold).

The structures were then assembled in larger structures (called
$detections$ in the following) using a friends-of-friends algorithm, as in
Mazure et al. (2007). We assigned to a $detection$ the redshift of its highest
S/N component.

  We had made several experimentations for the Mazure et al. (2007)
  preliminary work, and found that the 0.1 redshift width of most of
  the studied slices was the best compromise between the redshift
  resolution and the possible dilution in the density signal due to
  photometric redshift uncertainties. We chose to keep the slice width
  larger than the maximal photometric redshift 1$\sigma$
  uncertainty. For the wide fields, assuming the worse possible
  photometric redshift statistical error of 0.043$\times$$(1+z)$ (for
  i'=23, see Coupon et al. 2009) leads to a 1$\sigma$ error of 0.09 at
  z=1.15 (upper limit in redshift for the wide field analyses). For
  the deep fields, assuming a redshift statistical error lower than
  0.028$\times$$(1+z)$ leads to a 1$\sigma$ error of 0.07 at
  z=1.5. Both values are lower than the 0.1 slice width we choose.

  By definition of a Gaussian function, $\sim 32 \%$ of the objects will 
  have a true redshift
  differing by more than 1$\sigma$ from the most likely photometric
  redshift. This means that in the worse case (at the limiting
  magnitude and for the higher allowed redshift bin), slightly less
  than 30$\%$ of the objects (the slice width is slightly lower than the
  1$\sigma$ value) will be assigned to a wrong redshift
  slice (mostly in the immediately higher or lower redshift slices).
  At lower redshifts and for brighter magnitudes, the percentage of
  such lost objects is low and is not a concern regarding our
  analysis. If we are close to the study limitations, it is then
  likely that the lost objects will be numerous enough to still be
  detected as part of a structure in the adjacent slices. The
  friends-of-friends algorithm described earlier will therefore
  associate these structures shifted in redshift to their true parent
  structure, therefore not significantly penalizing our analysis.

\subsection{Modified Millennium catalogs}

With this method, we obtained catalogs of galaxy cluster candidates in
the various fields for a given significance level. In order to assess
our detection levels we applied the same method to a modified version of
the Millennium numerical simulation (e.g. Springel et al. 2005,
http://www.mpa-garching.mpg.de/galform/virgo/millennium/ ), as follows:

- We started from semi-analytic galaxy catalogues obtained by applying
the prescriptions of De Lucia $\&$ Blaizot (2007) to the dark-matter
halo merging trees extracted from the Millennium simulation (Springel
et al. 2005). The Millennium run distributes
particles in a cubic box of size 500 h$^{-1}$ Mpc. The simulation was 
built with a
$\Lambda$CDM cosmological model. For details on the semi-analytic model, 
we refer to De
Lucia $\&$ Blaizot (2007) and reference therein. Note that this model
uses the Bruzual $\&$ Charlot (2003) population synthesis model and a
Chabrier (2003) Initial Mass Function (IMF) to assign luminosities to
model galaxies. The $1\times1$ deg$^2$ light cones were generated with
the code MoMaF (Blaizot et al. 2005) and are complete in apparent
magnitude up to I$_{\rm AB}$=24. We used the 4 cones that had the most massive
structures, more adequate for investigating cluster search.

- For each simulated galaxy, magnitude errors were computed in order
to reproduce the (magnitude, magnitude error) mean relation obtained
from the CFHTLS catalogues in all photometric bands.  Magnitudes were
then recomputed in order to reproduce the spread of the (magnitude,
magnitude error) diagram assuming a gaussian distribution.

- We took into account occultation effects of background galaxies by
foreground galaxies. For a given galaxy, we searched for brighter
galaxies located in the foreground and large enough to occult the given
object (included within the disk size of the foreground galaxy as
defined in the Millennium simulation). If such occulting objects were
found, we removed the occulted objects from the simulation. This removed
8$\%$ of the objects of the Millennium simulation.

- We took into account possible lensing effects which could potentially
re-include occulted galaxies. For this, we estimated the image
displacement of a galaxy due to a foreground massive object. Assuming an
isothermal gravitational potential for the lens galaxy, the deflexion
of a background object is then given by:

  $$(\sigma _v / 186.5\ km\ s^{-1})^2 \times ({\rm D}_{ls}/{\rm D}_{os})$$

\noindent
in arcsec, where $\sigma _v$ is the velocity dispersion of the lens,
D$_{ls}$ the lens-source distance, and D$_{os}$ the observer-source
distance.

We computed $\sigma _v$ from the values of M$_{200}$ and r$_{200}$ given
in the Millennium simulation.

Then, if the deflexion amplitude was larger than the occulting object
disk radius, we re-included the lensed galaxy considered. This affected
less than 1$\%$ of the Millennium simulation objects and the effect was
therefore minor.

- We added noise to the true redshift values of the Millennium
simulation in order to mimic the photometric redshift distribution
computed in the CFHTLS Wide and Deep fields. This means that we
produced one modified Millennium simulation in order to compare with
the CFHTLS Deep fields and another one to compare with the CFHTLS Wide
fields. The process was simply to compute the basic gaussian 1$\sigma$ error of the
CFHTLS photometric redshifts as a function of redshift, as a function
of magnitude, and as a function of magnitude uncertainty. This
produced a cubic grid of $\sigma$'s with a resolution of 0.15 in
redshift, 0.5 in magnitude, and 0.2 in magnitude uncertainty. This grid
resolution is a good compromise between the computing time and the quality
of the photometric redshifts. We then applied this grid to the
true Millennium redshifts. These Millennium object redshifts were
re-shuffled within a $\sigma$ characterized gaussian function
according to the object redshift, magnitude, and magnitude error.
Fig.~\ref{fig:gruge} shows the resulting relations between true and
photometric-like redshifts in the Millennium simulation .

\begin{figure}[hbt]
\centering 
\caption[]{True redshift values for the modified Millennium simulation 
versus photometric
redshifts according to the CFHTLS criteria. Upper figure: Wide parameters,
lower figure: Deep parameters.}  \label{fig:gruge}
\end{figure}

- In order for the clusters detected in the Millennium simulations to be
comparable to those in the Deep and Wide CFHTLS data, the catalogs of
galaxies in the Millennium simulation were cut at R=25 and
R=23 respectively. As previously described, galaxy catalogs were
created in slices of photometric redshifts, density maps were computed,
structures were identified with SExtractor, and the significance
level was computed for each of them.

  We note that several artificial galaxy concentrations in
  photometric redshifts appear in Fig.~\ref{fig:gruge}, mainly for the
  Wide survey characteristics. This is simply due to the
  characteristics of the CFHTLS data.  We need to reproduce these
  biases in the Millenium simulations in order to properly quantify our
  false cluster detection rate in the CFHTLS data.

\subsection{Detection rate assessments}

We now need to estimate the
$detection$ rate of our cluster detection method in the CFHTLS. This is
complicated by the fact that the Millennium simulation has a very high
spatial resolution compared to our galaxy density maps. This results in
a non negligible multiplicity of Millennium halos in a single $detection$
made with our technique. For a given
$detection$, we can usually find more than one halo in the Millennium
simulation. We must therefore investigate each of our $detections$ in
the Millennium simulation (152 for the Wide parameters and 179 for the
Deep parameters respectively) to check how many Millennium halos can be
associated.

For this purpose, and for a given $detection$:

  - we computed the density of Millennium halos in the $detection$ as a
  function of their mass.

  - for a given minimal mass, we compared this density to the mean density of
Millennium halos included in the redshift range spanned by the $detection$
(and more massive than the given mass) but not spatially included in the
$detection$.

If the ratio between these two densities was less than 0.1, we assumed
that the density of halos present in a $detection$ was significantly
different from the mean density at the same redshift, and therefore we
had a significant mass concentration in our $detection$, i.e. our
$detection$ was real.  This means that we only keep $detections$
located in the 10$\%$ highest Millenium halo density regions.  We did
this exercise for several minimal Millenium halo masses (5 10$^{12}$,
8 10$^{12}$, 10$^{13}$, 2~10$^{13}$, 4 10$^{13}$, and 6 10$^{13}$
M$_\odot$).  For clarity, we choose to express this with a
significance parameter
$$p = 100. \times (1. - {\rm density ratio}) $$.
A $p$ value lower than 90$\%$ means that a
$detection$ is fake.  We considered a $detection$ as real if none of
the values of $p$ (for various minimal masses) was lower than
90$\%$.

\begin{figure}[hbt]
\centering
\mbox{\psfig{figure=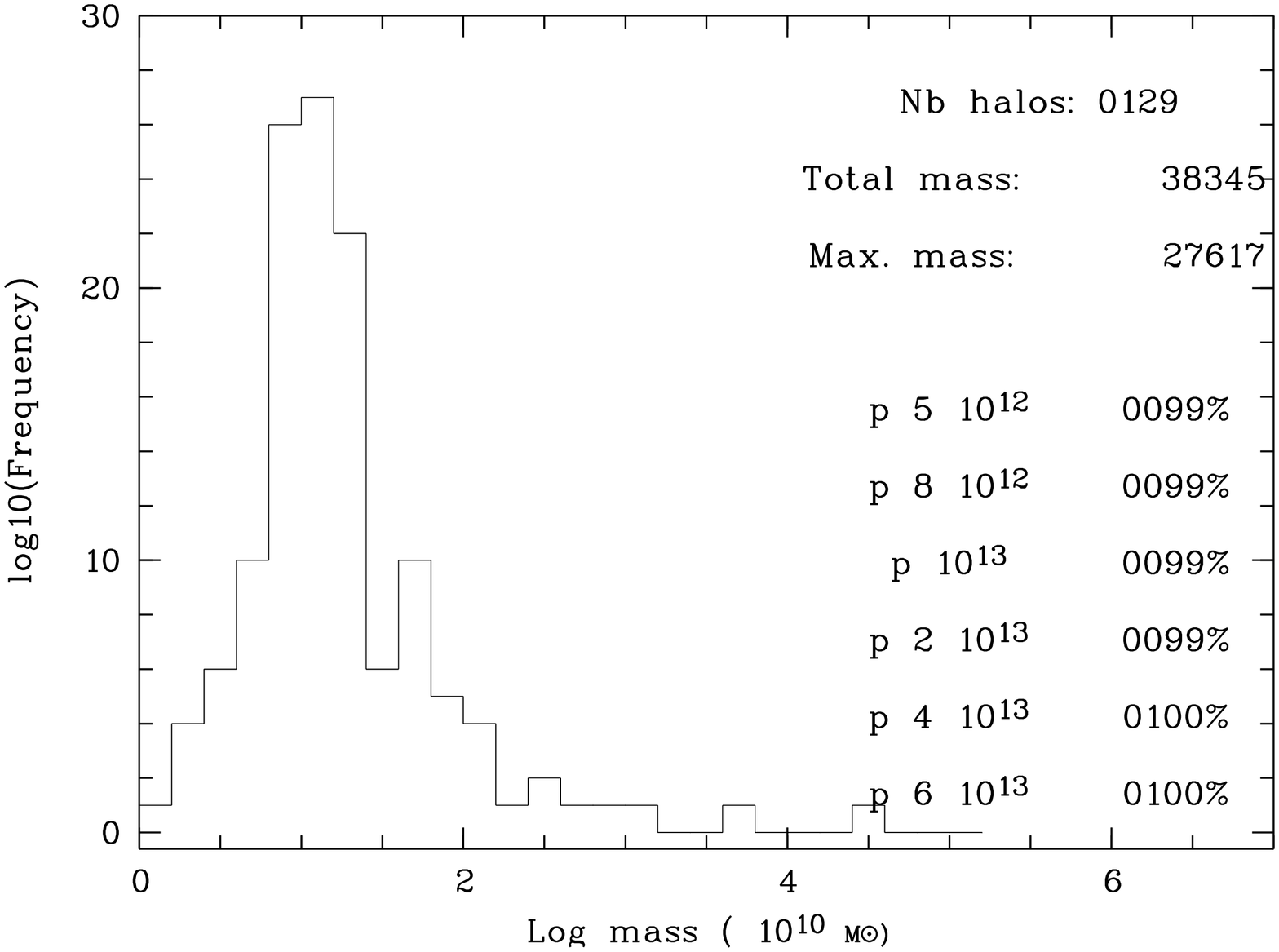,width=8.cm,angle=270}}
\mbox{\psfig{figure=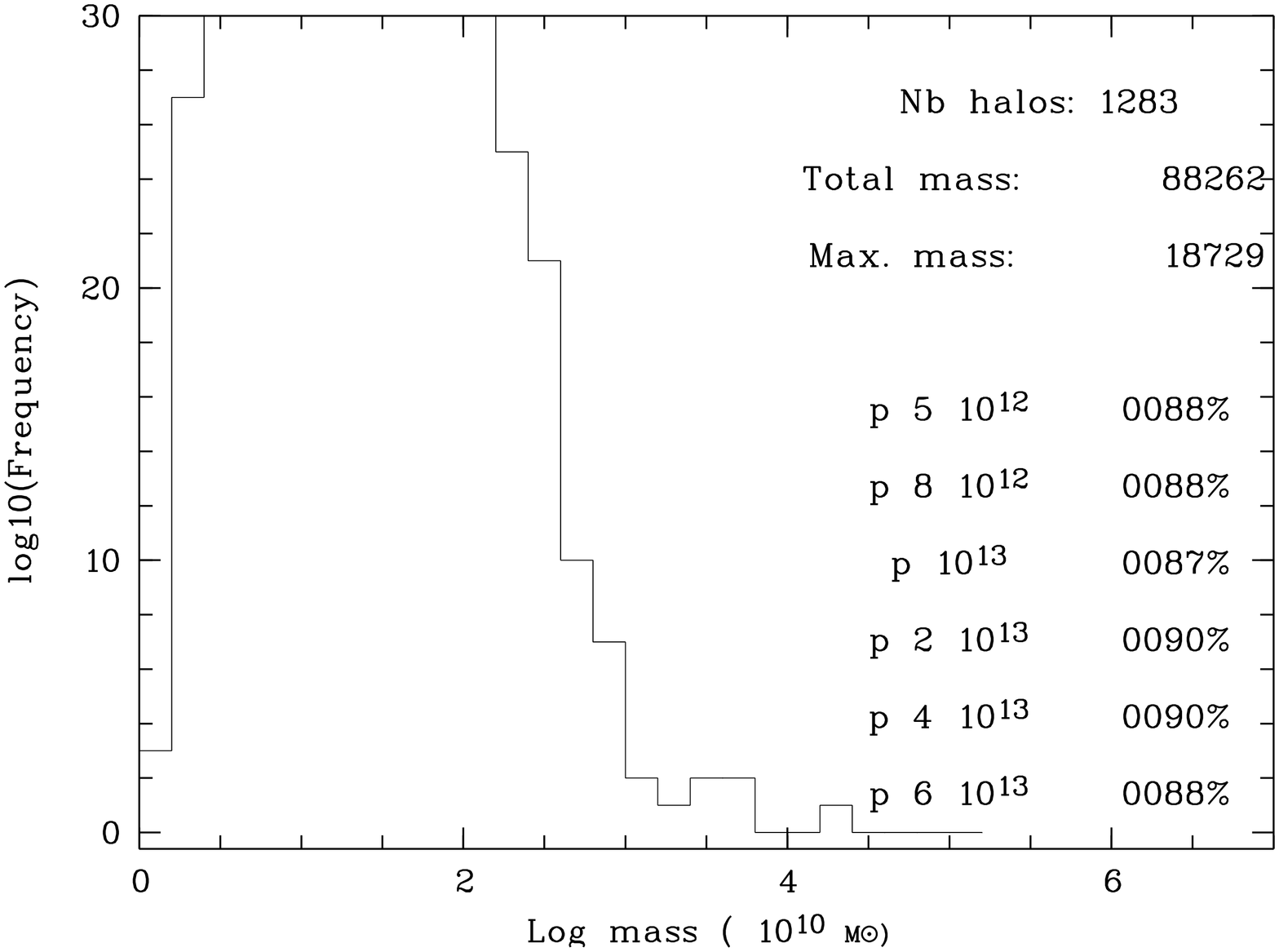,width=8.cm,angle=270}}
\caption[]{Mass histograms of the Millennium halos (Wide survey characteristics)
present in a real $detection$ (upper figure) and in a false $detection$ (lower
figure). Each figure gives the number of halos included in the $detection$, the
total and maximal mass of these halos (in log 10$^{10}$ M$_\odot$), and 6
significance parameters $p$ for 6 different minimal masses (see text).}
\label{fig:example1}
\end{figure}

We give in Fig.~\ref{fig:example1} examples of the mass histograms of
the Millennium halos present in two of our $detections$, one
considered as a real $detection$, and the other considered as false.

\begin{figure}[hbt]
\centering \mbox{\psfig{figure=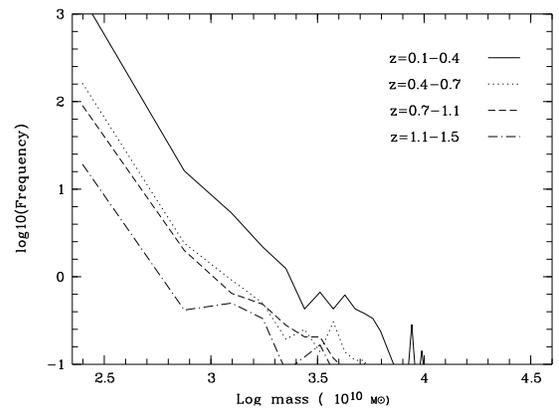,width=8.cm,angle=270}}
\caption[]{Mass distributions of all the Millennium halos (CFHTLS Wide
survey characteristics) included in real $detections$ as a
function of mass for various $detection$ redshift bins. }
\label{fig:example2}
\end{figure}

Fig.~\ref{fig:example2} gives the mean mass distributions of all the
Millennium halos included in all our real $detections$ (CFHTLS Wide
characteristics) as a function of mass for various $detection$ redshift
bins.

\begin{figure}[hbt]
\centering
\mbox{\psfig{figure=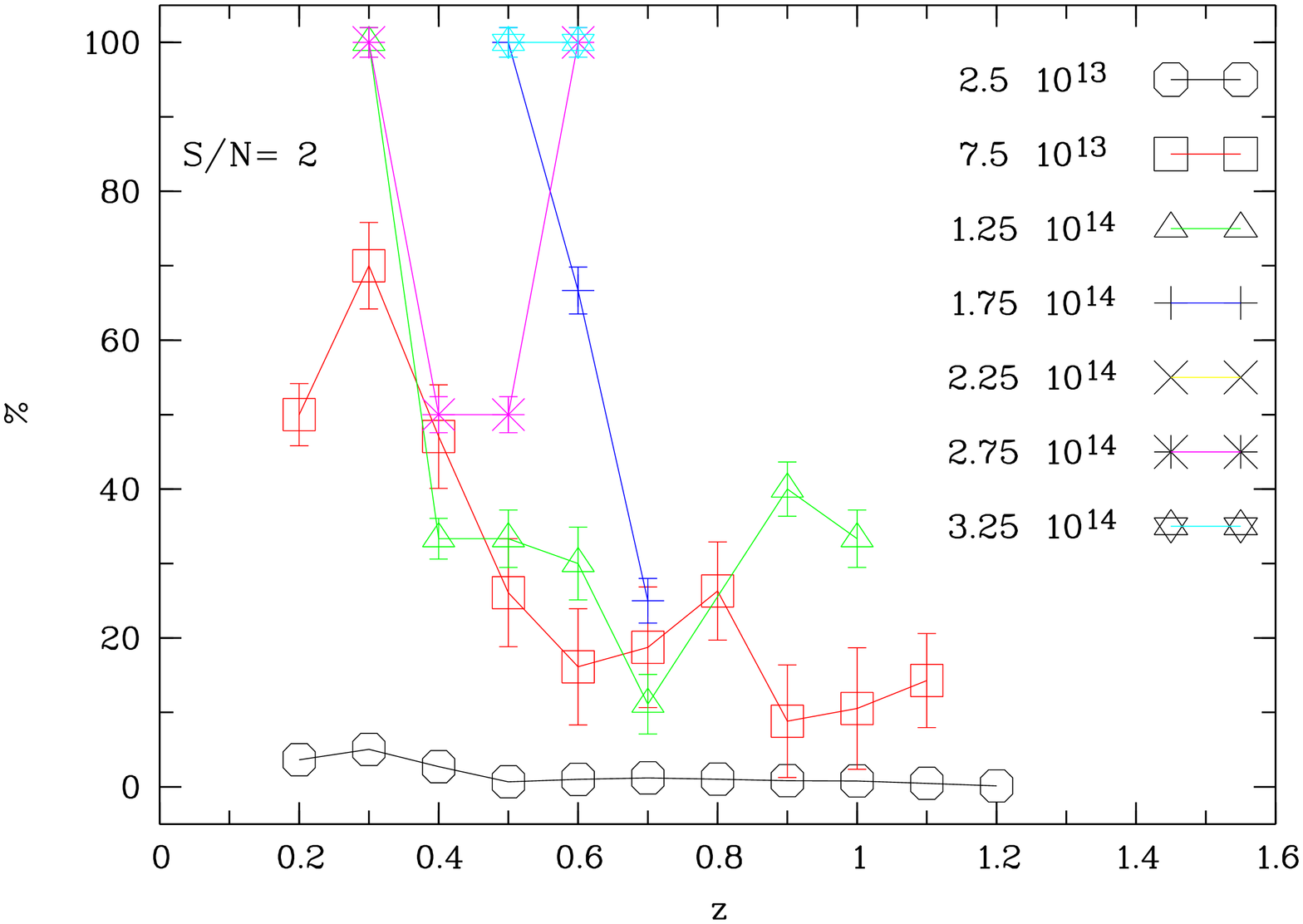,width=6cm,angle=270}}
\mbox{\psfig{figure=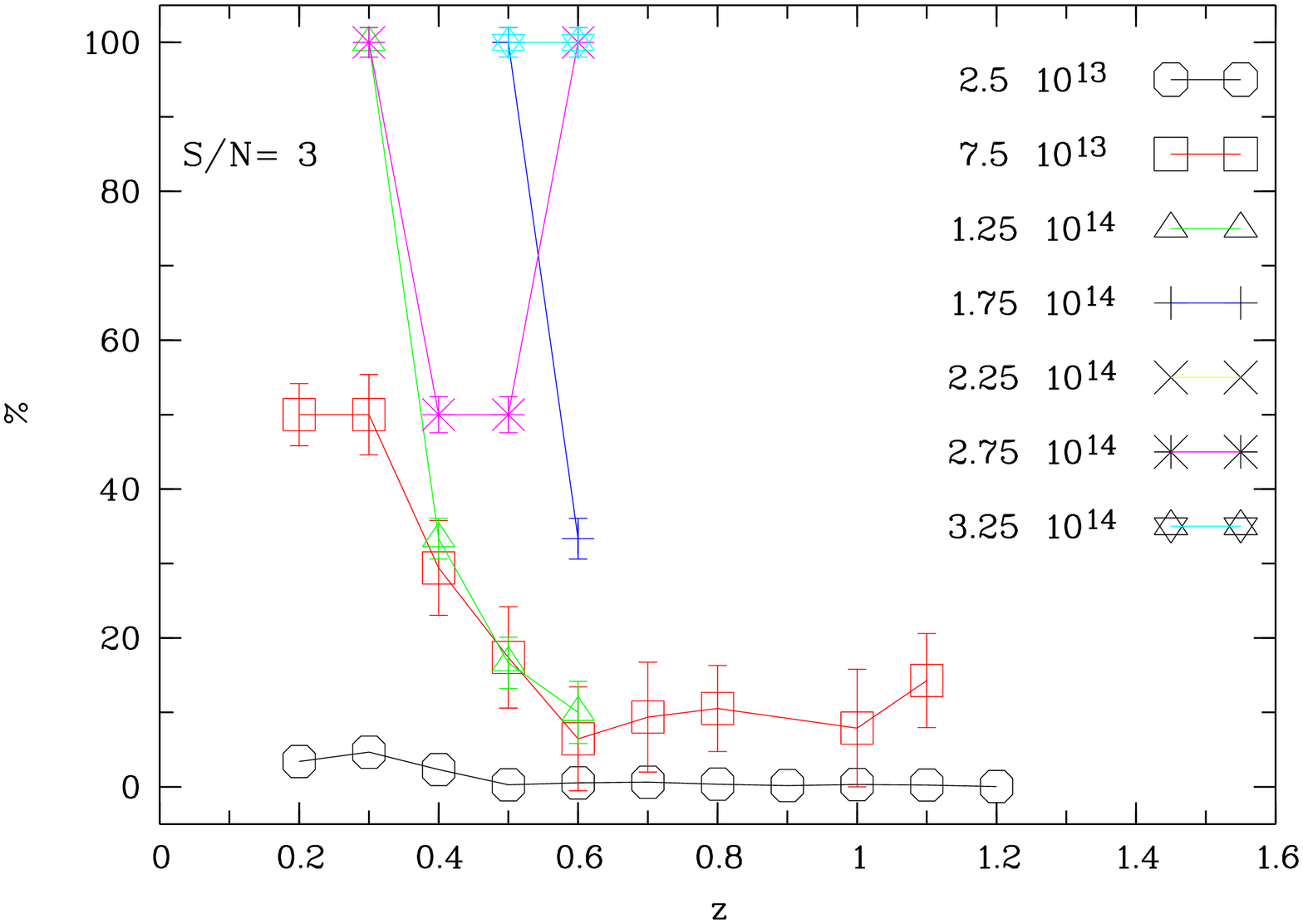,width=6cm,angle=270}}
\mbox{\psfig{figure=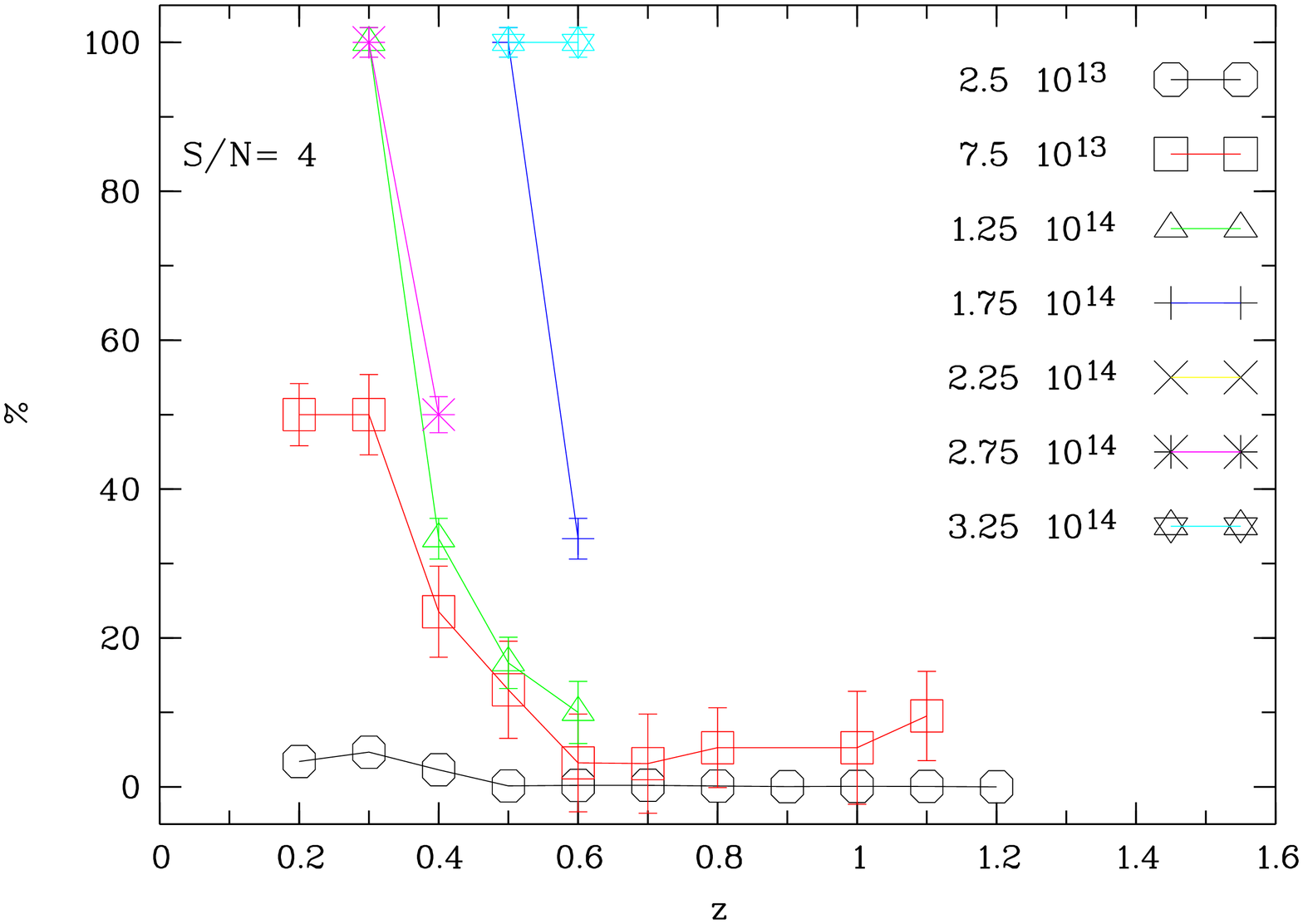,width=6cm,angle=270}}
\mbox{\psfig{figure=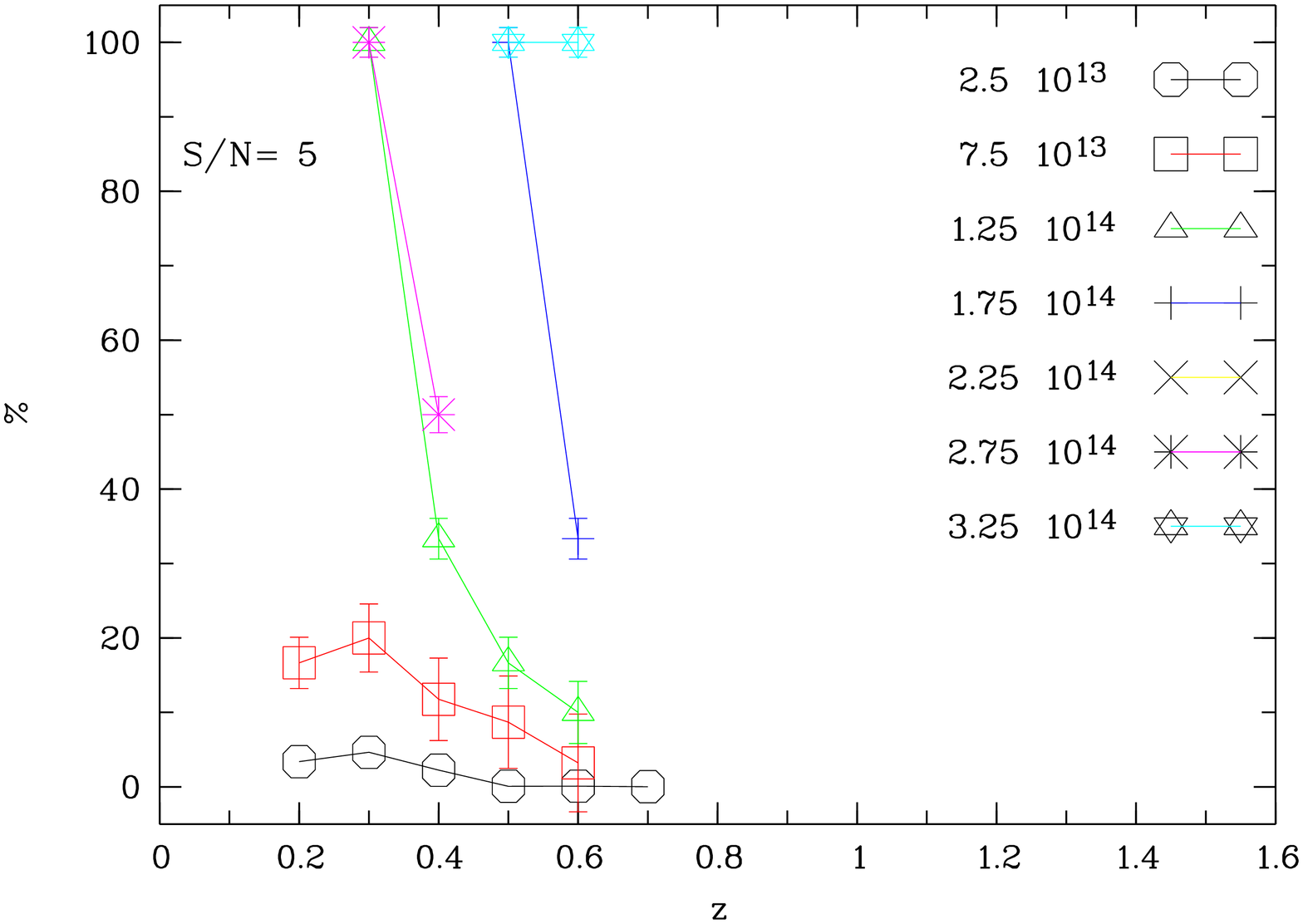,width=6cm,angle=270}}
\mbox{\psfig{figure=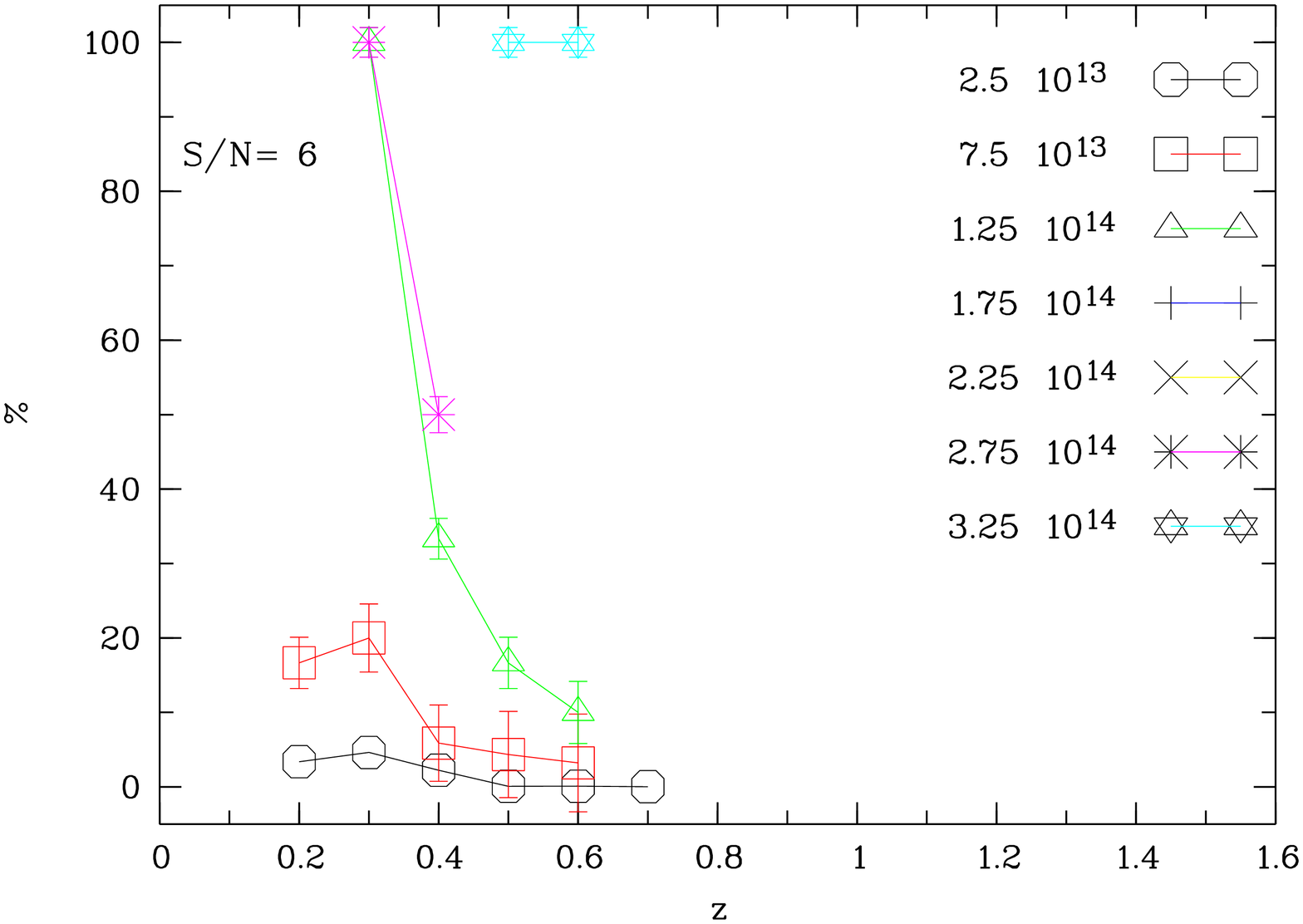,width=6cm,angle=270}}
\caption[]{Percentages of all Millennium halos included in real detections
  for the Wide survey characteristics as a function of redshift for
  different mass limits. }
\label{fig:detpw}
\end{figure}

\begin{figure}[hbt]
\centering
\mbox{\psfig{figure=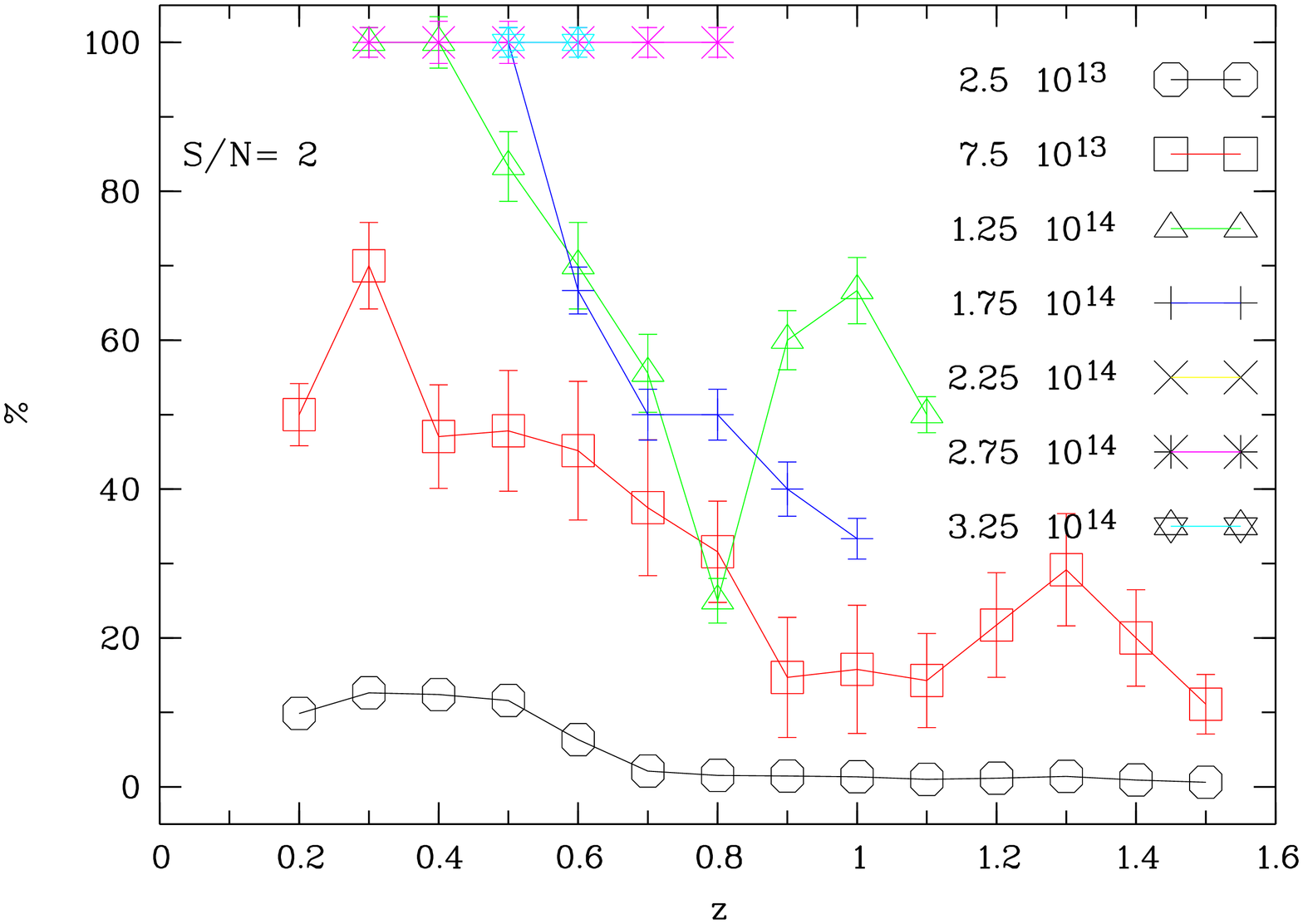,width=6.cm,angle=270}}
\mbox{\psfig{figure=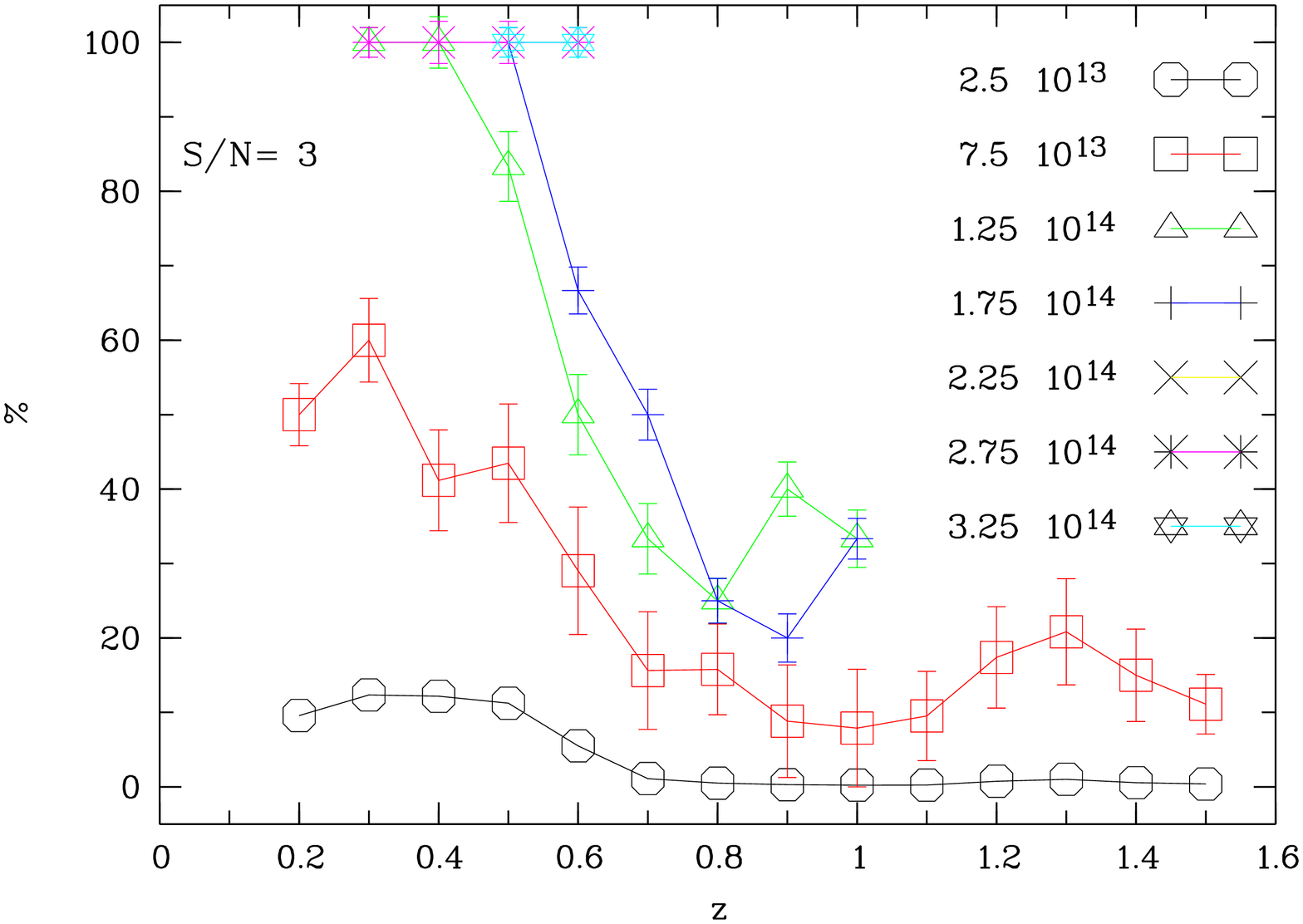,width=6.cm,angle=270}}
\mbox{\psfig{figure=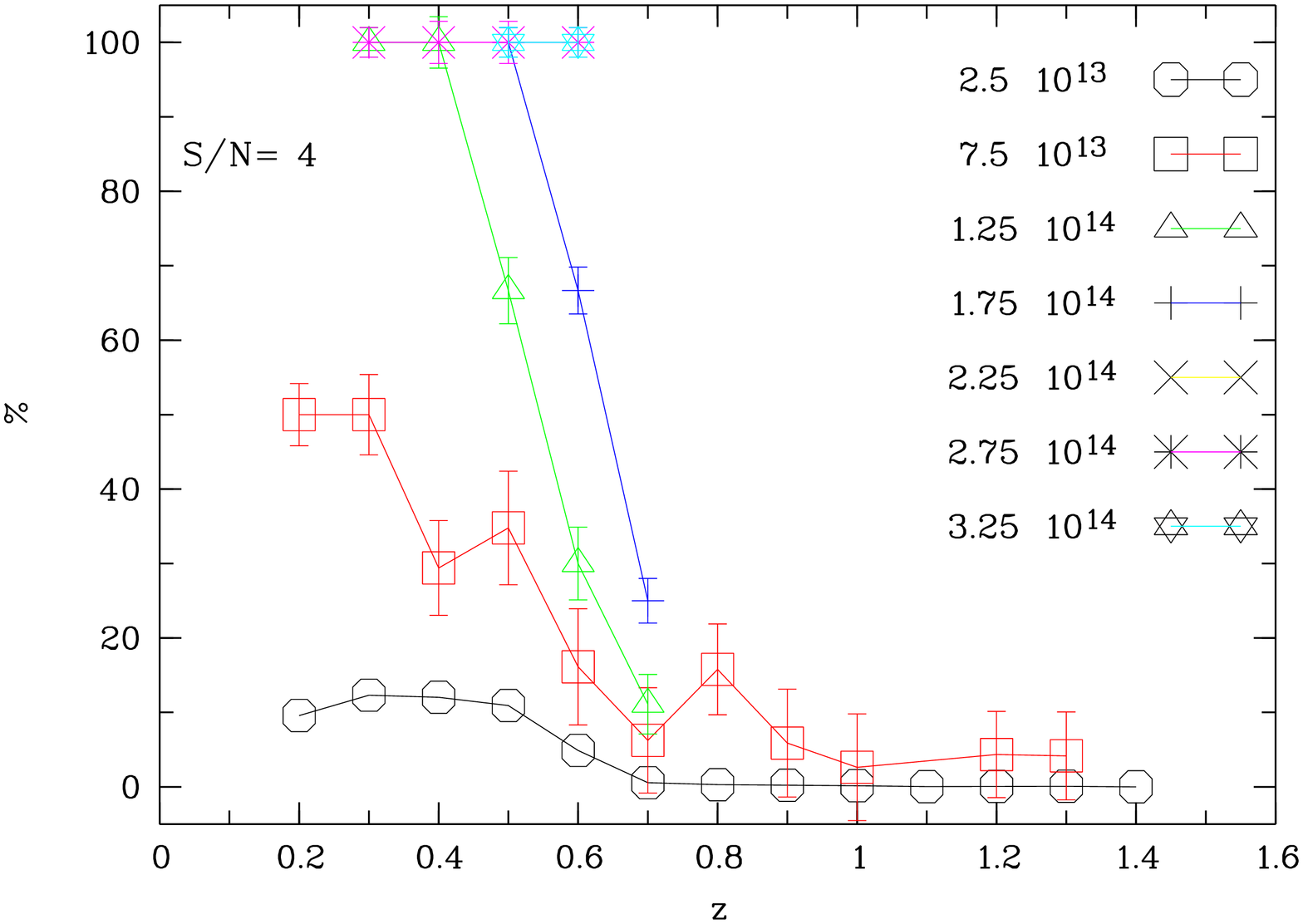,width=6.cm,angle=270}}
\mbox{\psfig{figure=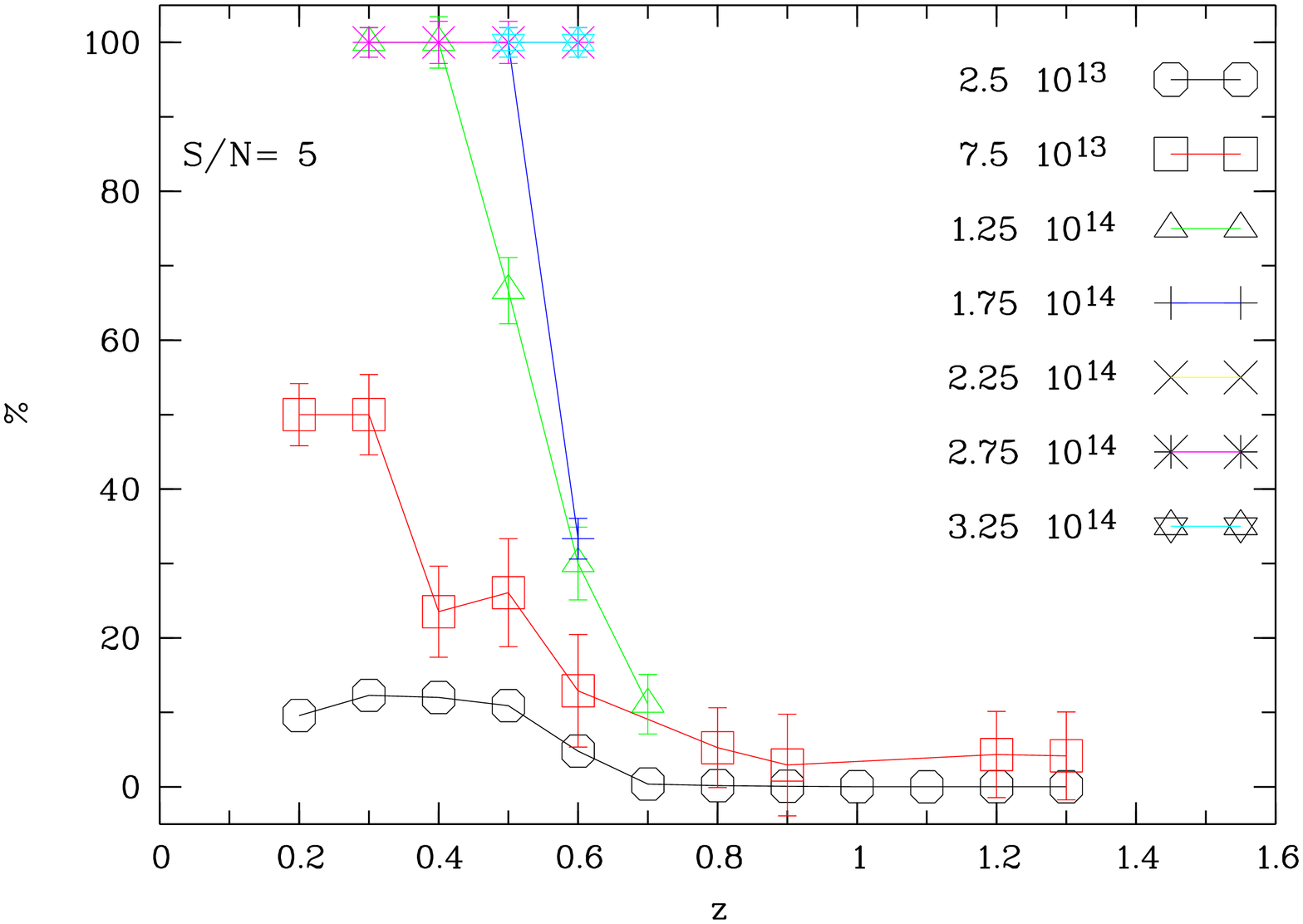,width=6.cm,angle=270}}
\mbox{\psfig{figure=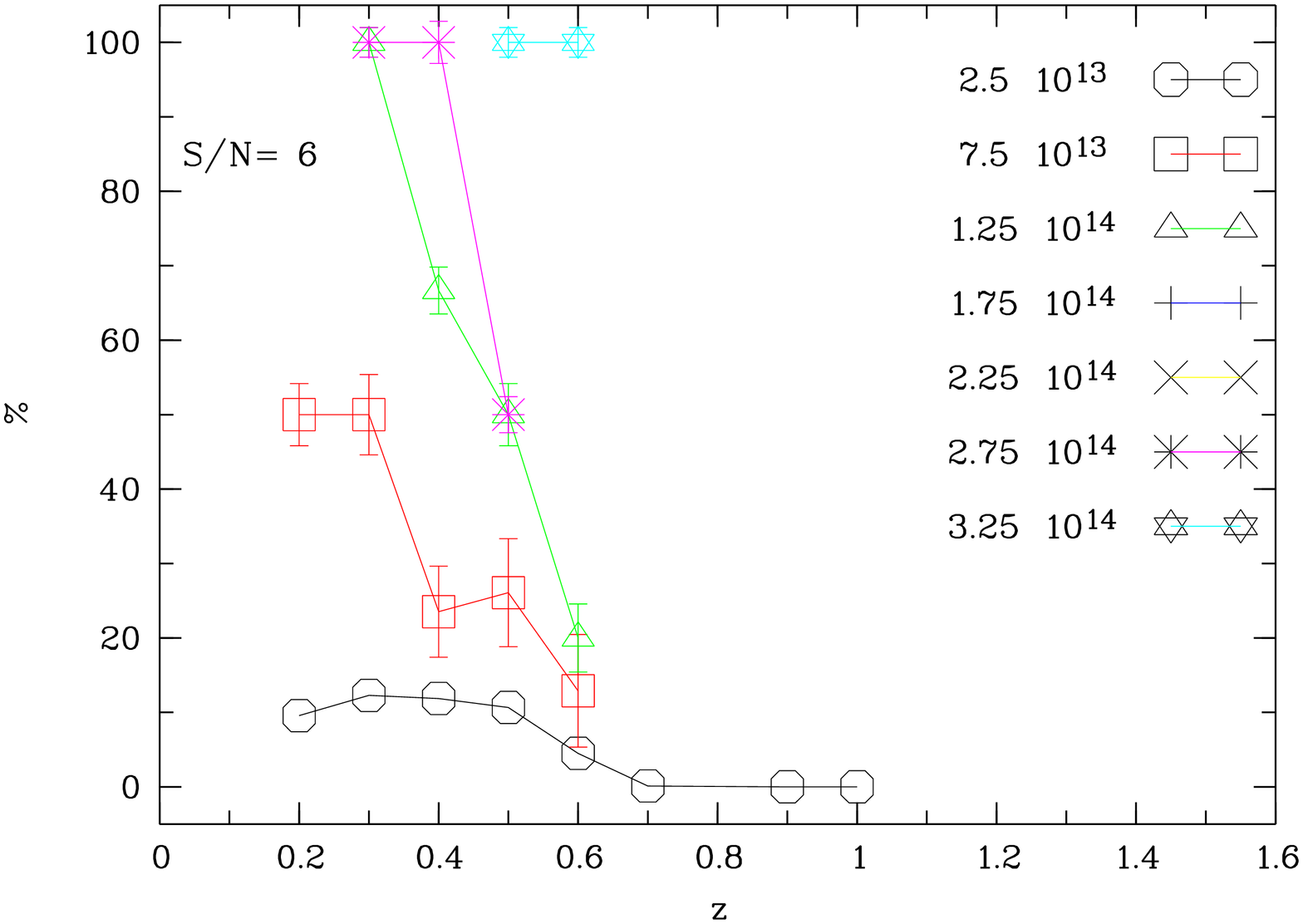,width=6.cm,angle=270}}
\caption[]{Percentages of all Millennium halos included in real detections
  for the Deep survey characteristics as a function of redshift for
  different mass limits. }
\label{fig:detpd}
\end{figure}

Considering the criterion described above, we are able to give success
rates in the detection process. Figs.~\ref{fig:detpw} and
\ref{fig:detpd} show the percentages of all Millennium halos identified with a
real $detection$ as a function of mass.

We see that for the Wide survey, we are only able to detect with a
success rate greater than $\sim$20$\%$ the Millennium halos more
massive than 7.5 10$^{13}$ M$_\odot$. The detection rates become quite
low at z$\geq$0.6-0.7. For the Deep survey, we also detect halos more
massive than 7.5 10$^{13}$ M$_\odot$ but up to z$\geq$0.9.

\begin{figure}[hbt]
\centering
\mbox{\psfig{figure=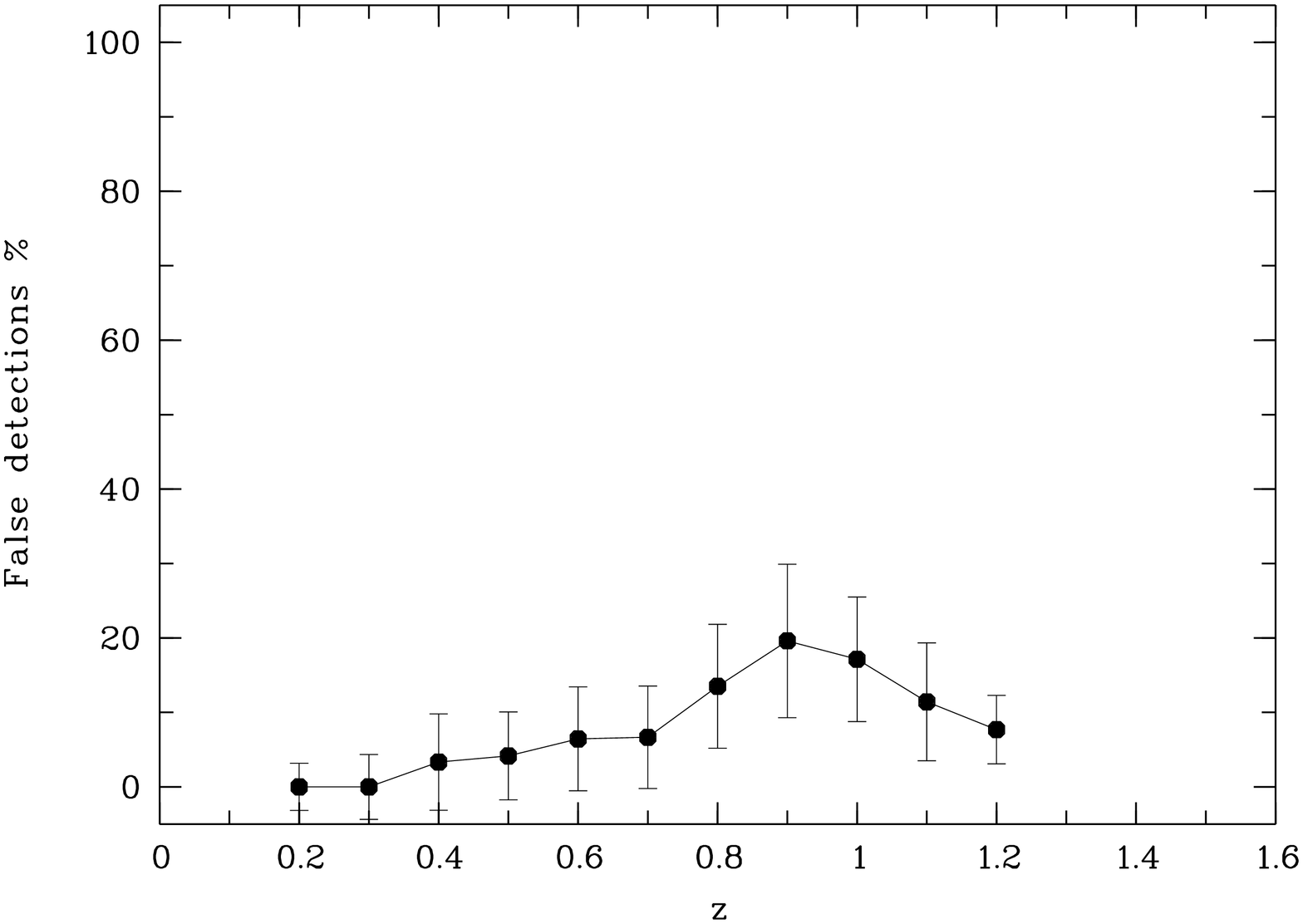,width=6.cm,angle=270}}
\mbox{\psfig{figure=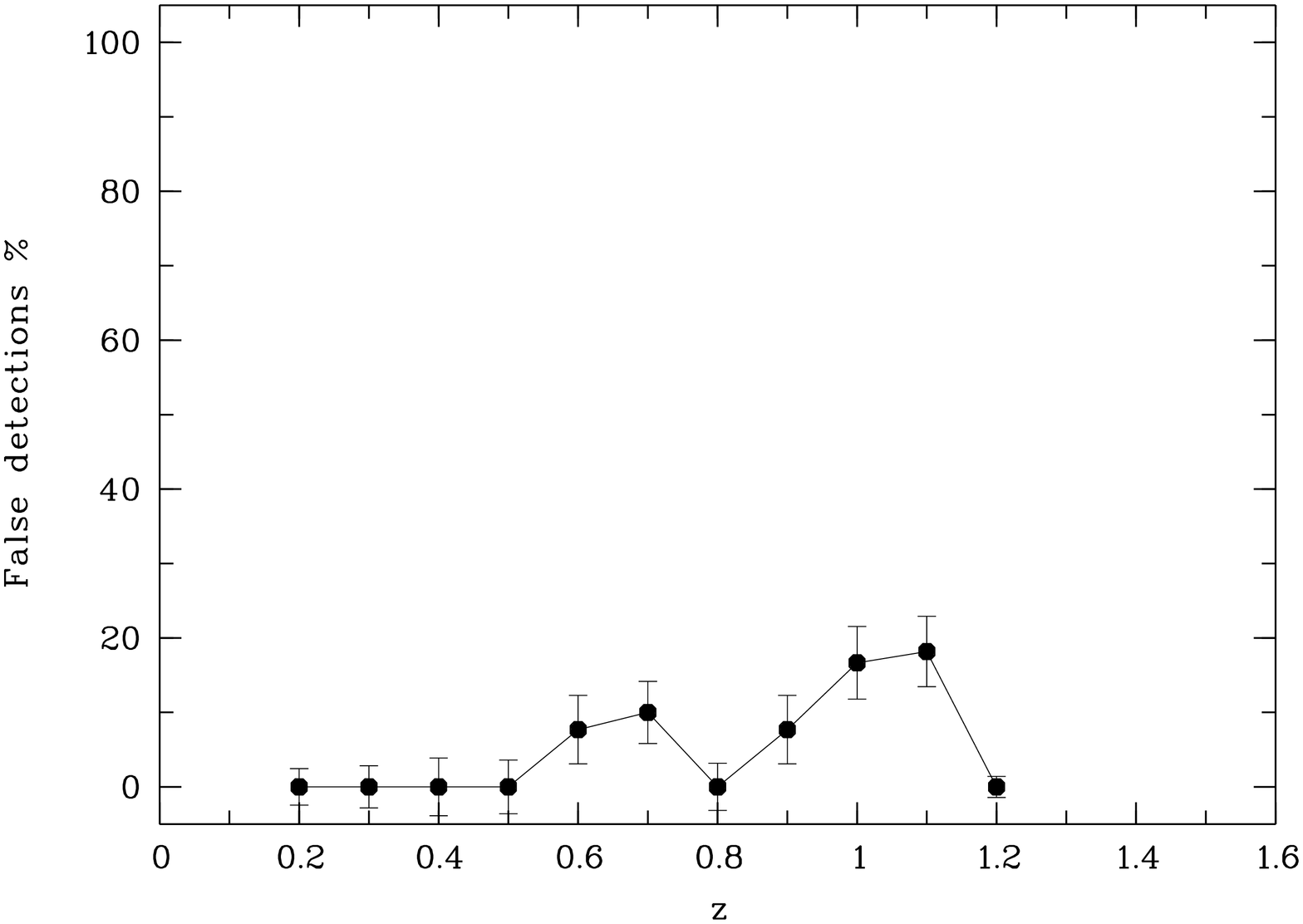,width=6.cm,angle=270}}
\mbox{\psfig{figure=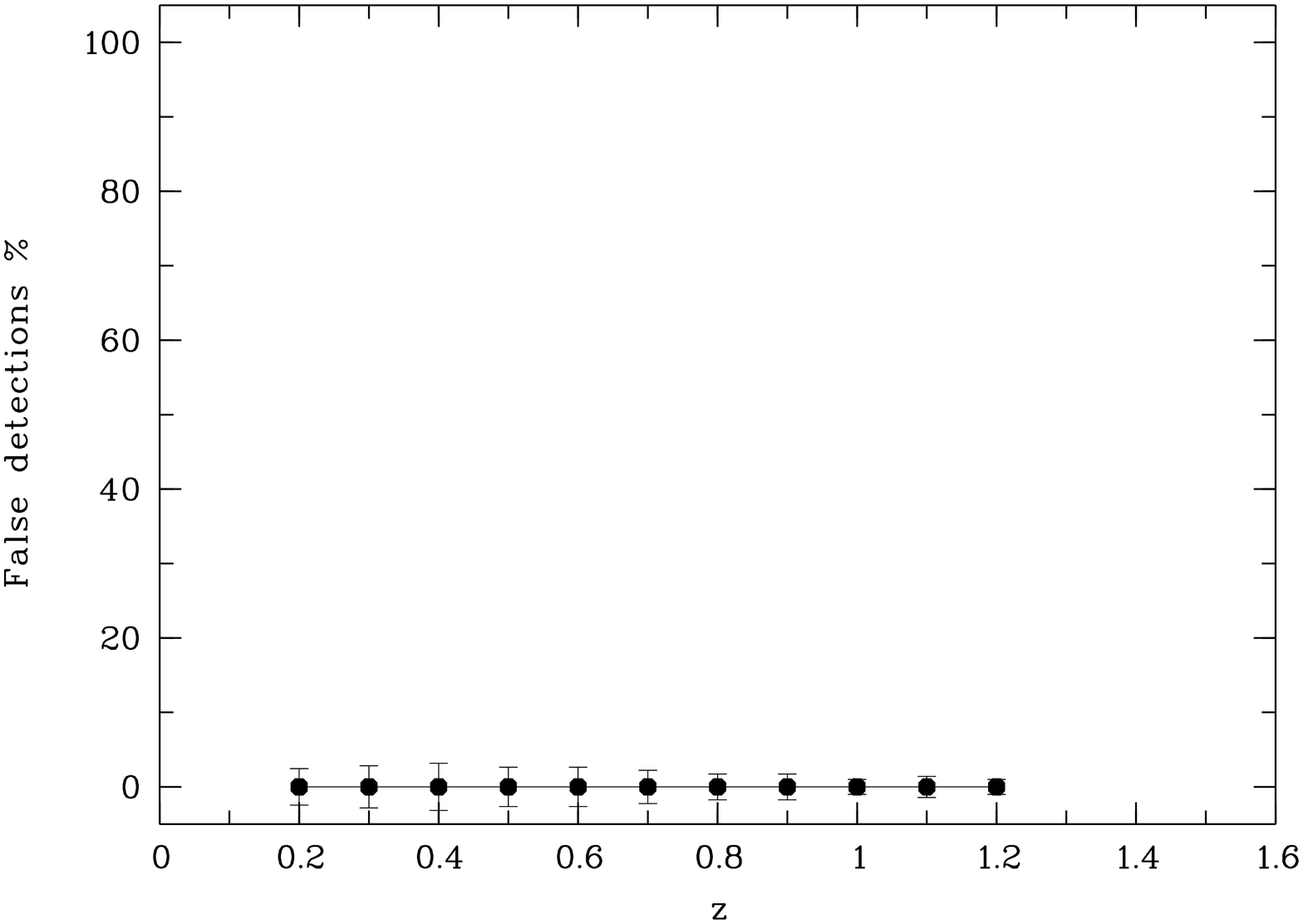,width=6.cm,angle=270}}
\mbox{\psfig{figure=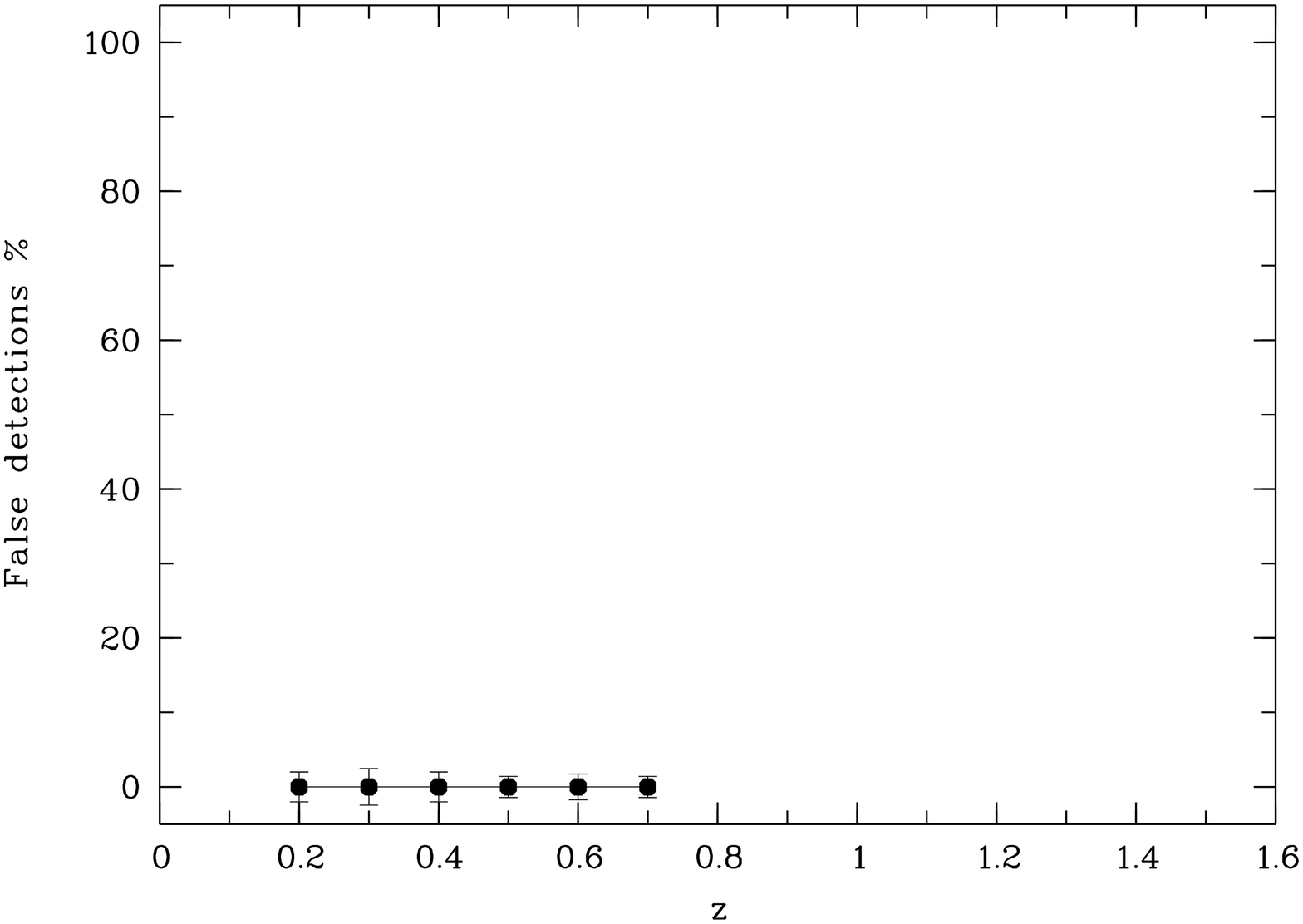,width=6.cm,angle=270}}
\mbox{\psfig{figure=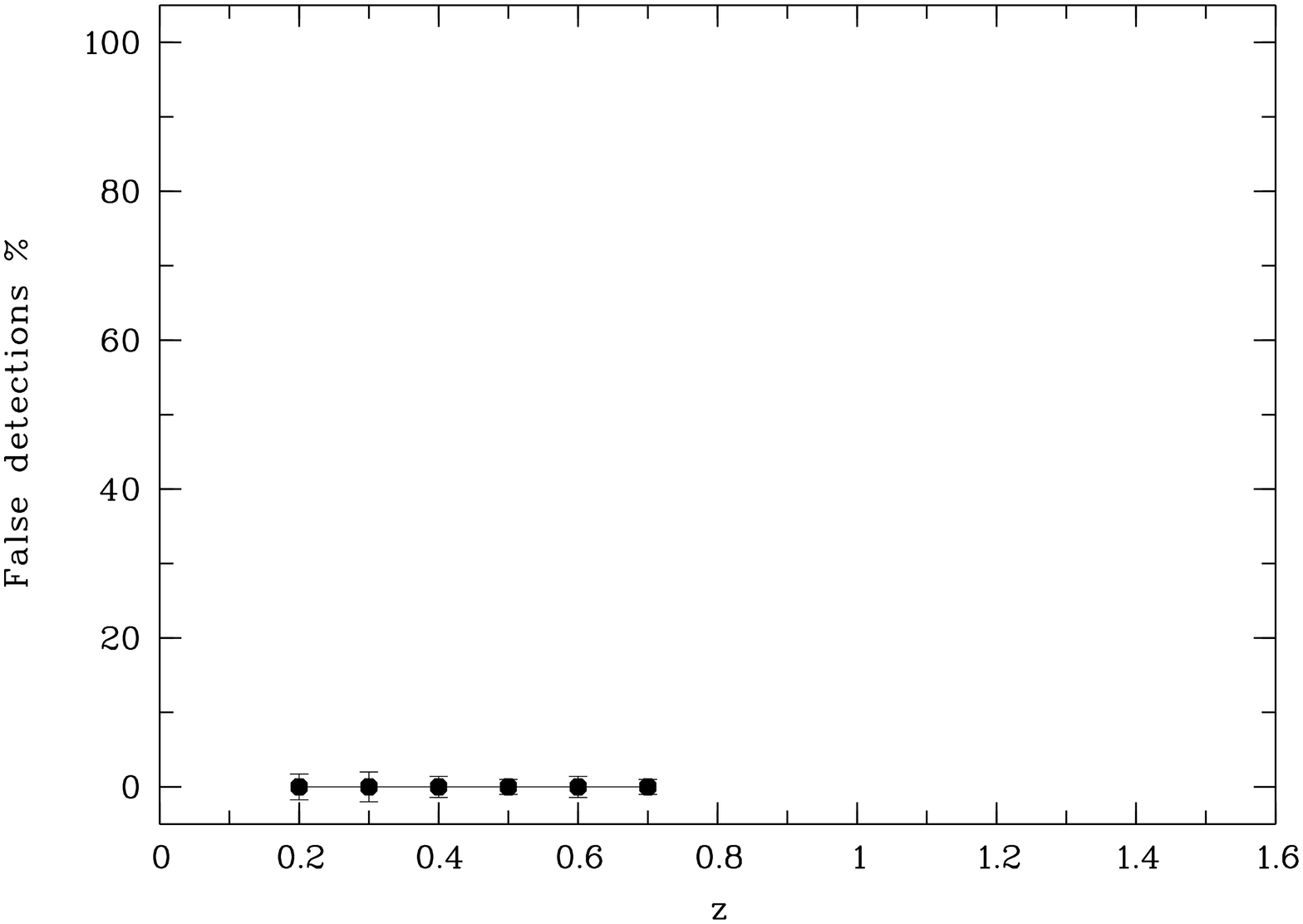,width=6.cm,angle=270}}
\caption[]{Percentages of fake $detections$ for the
  Wide survey characteristics as a function of redshift. From top
to bottom: S/N of 2, 3, 4, 5, and 6.}
\label{fig:fake23}
\end{figure}

\begin{figure}[hbt]
\centering
\mbox{\psfig{figure=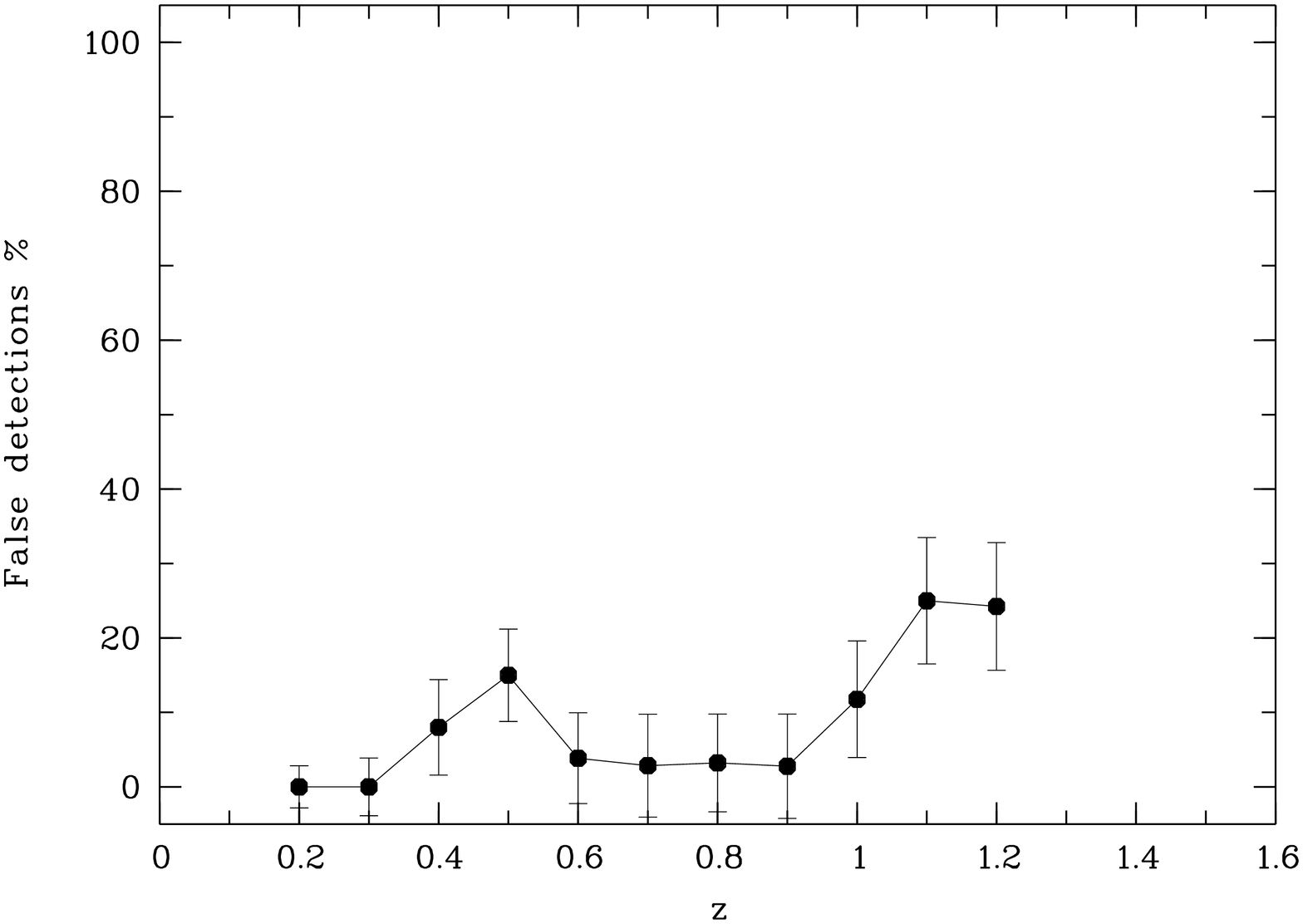,width=6.cm,angle=270}}
\mbox{\psfig{figure=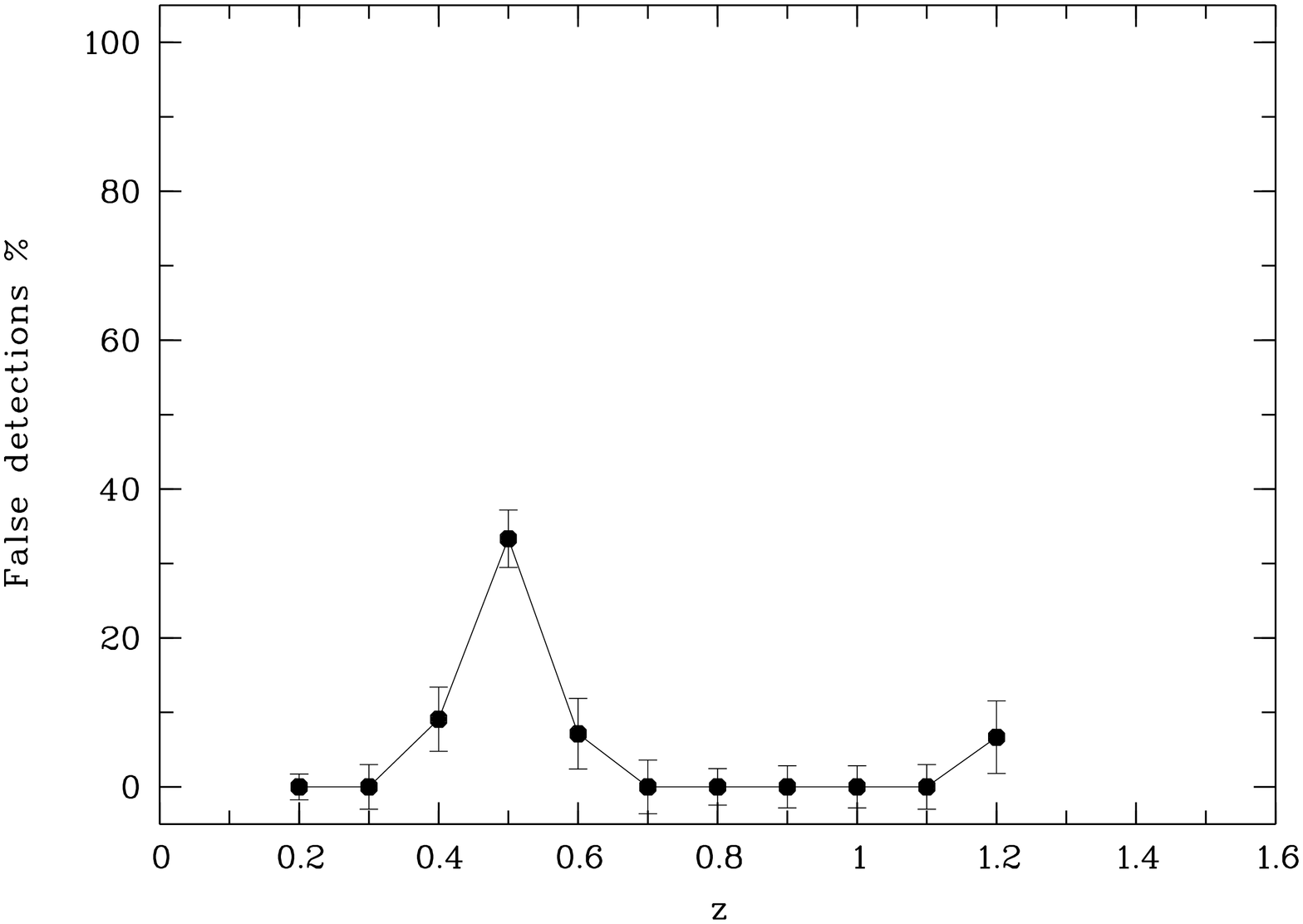,width=6.cm,angle=270}}
\mbox{\psfig{figure=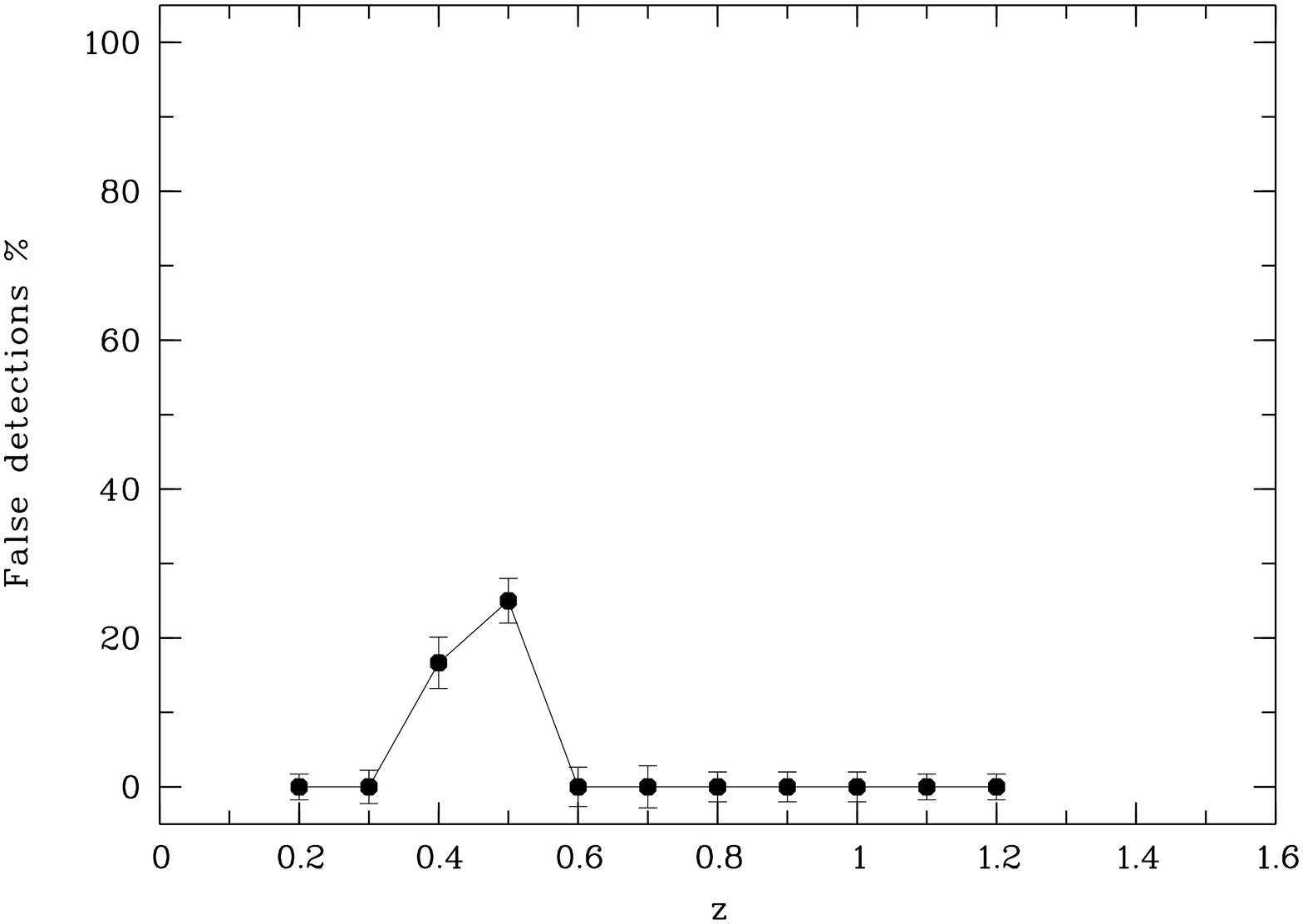,width=6.cm,angle=270}}
\mbox{\psfig{figure=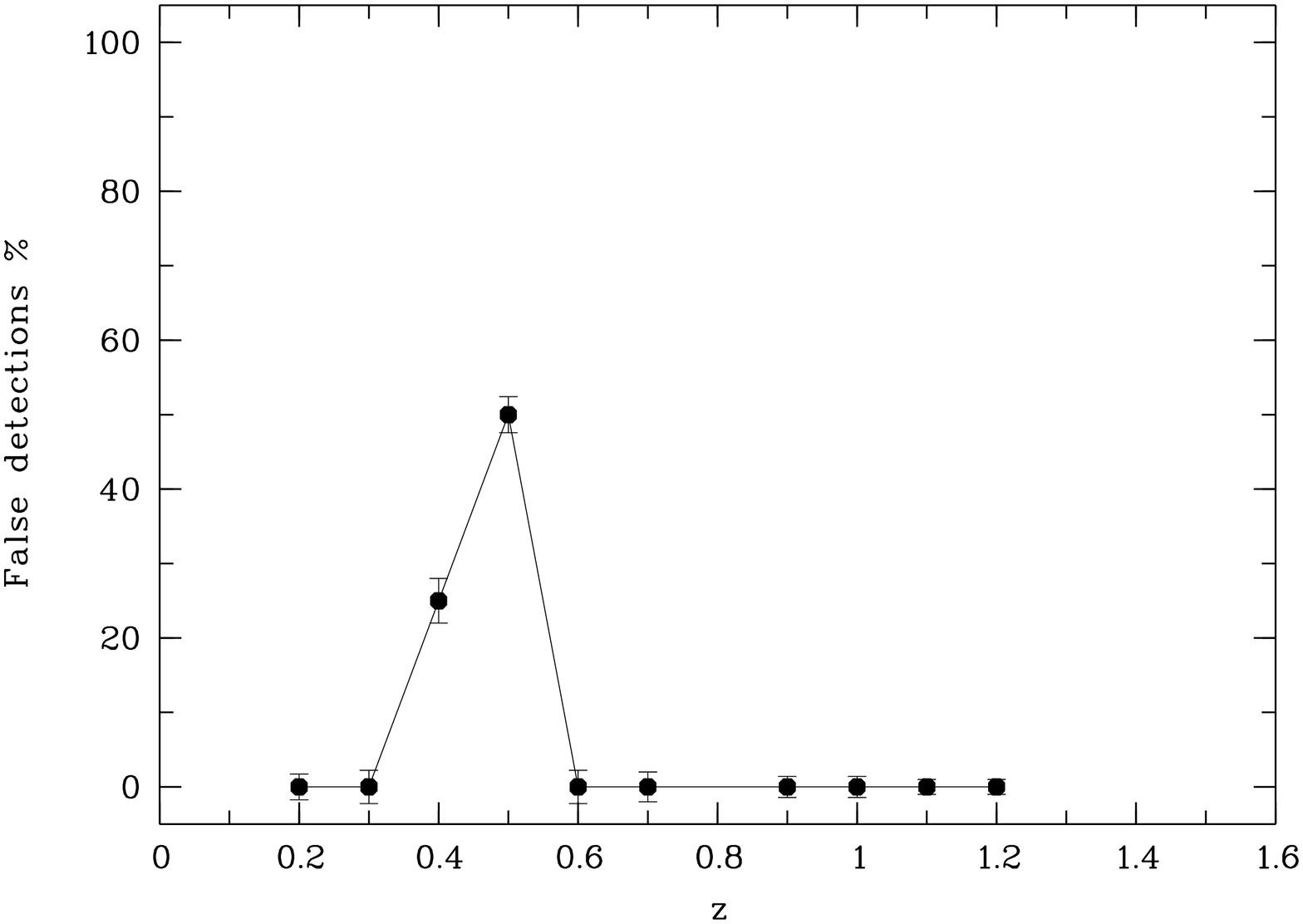,width=6.cm,angle=270}}
\mbox{\psfig{figure=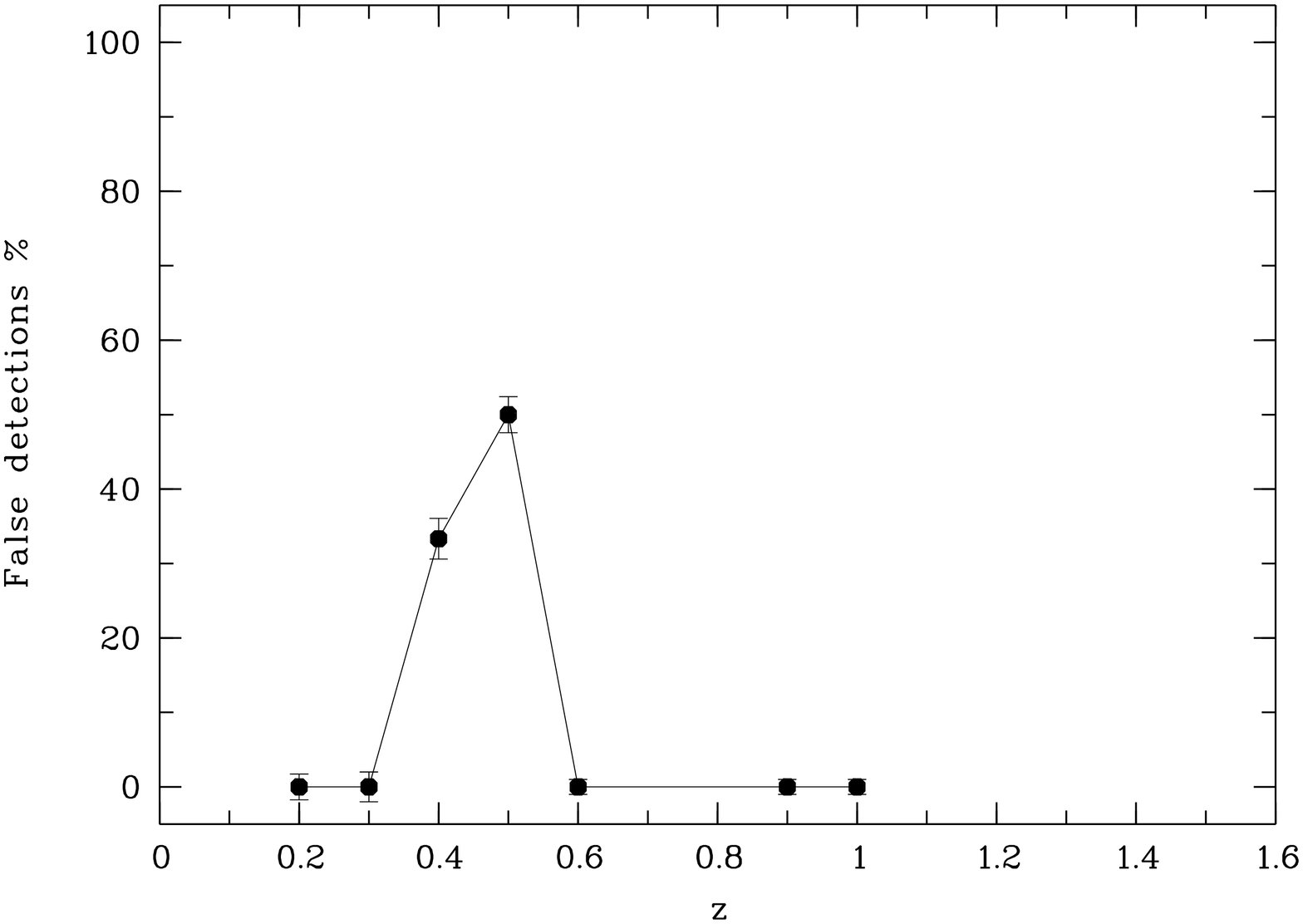,width=6.cm,angle=270}}
\caption[]{Percentages of fake $detections$ for the
  Deep survey characteristics as a function of redshift. From top
to bottom: S/N of 2, 3, 4, 5, and 6.}
\label{fig:fake25}
\end{figure}

We give in Figs.~\ref{fig:fake23} and \ref{fig:fake25} the percentages of
fake $detections$ (following the criterion previously exposed) as a
function of S/N detection and redshift for the Wide and Deep
surveys. For the Wide survey, false $detection$ rates are basically null
for S/N$\geq$4 and remain small for S/N$\leq$3 and z$\leq$0.8. For the
Deep survey, false $detection$ rates are small whatever the S/N and for
z$\leq$1. We note however an unexpected local increase of this rate at
z$\sim$0.5 which is perhaps due to degeneracies in photometric
redshift estimates producing artificial clustering signal for example because of
the discreteness of the templates.

As a compromise between detection rate and false detection rate, we
chose to not perform detections at S/N lower than 2. We could
  also have limited our catalogs to S/N$\geq$3 $detections$ to have a
  more robust sample. However, only about 10$\%$ of the S/N=[2;3[
  $detections$ are fake and this percentage decreases to 6$\%$ for the
  S/N=[3;4[ $detections$. At the same time, the number of $detections$
  is multiplied by $\sim$2.8 between S/N=[3;4[ and S/N=[2;3[. The
  number of real $detections$ therefore grows faster than the number
  of fake $detections$. Hence, considering S/N=[2;3[ $detections$
  allows to include numerous real clusters as well as to keep fake
  $detections$ to a level lower than 10$\%$. 

\begin{figure}[hbt]
\centering
\mbox{\psfig{figure=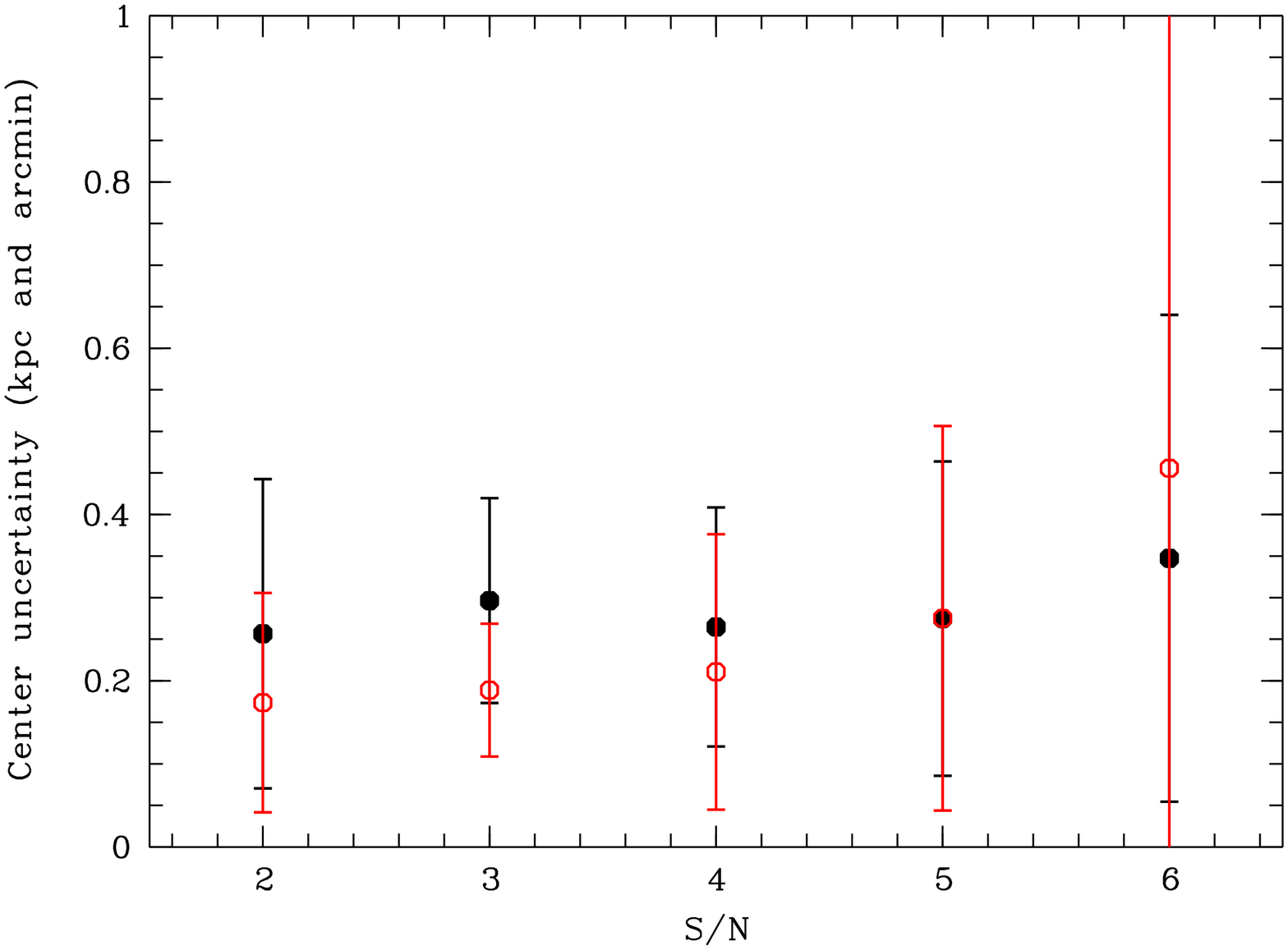,width=8.cm,angle=270}}
\mbox{\psfig{figure=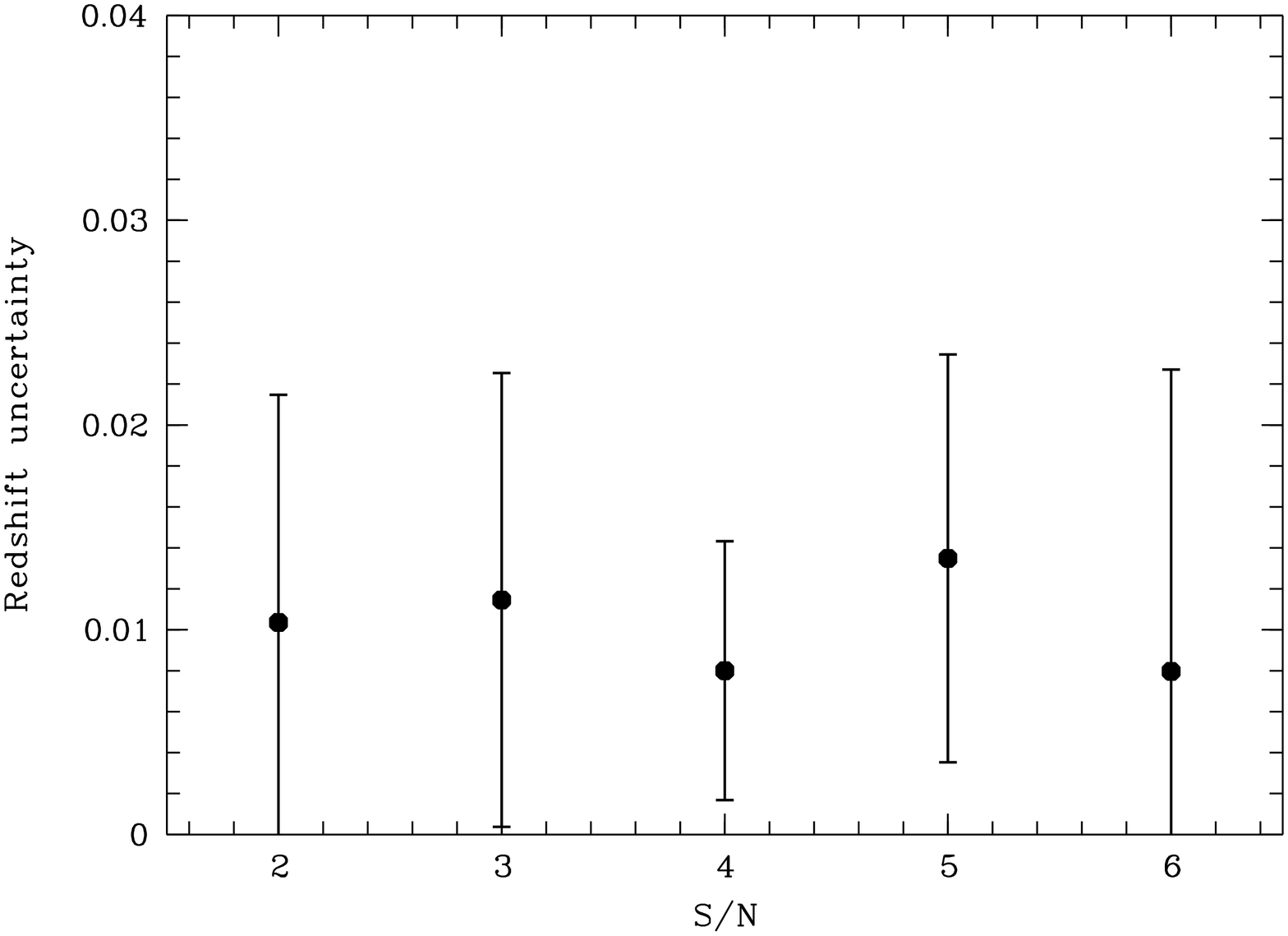,width=8.cm,angle=270}}
\caption[]{CFHTLS Wide-like Millennium simulation. Upper figure: shifts between true and estimated centers of the
  massive structures in the Millennium simulation as a function of the
  detection S/N. Open red circles with error bars are the values given
  in kpc, and filled black circles with error bars are the values given
  in arcmin. Lower figure: shifts between true and estimated redshifts of the
  massive structures in the Millennium simulation as a function of
  detection S/N.}
\label{fig:stat23}
\end{figure}

\begin{figure}[hbt]
\centering
\mbox{\psfig{figure=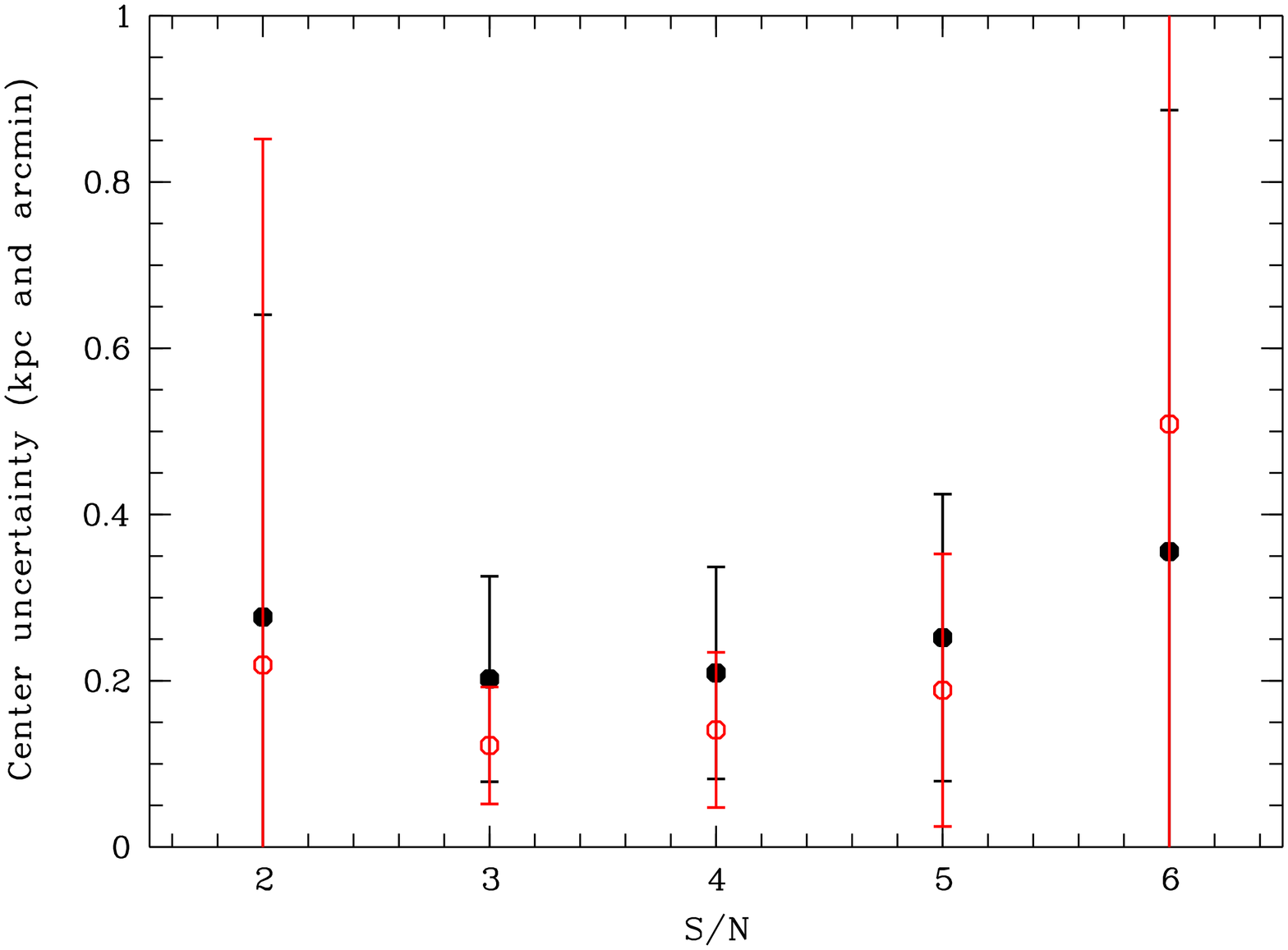,width=8.cm,angle=270}}
\mbox{\psfig{figure=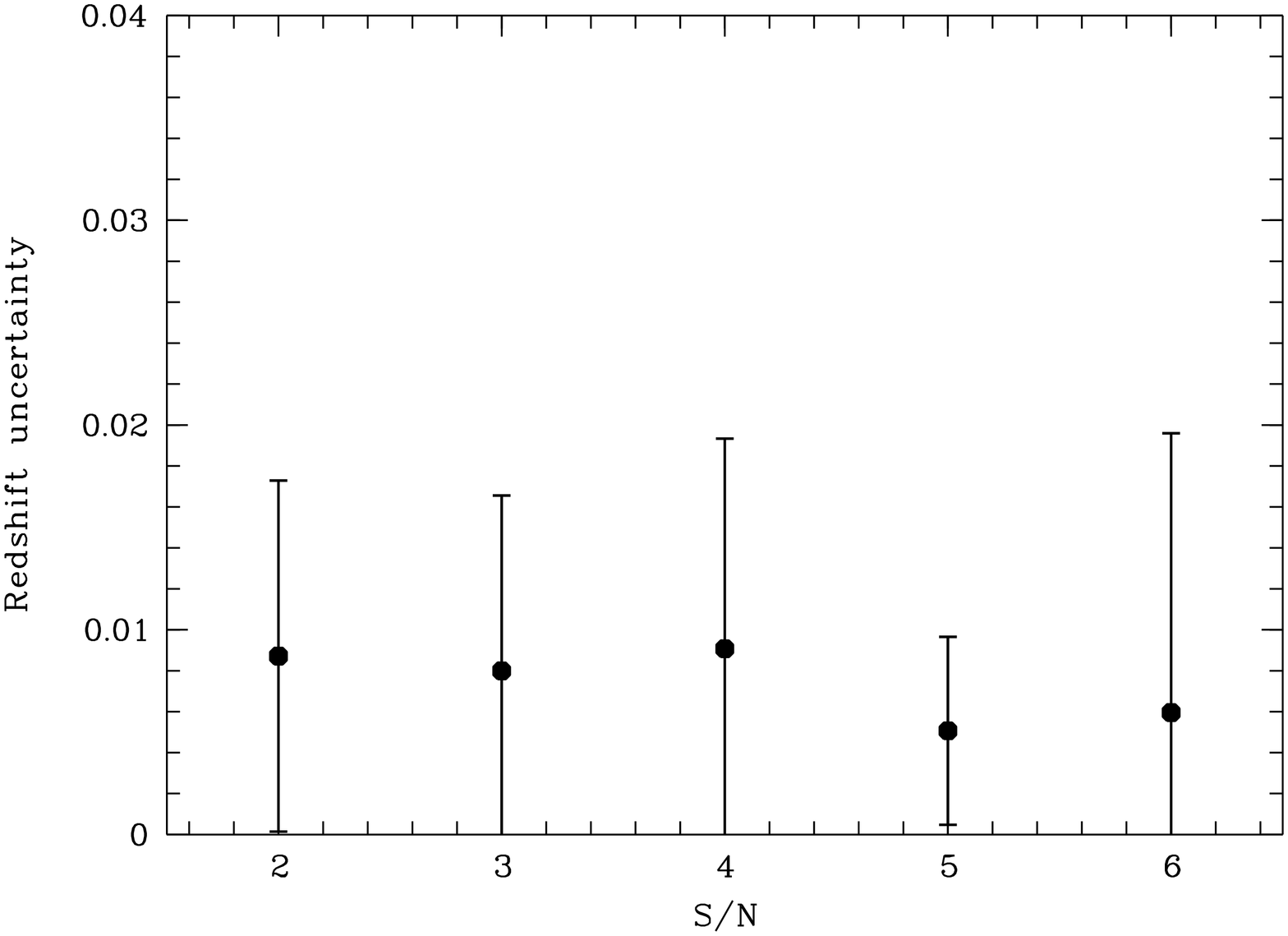,width=8.cm,angle=270}}
\caption[]{CFHTLS Deep-like Millennium simulation. Upper figure: shifts between true and estimated centers of the
  massive structures in the Millennium simulation as a function of the
  detection S/N. Open red circles with error bars are the values given
  in kpc, and filled black circles with error bars are the values given
  in arcmin. Lower figure: shifts between true and estimated redshifts of the
  massive structures in the Millennium simulation as a function of
  detection S/N.}
\label{fig:stat25}
\end{figure}

We now evaluate the coordinates and redshift precision of our
$detections$. Using the Millennium simulation, we detected 97 candidate
clusters at S/N=2, 27 at S/N=3, 16 at S/N=4, 4 at S/N=5, and 7 at
S/N=6, which turned out to be massive halos in the Millennium
simulation. With these detections, we estimated that the typical
uncertainty on the candidate cluster coordinates was smaller than
$\sim$1 arcmin or $\sim$0.5 kpc and smaller than 0.025 in redshift
(see Figs.~\ref{fig:stat23} and ~\ref{fig:stat25}). We note that
several massive Millennium halos can be identified with a single
$detection$ (see also next section). To compute the statistics given in
Figs.~\ref{fig:stat23} and ~\ref{fig:stat25}, we considered the
Millennium halo which is the closest to the considered
$detection$. We can note that the error bars tend to increase with
S/N. For the top figure, this can be explained by the fact that at
high S/N we detect massive clusters, which are often heavily
substructured and therefore the definition of their center is not
straightforward.

We also give a mass estimate based on the photometry. Each candidate
cluster is detected at a given $S/N$. This detection threshold is a
rough estimate of the cluster richness, being simply the net $flux$ of
the source, i.e. the number of galaxies in the source diminished of
the background level. The criterion is not precise enough (we only
took detection thresholds in steps of 1) and luminosity function based
richnesses would be better tracers of the total mass of the
structures. However, this criterion allows to give a minimal mass for
a detected structure. Table~\ref{tab:minmass} therefore gives the
relation between this detection threshold and the minimal cluster
mass. We clearly see that when the detection threshold increases, the
minimal mass also increases, both for the Deep and the Wide surveys.

\begin{table*}
\caption{Relation between the SExtractor detection threshold and the minimal
  and mean (over all the associated Millennium halos) cluster masses.}
\begin{center}
\begin{tabular}{rrrrr}
\hline
SExtractor detection threshold & Minimal mass    & Mean mass       & Minimal mass    & Mean mass       \\
                              & Wide (M$_\odot$) & Wide (M$_\odot$) & Deep (M$_\odot$) & Deep (M$_\odot$) \\
\hline
2 & 1.0 10$^{13}$ & 1.3 10$^{14}$ & 0.4 10$^{13}$   & 1.4 10$^{14}$ \\
3 & 1.3 10$^{13}$ & 1.8 10$^{14}$ & 0.8 10$^{13}$   & 2.1 10$^{14}$ \\
4 & 3.3 10$^{13}$ & 1.8 10$^{14}$ & 2.4 10$^{13}$   & 1.9 10$^{14}$ \\
5 & 3.5 10$^{13}$ & 1.3 10$^{14}$ & 7.7 10$^{13}$   & 3.5 10$^{14}$ \\
6 & 5.5 10$^{13}$ & 12.6 10$^{14}$ & 6.3 10$^{13}$  & 10.3 10$^{14}$ \\
\hline
\end{tabular}
\end{center}
\label{tab:minmass}
\end{table*}

\subsection{Level of substructuring}

\begin{figure}[hbt]
\centering
\mbox{\psfig{figure=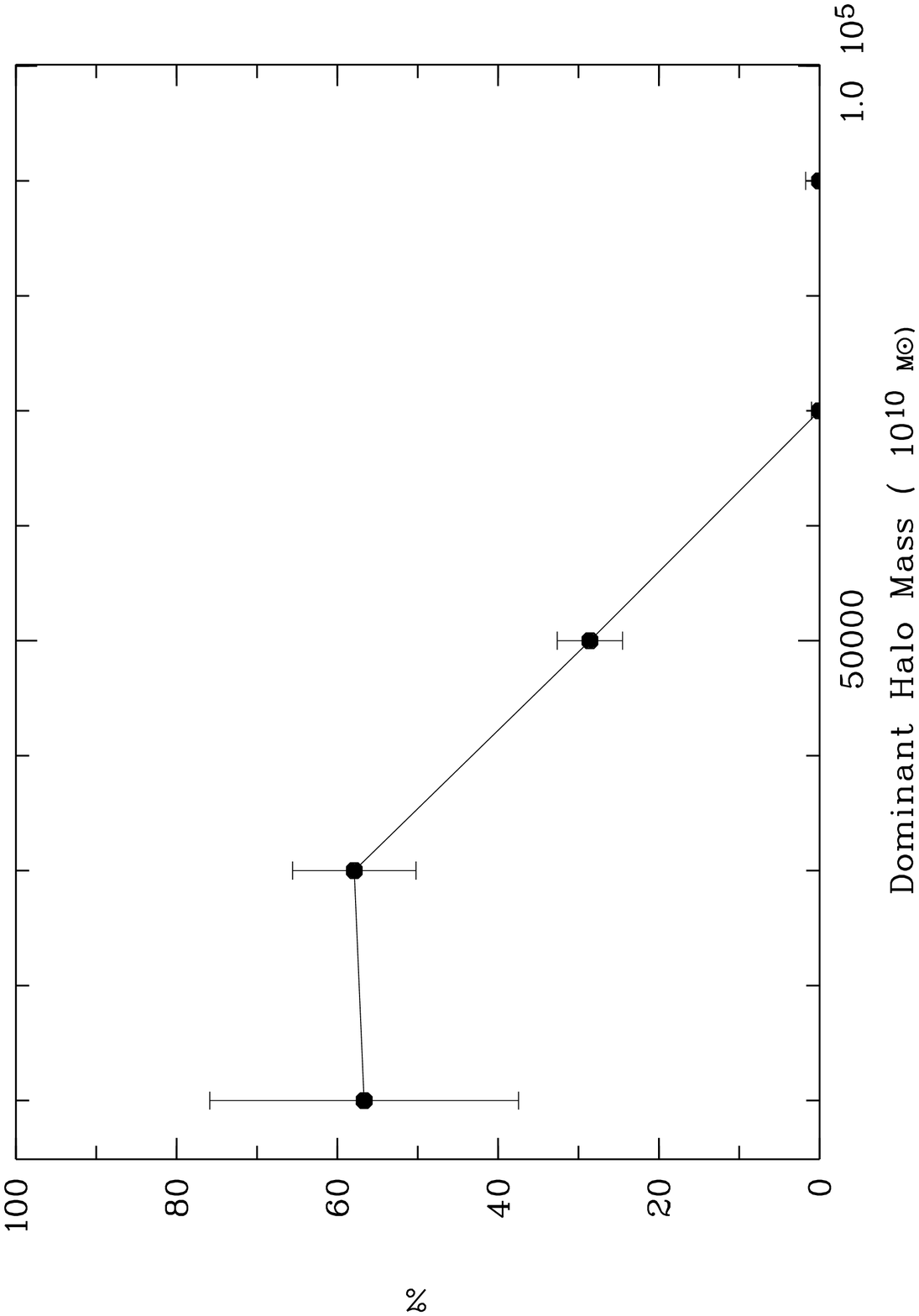,width=8.cm,angle=270}}
\mbox{\psfig{figure=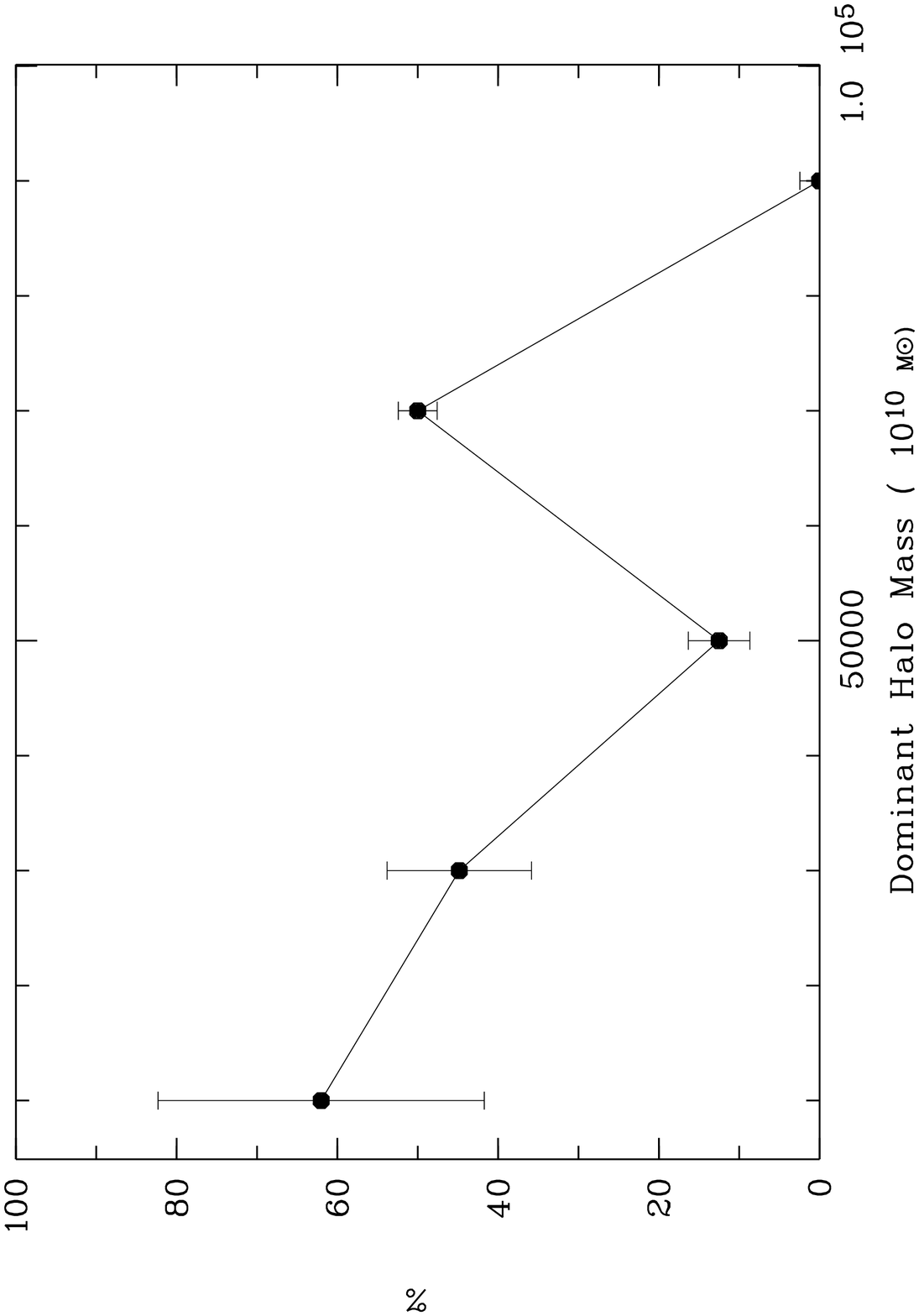,width=8.cm,angle=270}}
\caption[]{Percentage of $detections$ (S/N$\geq$2) with a ratio between the total included
  mass and the most massive halo mass lower than 3, as a function of  total
mass. Upper figure: Wide survey characteristics, lower figure: Deep survey characteristics.}
\label{fig:multi}
\end{figure}

We now ask the question of the substructure level of our
$detections$. As already explained, for a single $detection$, we have
most of the time several attached Millennium halos. Each of these
halos can be considered as a potential substructure of the
$detection$. The question is to know what is this level of
substructure. We therefore chose to compute for a given $detection$
the ratio of the total mass included in our detection to the mass of
the most massive Millennium halo included in the $detection$. We plot
in Fig.~\ref{fig:multi} the percentage of $detections$ (S/N$\geq$2)
for which the mass of the most massive included halo is at least 1/3
of the total mass (therefore with a low expected substructure
  level), as a function of the total mass. We see that in the Wide
survey, halos with total mass lower than 5 10$^{14}$ M$_\odot$ are not
strongly substructured while more massive $detections$ are strongly
substructured. In the Deep survey, the general tendency is similar.

When we investigated the potential effect of the S/N threshold on the
substructure level, we did not find any significant tendency for S/N$\geq$3.

\begin{figure}[hbt]
\centering
\mbox{\psfig{figure=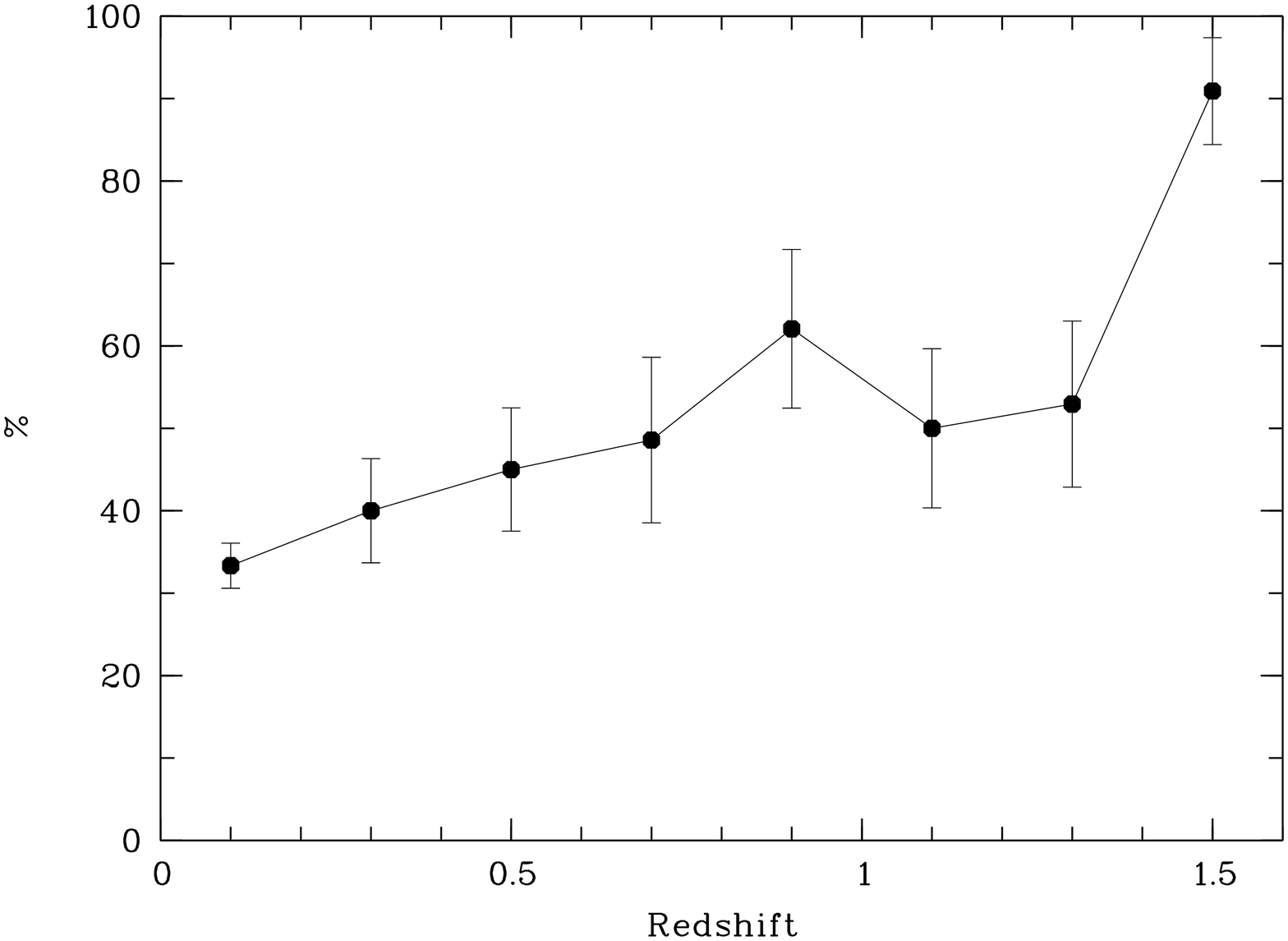,width=8.cm,angle=270}}
\caption[]{Percentage of $detections$ (S/N$\geq$2) with a ratio between the total included
  mass and the most massive halo mass lower than 3, as a function of redshift
 for the Deep survey characteristics.}
\label{fig:redshift}
\end{figure}

  According to the hierarchical structure growth assumed in the
  Millennium simulation, we could also expect a higher level of
  substructures with increasing redshift. We do not detect a very
  clear tendency in the Millenium Wide-like catalog, but
  Fig.~\ref{fig:redshift} (S/N$\geq$2) seems to show a reverse
  behaviour for the Millenium Deep-like survey: high redshift
  $detections$ appear less substructured than nearby ones.  However
  this is at least partially a selection effect explained by the fact
  that high redshift $detections$ are preferentially low mass
  structures, less substructured than high mass $detections$ by
  definition.

\section{Spatial and redshift $detection$ distributions}

\subsection{$Detection$ counts}

Tables~\ref{tab:candD2}, \ref{tab:candD3}, \ref{tab:candD4},
\ref{tab:candW1}, \ref{tab:candW3}, and \ref{tab:candW4}, give the
$detections$ with their coordinates, redshift, and S/N
for the Deep and Wide fields.

\begin{figure}[hbt]
\centering
\mbox{\psfig{figure=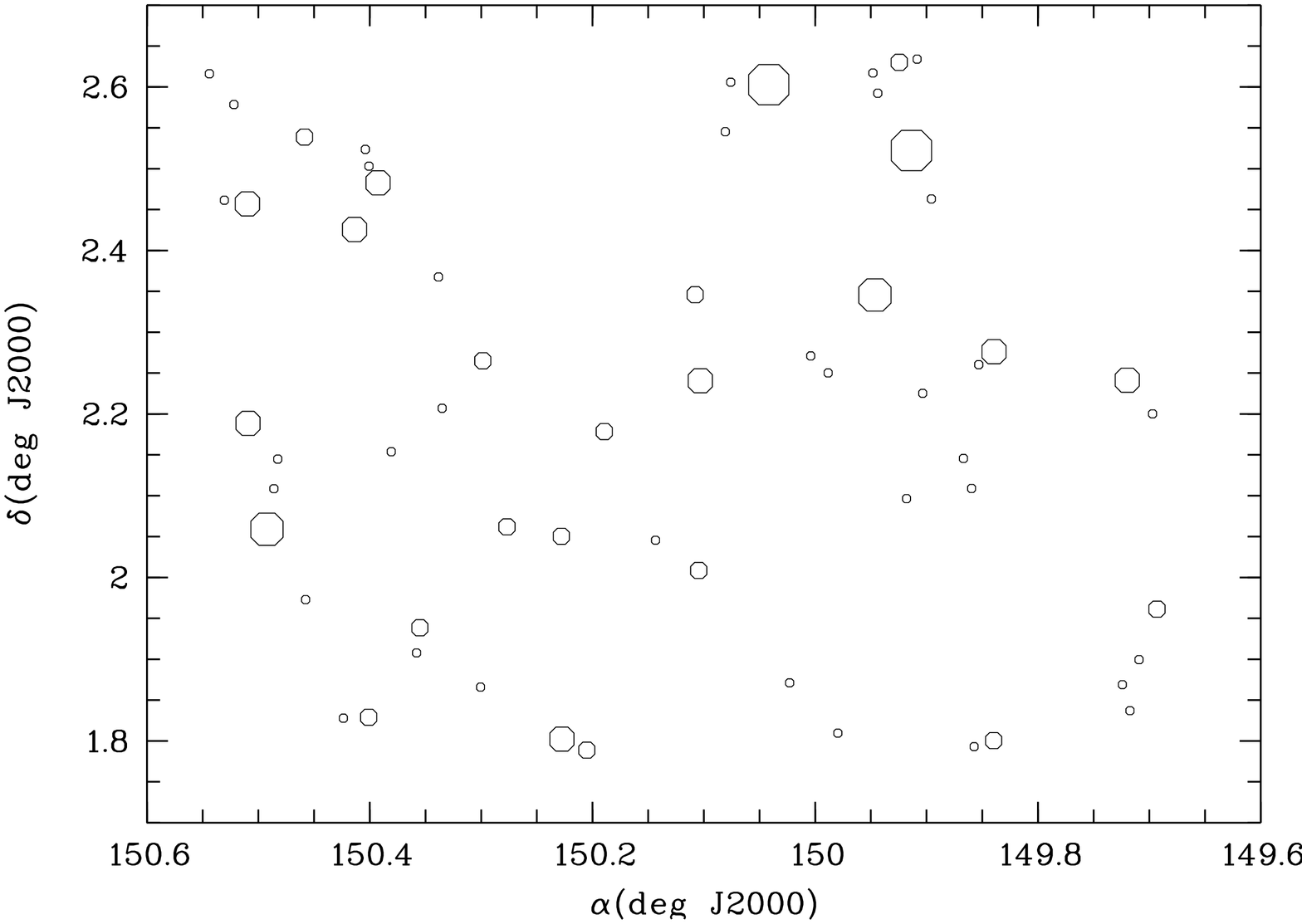,width=8.cm,angle=270}}
\mbox{\psfig{figure=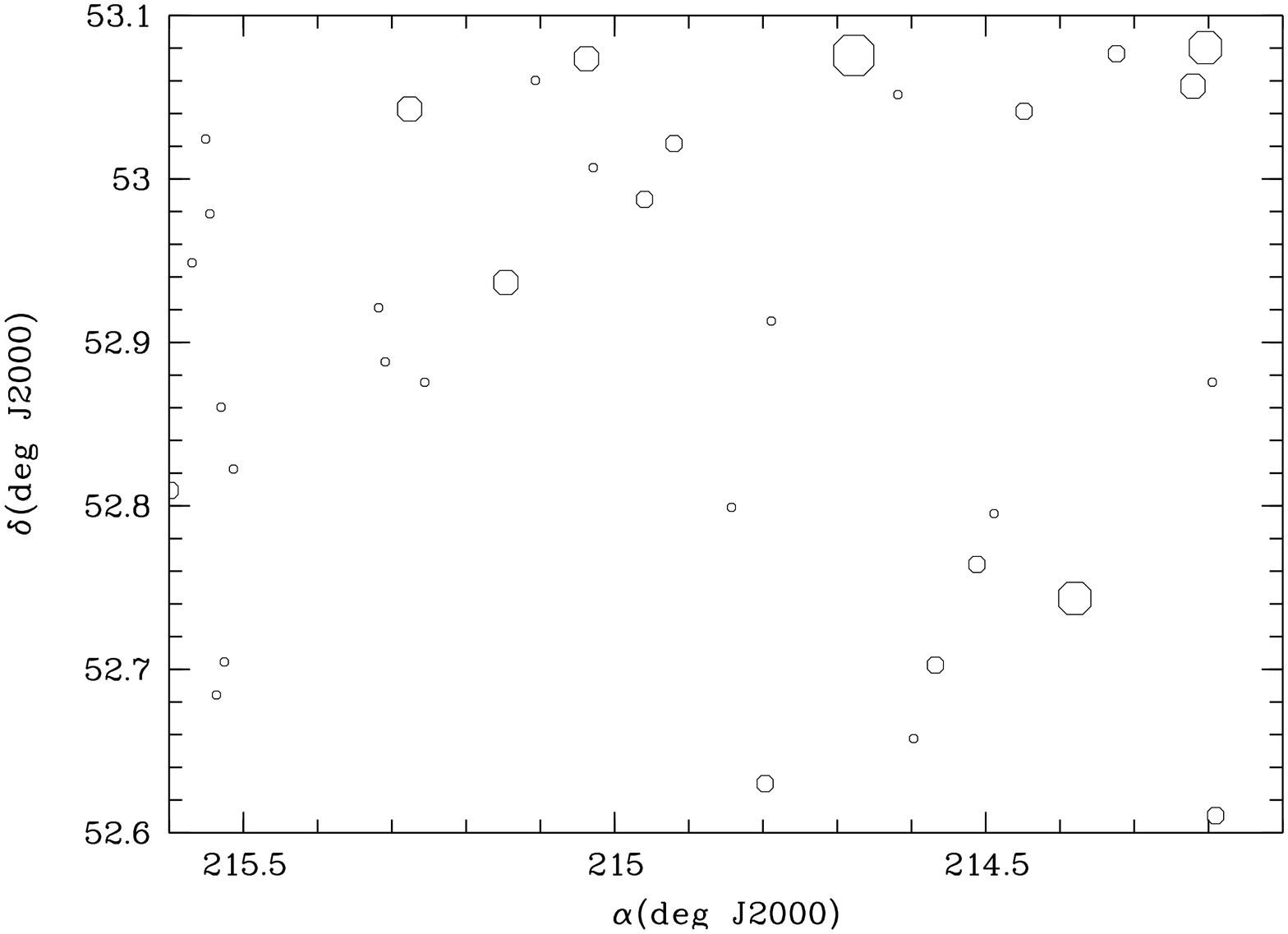,width=8.cm,angle=270}}
\mbox{\psfig{figure=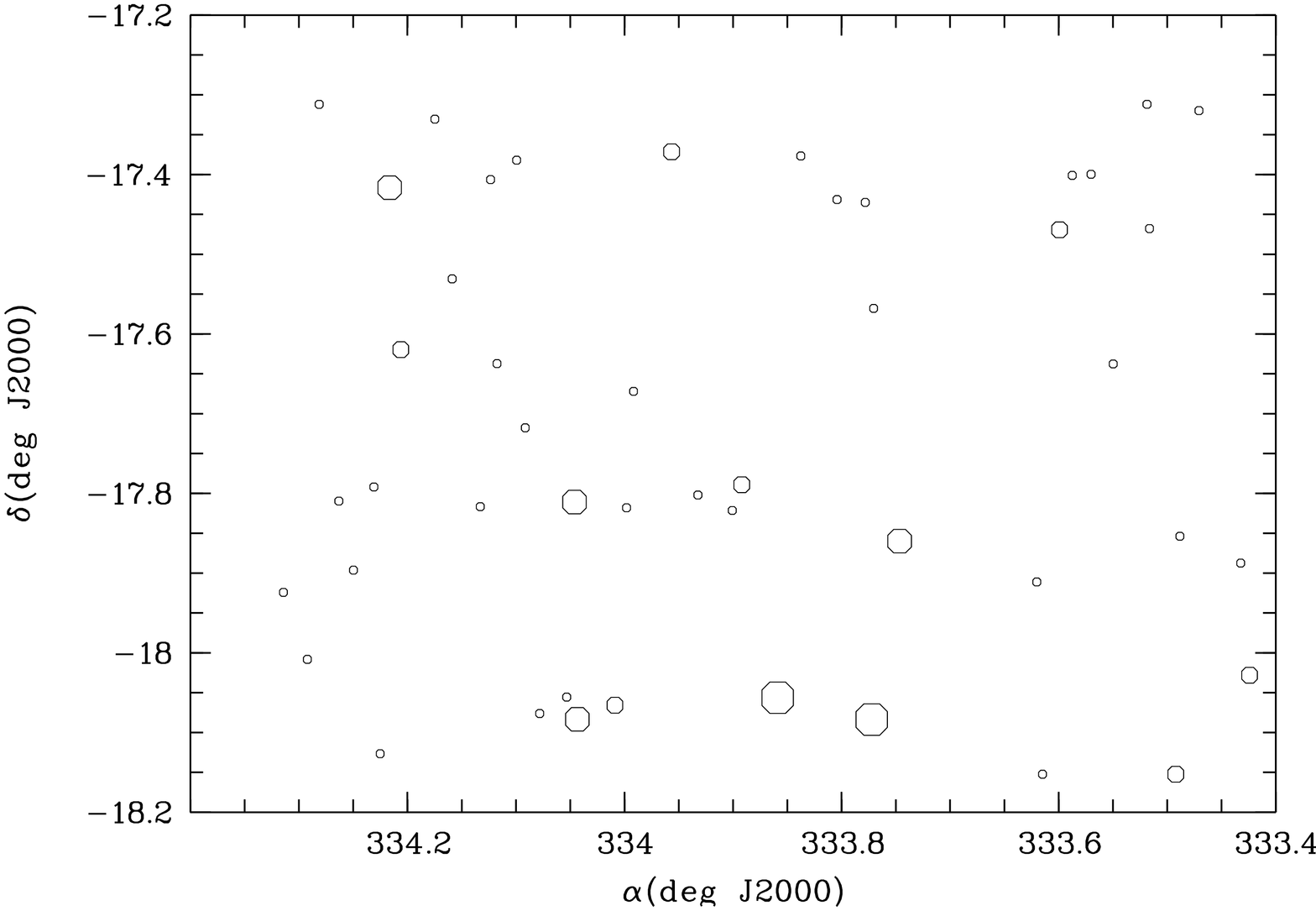,width=8.cm,angle=270}}
\caption[]{Spatial distribution of the detected structures in the
  three searched CFHTLS Deep fields. From top to bottom: D2, D3, and
  D4. The symbol sizes increase with the S/N of the detection. }
\label{fig:distribadD}
\end{figure}

\begin{figure}[hbt]
\centering
\mbox{\psfig{figure=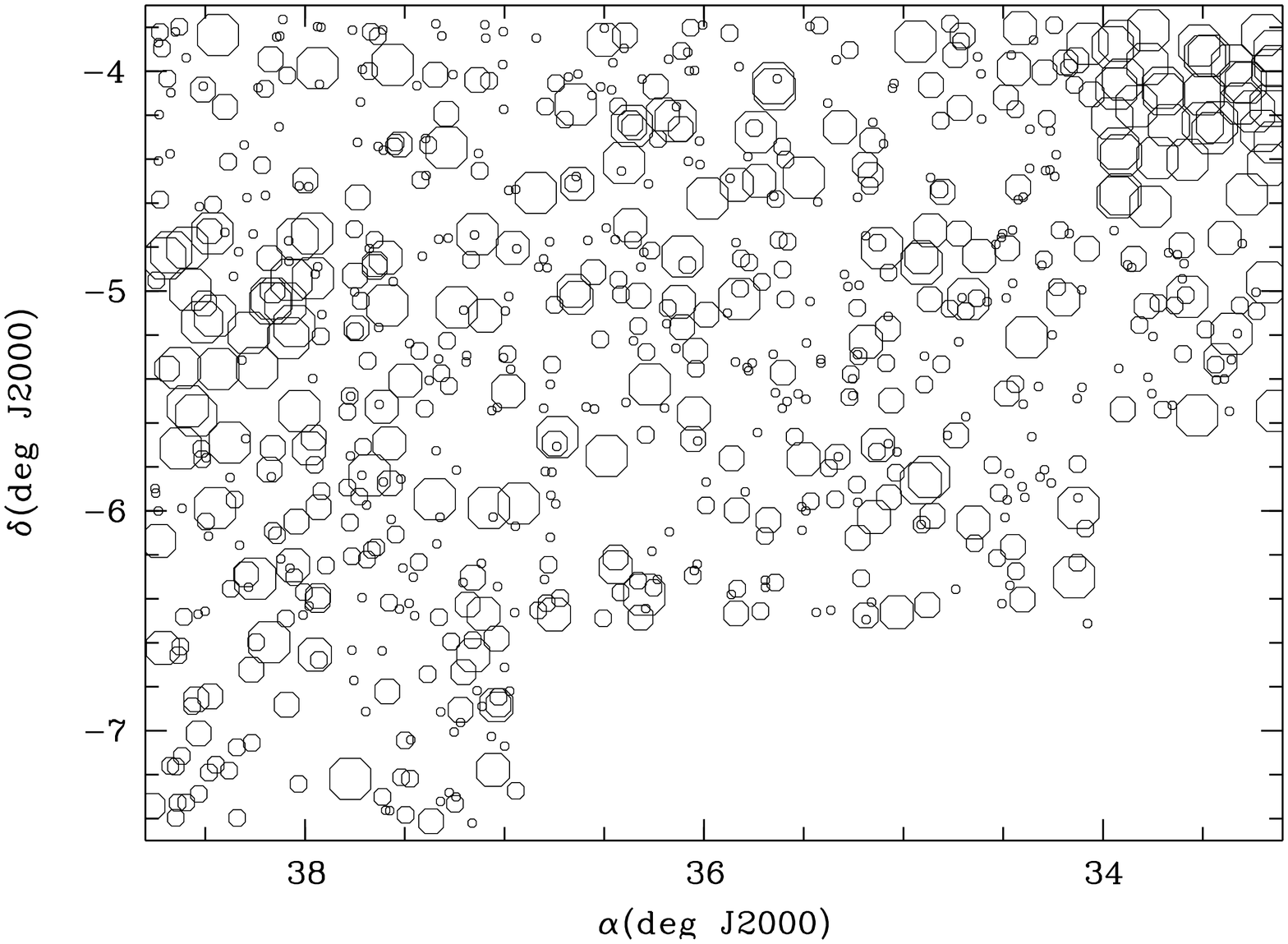,width=8.cm,angle=270}}
\mbox{\psfig{figure=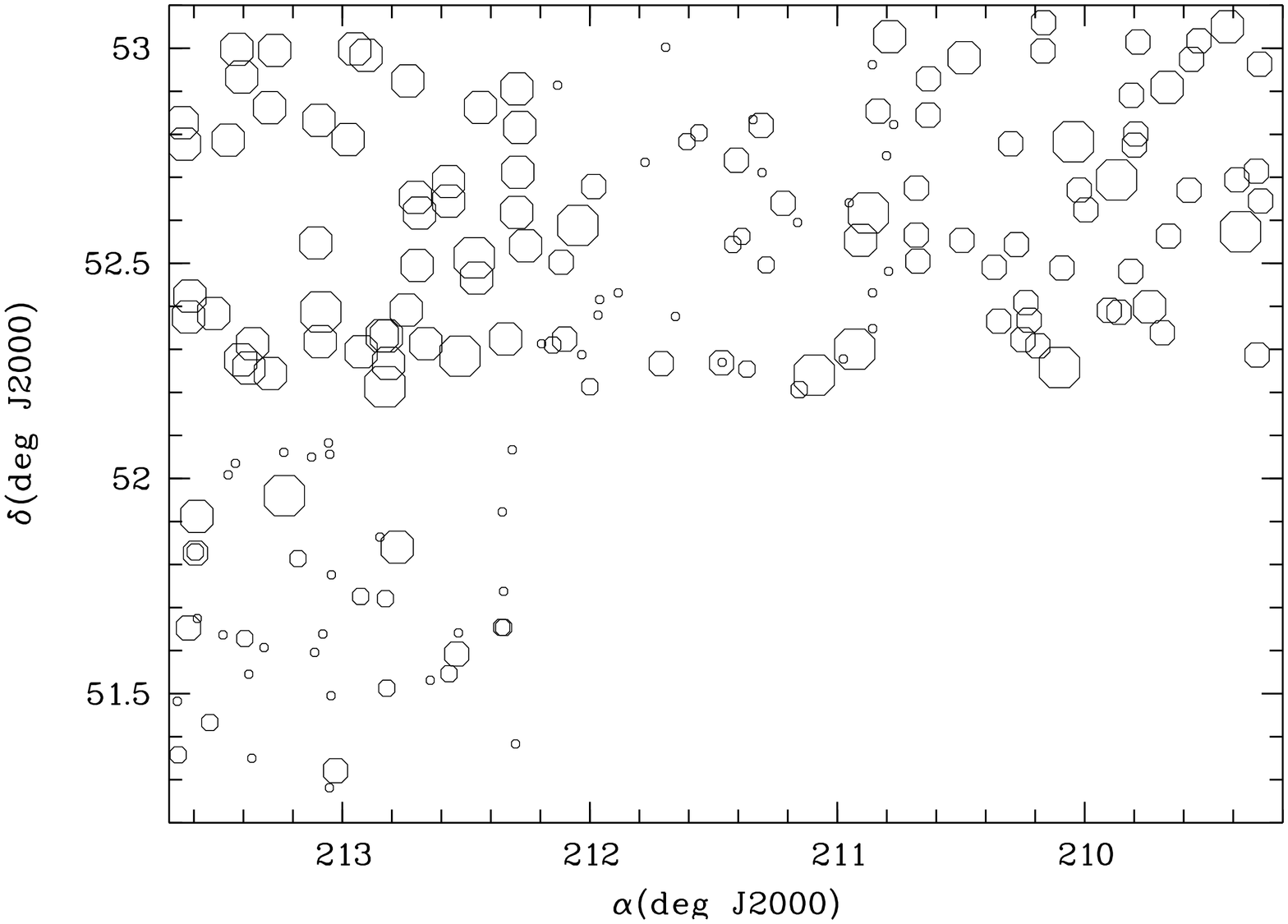,width=8.cm,angle=270}}
\mbox{\psfig{figure=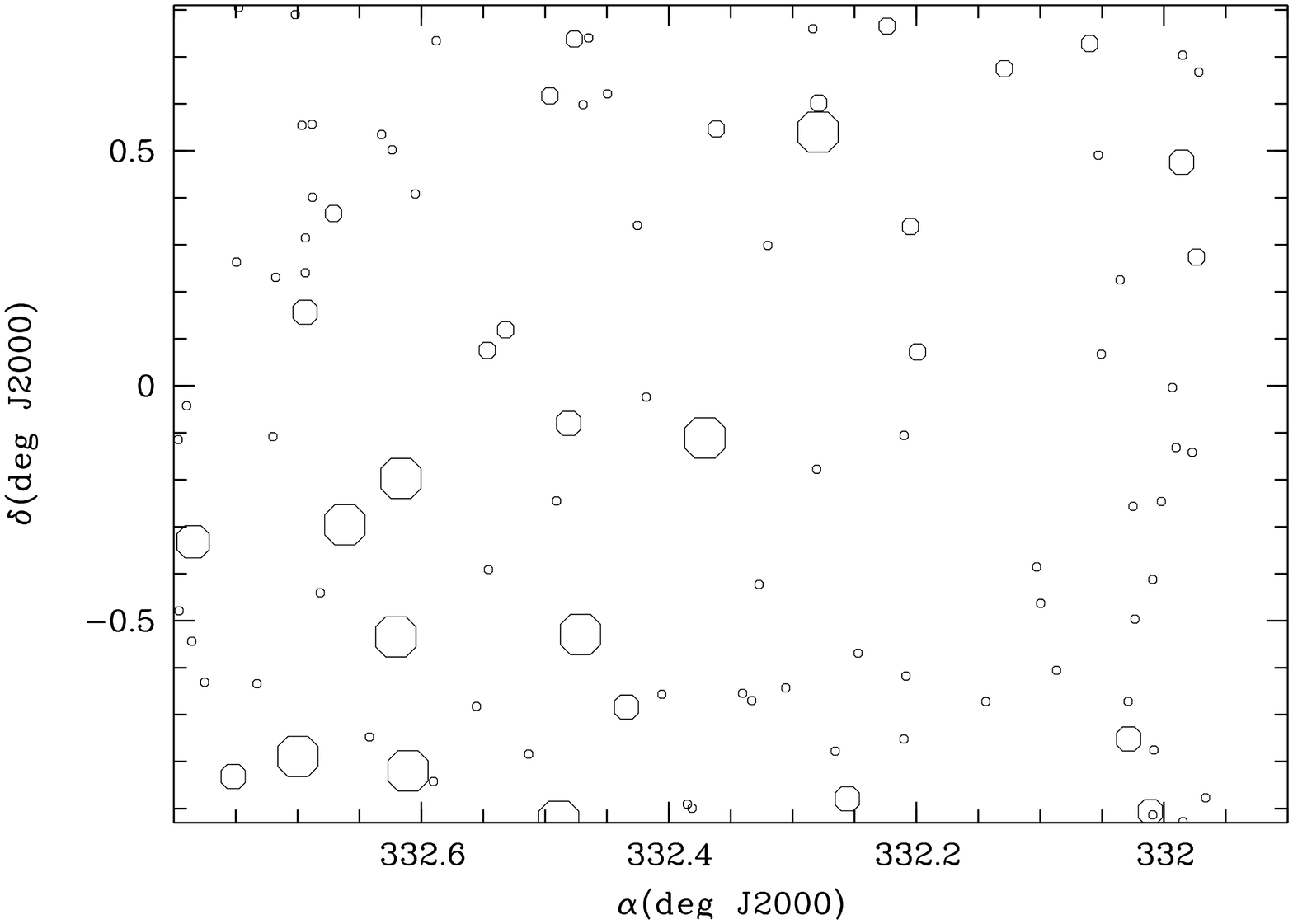,width=8.cm,angle=270}}
\caption[]{Spatial distribution of the detected structures in the
  three searched CFHTLS Wide fields. From top to bottom: W1, W3, and
  W4. The symbol sizes increase with the S/N of the detection. Note
  the different sizes of the three fields, W1 being the largest
  one. Only W4 covers a full rectangle.}
\label{fig:distribadW}
\end{figure}

\begin{figure}[hbt]
\centering
\mbox{\psfig{figure=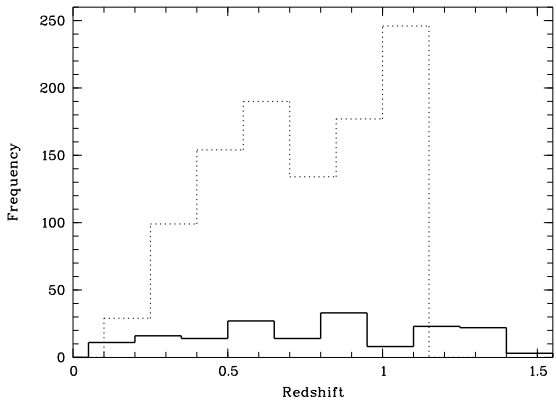,width=8.cm,angle=270}}
\caption[]{Redshift distribution of the detected structures (S/N$\geq$2) in the searched
  CFHTLS fields (thin dotted lines: Wide fields, thick continuous line: Deep
  fields).These histograms
are not corrected for detection efficiency.}
\label{fig:distribz}
\end{figure}

\begin{figure}[hbt]
\centering
\mbox{\psfig{figure=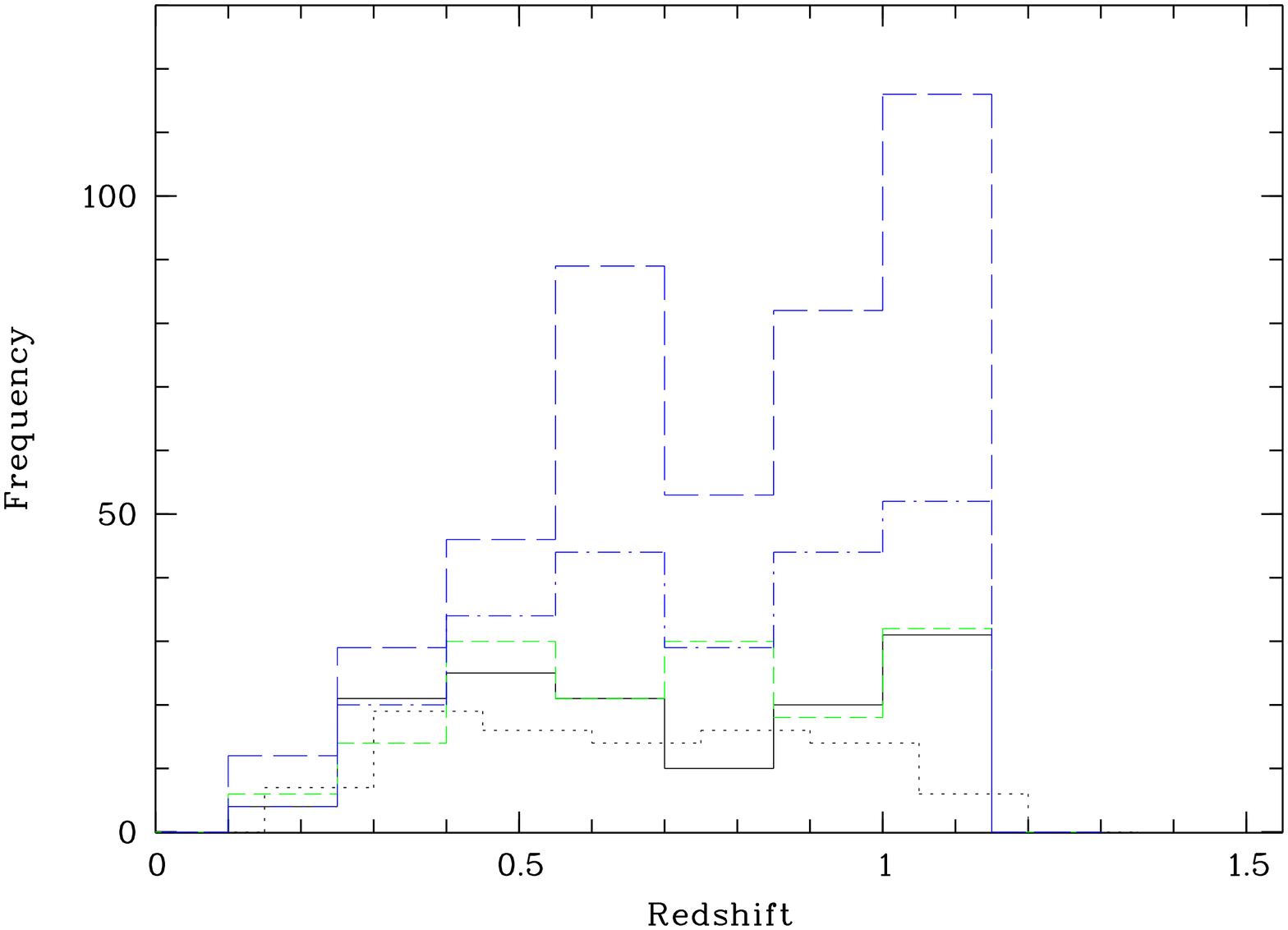,width=8.cm,angle=270}}
\mbox{\psfig{figure=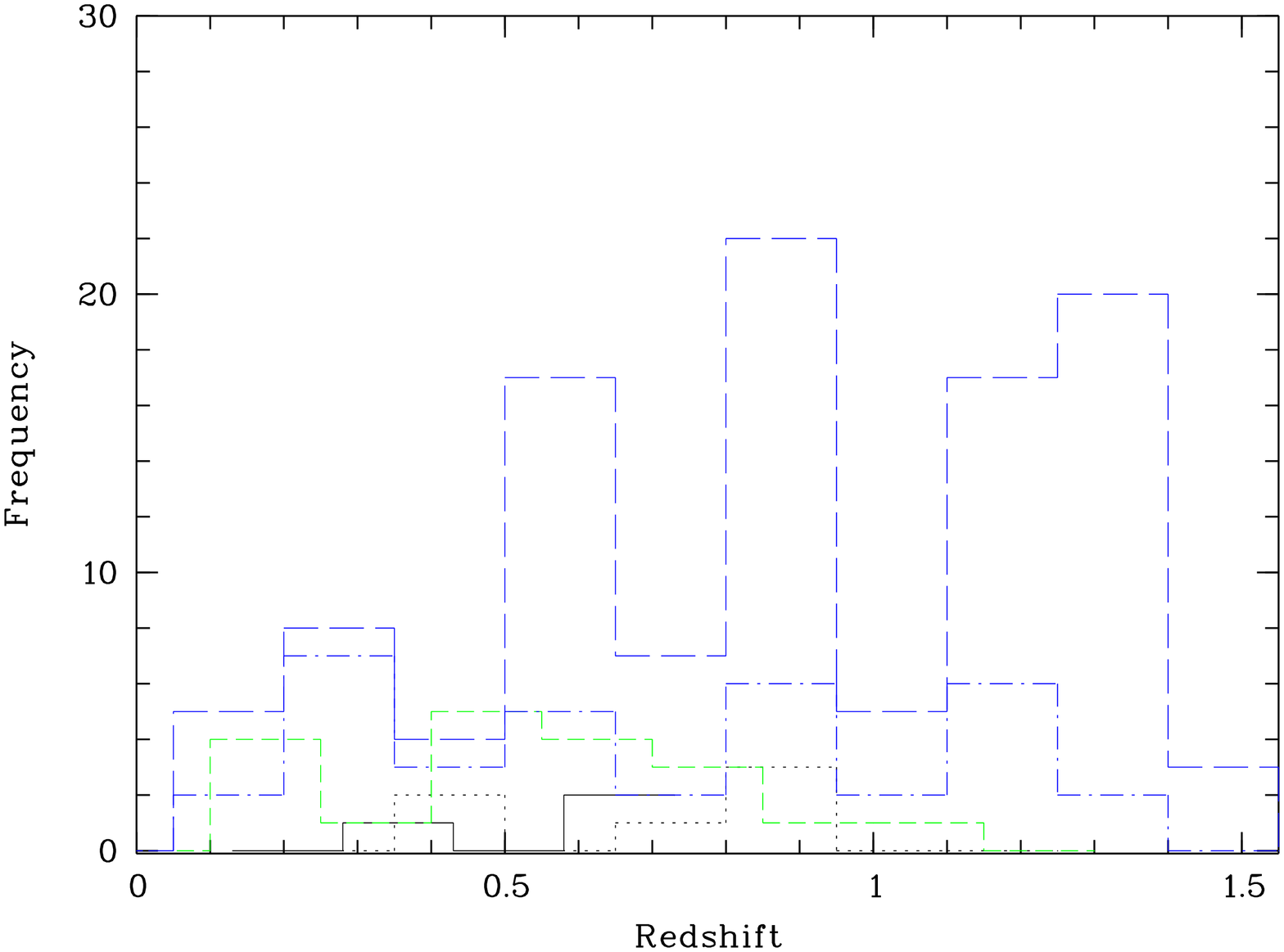,width=8.cm,angle=270}}
\caption[]{Redshift distribution of the detected structures in the
  searched CFHTLS fields as a function of the detection
  S/N. Continuous black line: S/N=6, dotted black line: S/N=5, dashed green
  line: S/N=4,
  dot-dashed blue line: S/N=3, long-dashed blue line: S/N=2. Upper figure: wide
  fields, lower figure: deep fields. These histograms are not
  corrected for detection efficiency.}
\label{fig:distribz2}
\end{figure}

\begin{figure}[hbt]
\centering
\mbox{\psfig{figure=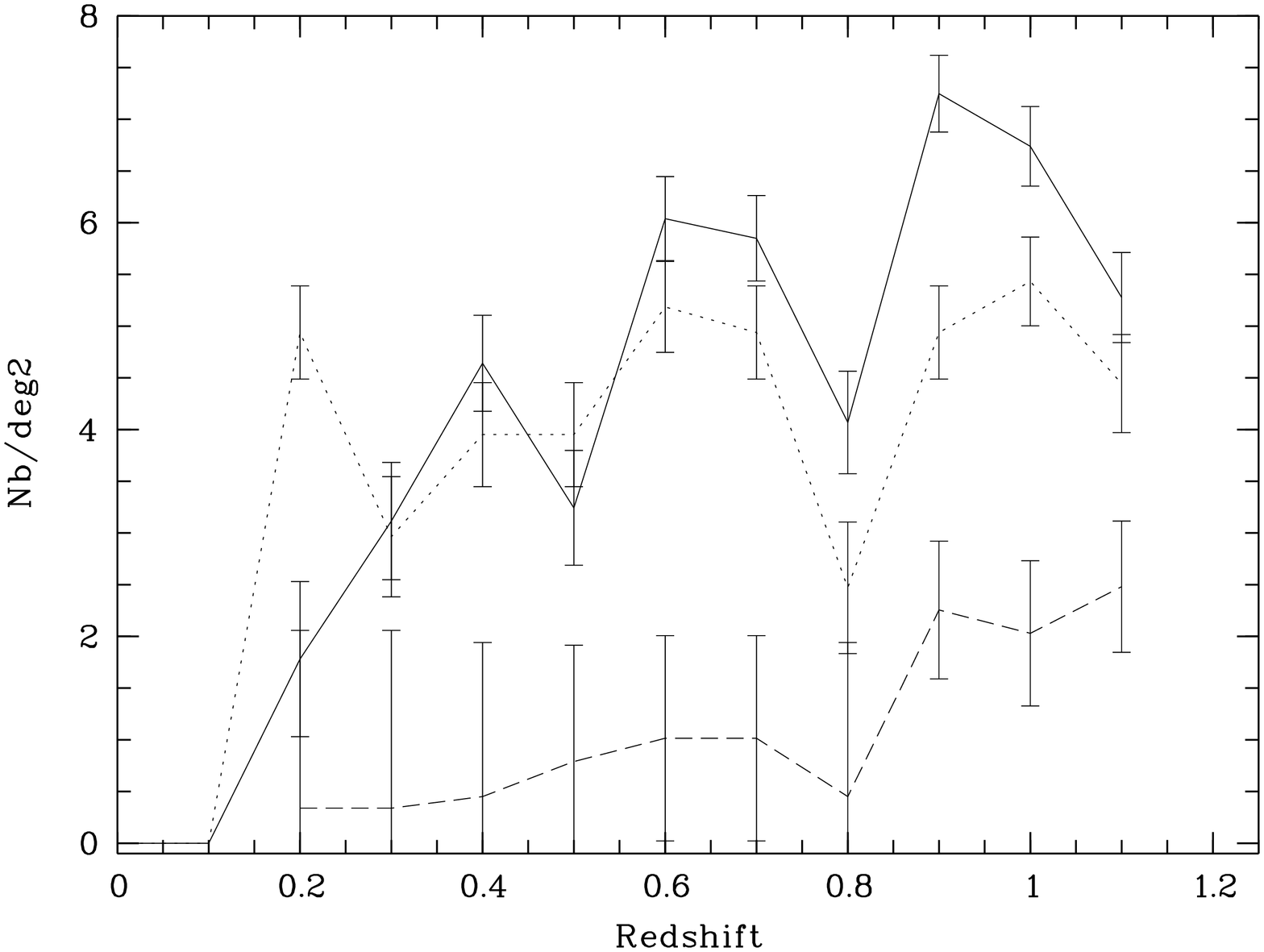,width=8.cm,angle=270}}
\caption[]{Candidate cluster (S/N$\geq$2) density (per deg$^2$) as a
  function of redshift. Continuous line: W1 field (computed over
  15.73~deg$^2$), dotted line: W3 field (3.24~deg$^2$), dashed line:
  W4 field (8.87~deg$^2$). Error bars are Poissonian. These curves are
  not corrected for detection efficiency.}
\label{fig:distribz3}
\end{figure}

We show in Figs.~\ref{fig:distribadD} and ~\ref{fig:distribadW}
the spatial distributions and in Figs.~\ref{fig:distribz},
~\ref{fig:distribz2}, and ~\ref{fig:distribz3} the redshift
distributions of our detections.

We see in Fig.~\ref{fig:distribz2} a regular increase in the number of
$detections$ as a function of S/N, in good agreement with the expected behaviour
of the detection method.

We see in Fig.~\ref{fig:distribz3} that the W4 field provides significantly
fewer detections (S/N$\geq$2) than W1 and W3. This is not due to a galaxy
catalog incompleteness (see Coupon et al. 2009). As seen in Table~1
the numbers of galaxies per deg$^2$ for the W1 (250292 gal/deg$^2$), W3
(272147 gal/deg$^2$), and W4  (263685 gal/deg$^2$) fields are similar.
Coupon et al. (2009) also find slightly higher uncertainties in the W4
photometric redshift estimates, with a level
of catastrophic errors  $\sim$35$\%$ higher than in the W1 field.
This could have an effect on the
cluster detection level. However, this difference in cluster density  probably
means that the W4 field is intrinsically poor in terms of structures and that
this field probably does not include many massive large-scale structures
due to cosmic variance.

\subsection{Angular correlation function}

The goal of this subsection is to consider the angular correlation of the distribution
of our cluster candidates as a test of consistency of the catalog. If real,
$detections$ should not be randomly distributed.

The spatial correlation function of
clusters has been known for a long time to behave as a power-law
(Bahcall $\&$ Soneira 1983, Nichol et al. 1992):

$$\xi_{cc}(r) = (r/R_{0})^{-\gamma}$$
\noindent
where the correlation length $R_0$ depends on cluster richness and the slope
is $\gamma$ $\sim 1.8$. The angular correlation function is then also expected
to show a power law behaviour:

$$\omega_{cc}(\theta) = (A_\omega (\theta))^{-\delta}$$
\noindent
with $\delta$=$\gamma$-1.

To estimate the angular correlation of the distribution
of our cluster candidates, we limited our sample
to the 0.4-0.8 redshift range inside the W1 CFHTLS field in order to both take
advantage of the large contiguous coverage of this field and to avoid
the redshift
range where the W1 field did not provide a high enough detection rate.

The angular two-point correlation function $\omega(\theta)$
represents the excess probability
for an object (here a {\it detection}) to have a neighbour located at an
angular separation $\theta$ with respect to a random distribution of points
(Peebles 1980).

We compute the angular correlation using the estimator of Landy \& Szalay (1993):

$$\omega(\theta)=\frac{DD(\theta) - 2DR(\theta) + RR(\theta)}{RR(\theta)}$$
\noindent
where DD, RR and DR are respectively the normalised number of data-data,
random-random and data-random pairs with an angular separation $\theta$
and $\theta_i\le \theta\le \theta_0 + i\times \Delta\theta$.
We generate a random catalog of 10000 points with the same geometry and masked as the data.
Since the measurement is noisy, we consider several logarithmic binnings
and zero points $\theta_0$. As a consequence, the measurement at a given
angular scale is strongly correlated with the others but the combined
measurement gives us at least a qualitative trend, to answer the question
of whether or not our sample is randomly spatially distributed.

We do not correct our measurement
for the integral constraint due to the finite size of the field (e.g.
Cappi \& Maurogordato 1995), as the
scope of this work is not the clustering analysis itself, but the use of
$\omega(\theta)$ as a test of consistency.
Our measurement is then a (moderate) underestimate of the real angular
correlation function.

We estimate the uncertainty on each data point
considering only Poisson errors on the data-data pairs.
In the case of the Landy \& Szalay estimator, they are given by :

$$\delta\omega=\frac{1+\omega(\theta)}{\sqrt(DD(\theta))}$$

\begin{figure}[hbt]
\centering
\mbox{\psfig{figure=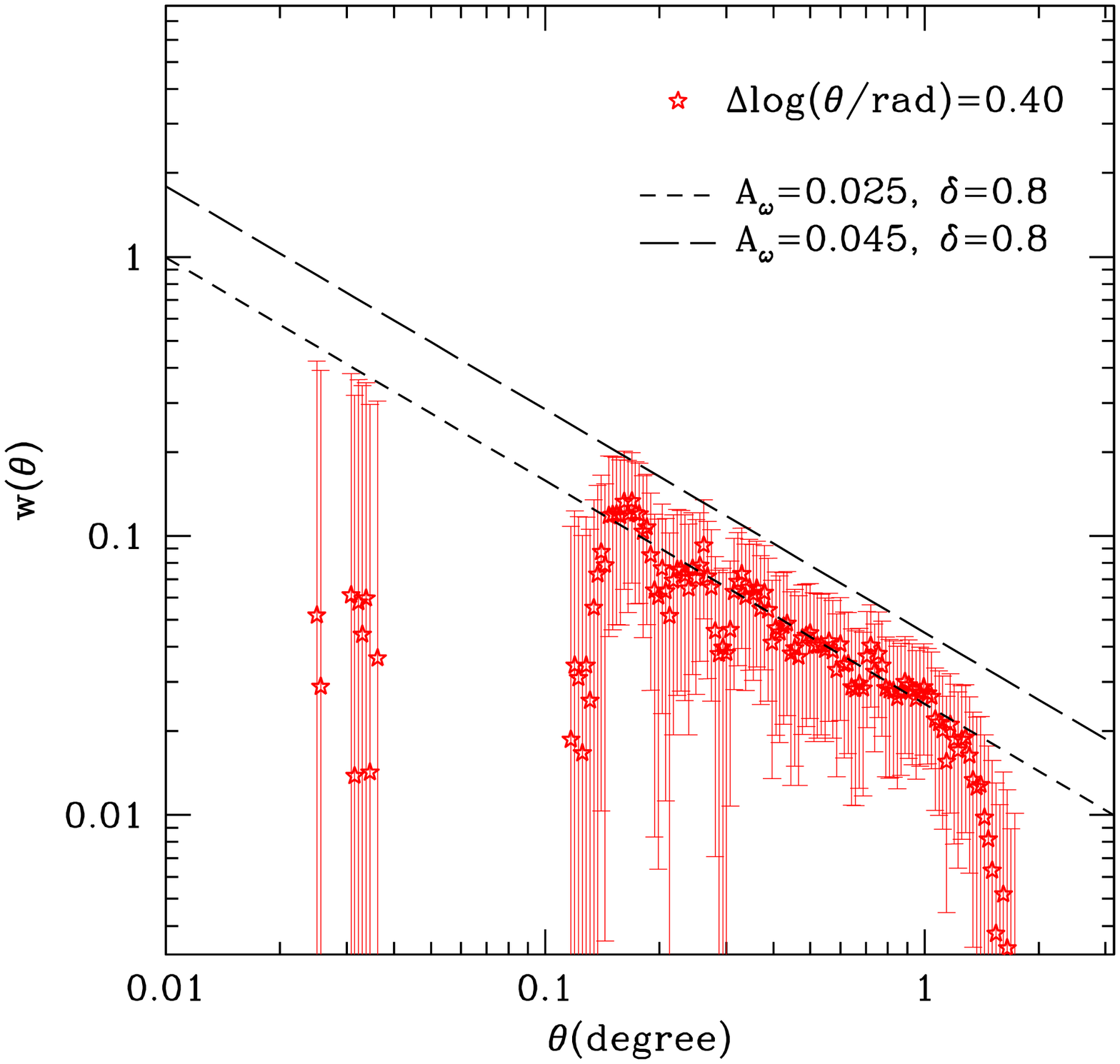,width=8.cm,angle=0}}
\caption[]{Angular correlation function of our $detections$ in the W1 CFHTLS
  field for z=[0.4,0.8] for S/N=3. Error bars are Poissonian. The two straight
  lines are two power laws (see text).}
\label{fig:corr2}
\end{figure}

We show in  Fig.~\ref{fig:corr2} the combined angular 2-point correlation function
in the W1 field for the S/N=3 {\it detection} sample.
We compare our measurements to two power laws conventionally defined as:

$\omega(\theta)=A_\omega (\theta/1^o)^{-\delta}$
\noindent

The slope is fixed to $\delta=0.8$ (a reasonable approximation of the
true slope) and the amplitude is arbitrarily chosen to guide the eye.

The sample of {\it detections} selected with S/N=3 shows a clear signal at
large angular scales.
A power law with amplitude $A_\omega=0.025$ appears to be a good approximation
on scales [0.15 -- 1] degrees.

At small angular scales, the signal is weaker, and results are not
significant, since they are based only on a small number of
$detections$, as shown by the large error bars.

This qualitative analysis shows that {\it detections} are dominated by
structures showing a realistic clustering when compared to other high
redshift analyses. Similarly to our results, Papovich (2008) found for
example an angular correlation function for his high redshift cluster
sample consistent with a power-law fit over the interval
[0.03,1.7]deg. The slope of this power law was found to be 1.1$\pm$0.1
in relatively good agreement with our slope of 0.8. Other works
  such as Bahcall et al. (2003 and references therein) or Brodwin et
  al. (2007) also show power-law angular correlation functions.

\subsection{Tracing the large scale structure of the Universe.}

The Millennium simulation, among others, put in evidence the increasing
filament constrast (bridges joining massive clusters) with decreasing
redshift. This is the well known
hierarchical behaviour of the Universe and could ultimately be used as
a cosmological test assuming that we are able to trace precisely this
filamentary structure as a function of redshift. However, detecting
large scale filaments is a very difficult task due to strong
superposition effects and to the fact that they are only weak X-ray
emitters (see e.g.  the Abell 85 filament, Bou\'e et al. 2008).  We
could in principle consider large scale galaxy surveys such as the CFHTLS
combined with photometric redshift computations to try to detect these
filaments. However, the galaxies populating these filaments can have
very low masses and are therefore likely to be very faint. We must
therefore assess the survey depth required to reach such a goal.

Rather than trying to detect individually the filaments present in the
CFHTLS survey, we chose a statistical approach based on the Minimal
Spanning Tree ($mst$ hereafter) technique. The $mst$ technique is a
geometrical construction issued from the graph theory (e.g. Dussert
1988) which allows to characterize quantitatively a distribution of points
(e.g. Adami $\&$ Mazure 1999). Very briefly, it is a tree joining all
the points of a given set, without any loop and with a minimal length;
each point is visited by the tree only once. The main aspect here is
the unicity of such a construction. For a given set of points, there
is more than one $mst$, but the histogram H of the length of the $mst$ edges is
unique. This is fundamental because it is then possible to
characterize completely a set of points with H. The details of the
procedure and of the normalisations are given in Adami $\&$ Mazure
(1999); we took into account the first three momenta (mean, sigma, and
skewness) of H to characterize this histogram.

We chose to compute the distance D in the (mean, sigma, skewness)
space between a given distribution of points and a uniform distribution for
which the mean, sigma, and skewness values are well known (see Adami $\&$
Mazure 1999). The distance D is simply given by:

\[
D =\sqrt{\sum_{i=1}^n(p_i-q_i)^2}
\]
\noindent
with $p_i$ and $q_i$ being successively the mean, dispersion and skewness of
the uniform distribution and of the considered distribution.

\begin{figure}[hbt]
\centering
\mbox{\psfig{figure=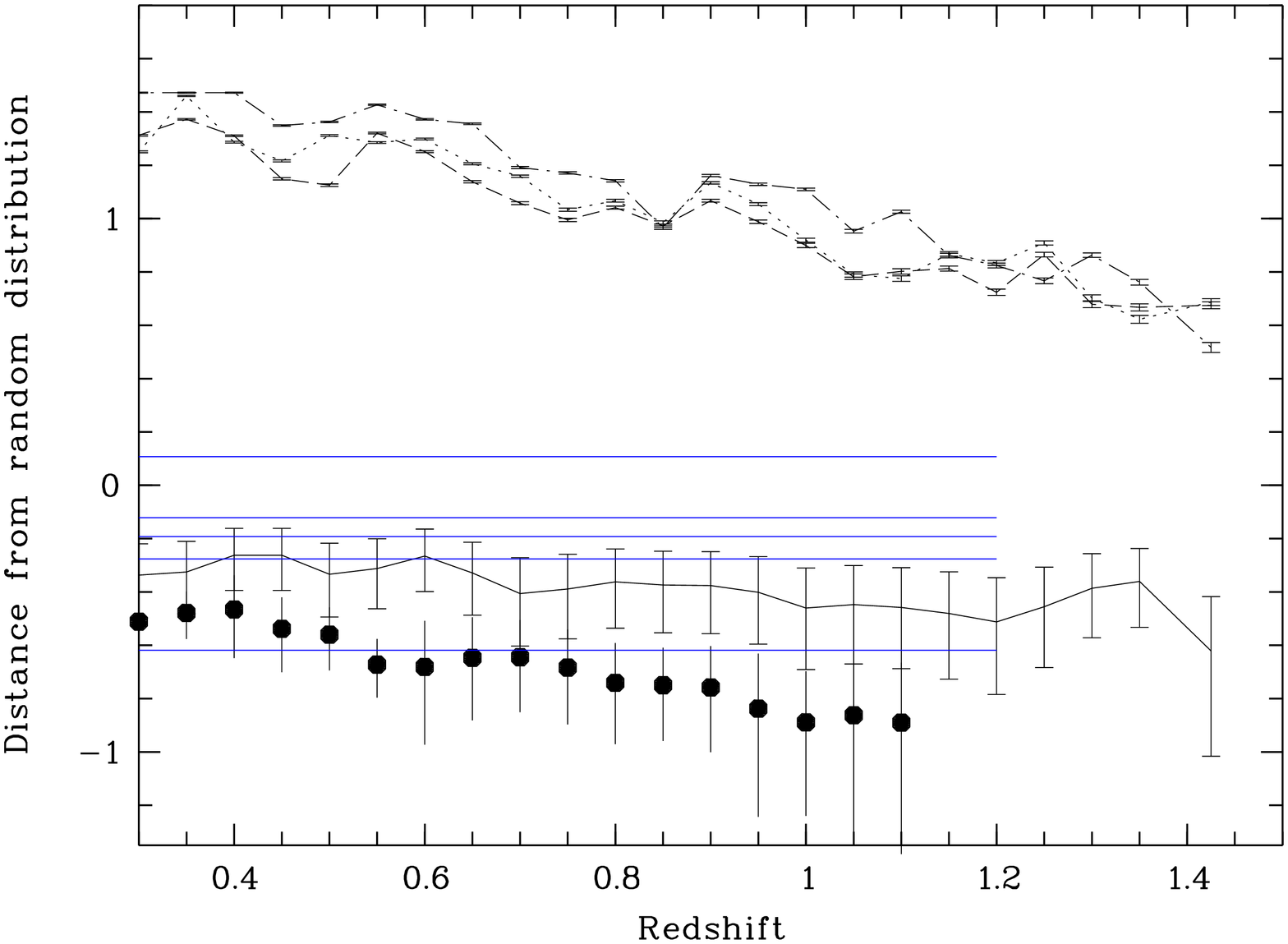,width=8.cm,angle=270}}
\caption[]{Variation of D as function of redshift. The continuous,
  dashed, dotted, and dash-dotted lines with error bars are for
  Millennium halos more massive than 10$^{10}$ M$_\odot$, 3 10$^{13}$
  M$_\odot$, 5 10$^{13}$ M$_\odot$ , and 10$^{14}$~M$_\odot$.
Unconnected filled circles with error bars show the
  CFHTLS W1 galaxies limited to i'=23. The five horizontal blue lines
  correspond to the mean D value for clusters detected in the W1 field
  (from top to bottom: S/N=6, 5, 4, 3, and 2). }
\label{fig:mstwide}
\end{figure}

\begin{figure}[hbt]
\centering
\mbox{\psfig{figure=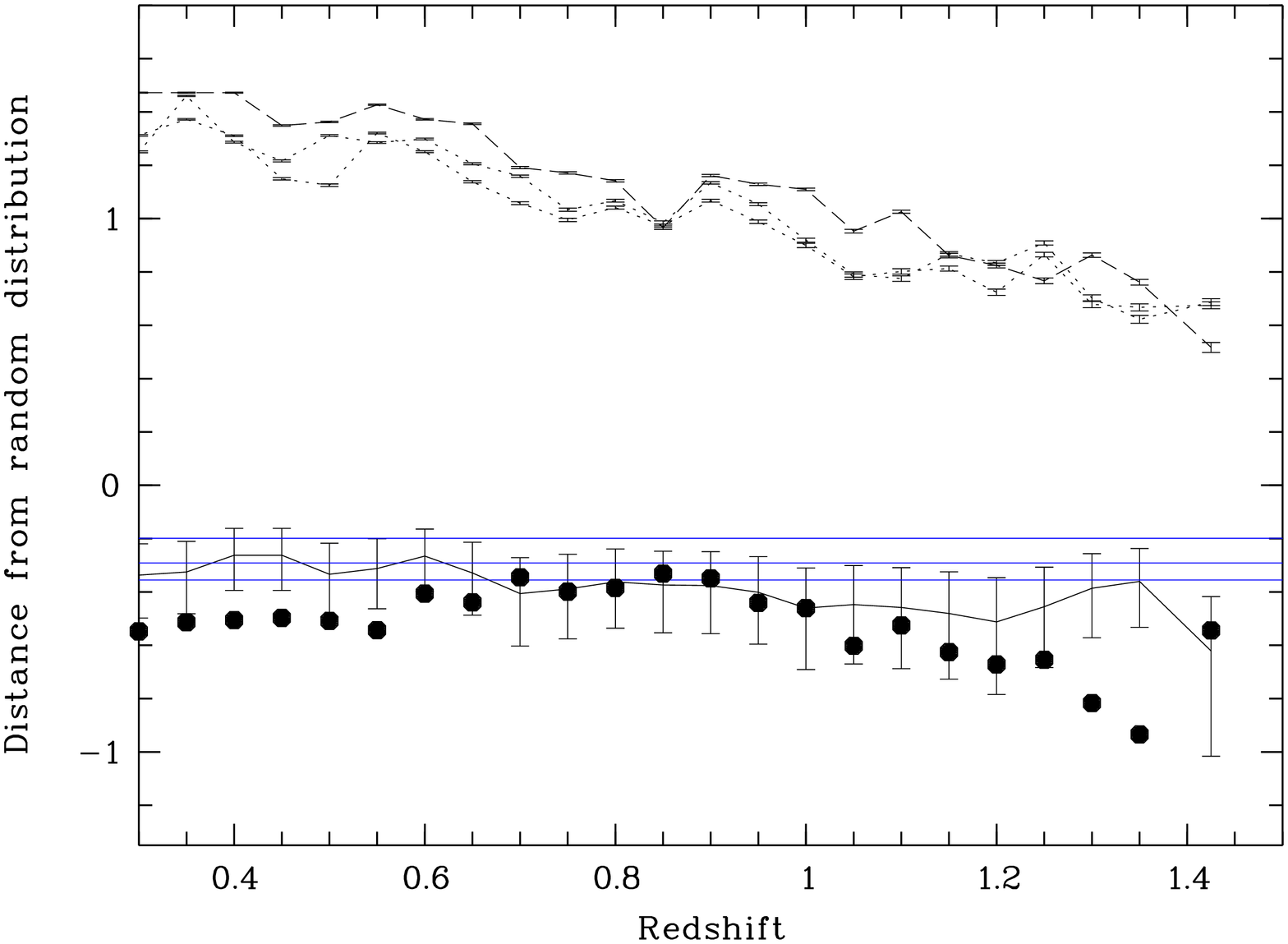,width=8.cm,angle=270}}
\caption[]{Variation of D as function of redshift. The continuous,
  dashed, dotted, and dash-dotted lines with error bars are for
  Millennium halos more massive than 10$^{10}$ M$_\odot$, 3 10$^{13}$
  M$_\odot$, 5 10$^{13}$ M$_\odot$ , and 10$^{14}$~M$_\odot$.
Unconnected filled circles with error bars show the
  CFHTLS D2 galaxies limited to i'=25. The four horizontal blue lines
  correspond to the mean D value for clusters detected in the D2 field
  (from top to bottom: S/N=5, 4, 3, and 2). S/N=5 and 4 are overlapping.}
\label{fig:mstdeep}
\end{figure}

We computed D as a function of redshift in the Millennium simulation.
Figs.~\ref{fig:mstwide} and ~\ref{fig:mstdeep} show these variations for
several halo classes: more massive than 10$^{10}$ M$_\odot$ ($\sim$galaxies),
more massive than 3 10$^{13}$ M$_\odot$ ($\sim$groups of galaxies), more
massive than 5 10$^{13}$ M$_\odot$ ($\sim$major groups of galaxies), and more
massive than 10$^{14}$ M$_\odot$ ($\sim$clusters of galaxies). On these
figures, a uniform distribution would be a horizontal line at minus infinity.

We first note that our curves all decrease, meaning that the more
distant the sample we consider, the closer to a uniform distribution it
is. This is not surprising as the main ingredient of the Millennium simulation
is a hierarchical gravity dominated Universe.

Second, we plot on these figures the distance D computed with the
CFHTLS galaxies in the W1 (Fig.~\ref{fig:mstwide}) and D2 fields
~(Fig.\ref{fig:mstdeep}) assuming the i'=23 and i'=25 magnitude
limitations and the computed photometric redshifts. If the CFHTLS
fields are good tracers of the large scale structure of the Universe
(and if the Millennium simulation assumed the correct cosmology),
distances D computed with the CFHTLS data should be included in the
Millennium curve for objects more massive than 10$^{10}$ M$_\odot$. We
show in Figs.~\ref{fig:mstwide} and ~\ref{fig:mstdeep} that the CFHTLS
Wide data become different from the expected Millennium behaviour at
z$\ge$0.5. This means that the CFHTLS Wide survey is not able to
recover properly the filamentary structure of the Universe above
z$\sim$0.5. The resulting galaxy distribution (penalized by galaxy
detection incompleteness) then becomes too close to a uniform
distribution. This also shows that the deep CFHTLS fields are good
datasets to achieve such a goal up to z=1.25.

Third, we add on these two figures the distances D computed for our
cluster $detections$ and for several values of the detection S/N. We first note that
the variation of D with redshift is not significant. We therefore
chose to show only the mean value of D over the spanned redshift range
(z=[0.1;1.2] for the W1 and z=[0.1;1.5] for the D2). We also find that
the higher the S/N (hence the more massive the cluster), the more
different from a random distribution the cluster catalogs are.
We finally note that the S/N$\geq$4
CFHTLS cluster $detections$ (so with masses greater than 3.3 10$^{13}$
M$_\odot$) show a different behaviour from Millennium halos more
massive than 3 10$^{13}$ M$_\odot$. The Millennium halos considered are more
clustered than the corresponding real clusters (in terms of mass).
We now investigate why we have such a difference.

\begin{figure}[hbt]
\centering
\mbox{\psfig{figure=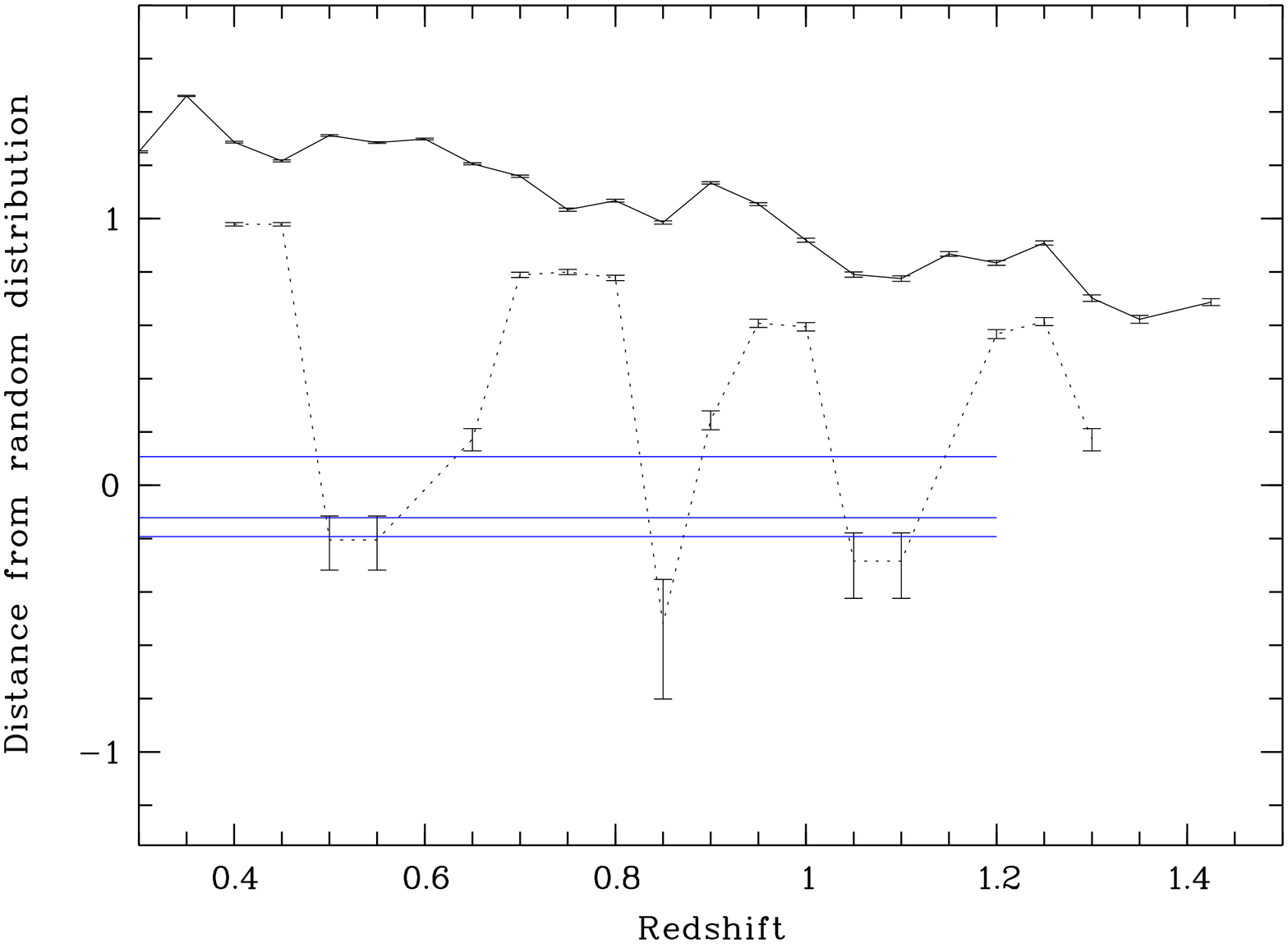,width=8.cm,angle=270}}
\caption[]{Variation of D as function of redshift. The continuous and dotted
 lines with error bars are for Millennium halos more massive
 than 3 10$^{13}$ M$_\odot$ with high and low levels of substructures. The
 three horizontal blue lines are the mean D value for clusters detected in
 the W1 field (from top to bottom: S/N=6, 5, and 4). }
\label{fig:mstwidesss}
\end{figure}

Considering low mass Millennium halos implies that highly spatially correlated halos will
be included in the $mst$ calculation. This will therefore artificially
increase the D value because part of these halos are in fact subhalos of more
massive structures. Our $detections$ do not include such objects by definition.
A way to not include these highly spatially correlated sub-halos is to
select only the less substructured Millennium halos.
Therefore, we redid the previous exercise selecting only
halos more massive than 3 10$^{13}$ M$_\odot$ and with a substructure
level lower than 20 subhalos included in the main halo. We generated
Fig.~\ref{fig:mstwidesss} where we show the D value before and after
considering low substructure level halos. We clearly see that removing
halos with a high level of substructure makes the Millennium and CFHTLS D
values compatible. This suggests that the Millennium simulation may
exhibit higher than normal substructure levels in massive clusters.

\section{Literature assessments of our detections}

We computed in previous sections statistical assessments of our detections based on
simulations. We now try to compare our detections with literature data,
i.e. known clusters in the surveyed areas.

\subsection{Internal assessment of our cluster detections.}

Among the searched CFHTLS fields, the W1 and D1 fields overlap (see
Mazure et al. 2007), allowing us to compare detections based on Deep
and Wide CFHTLS data. Wide field data exhibit lower detection rates
compared to Deep fields, so we do not expect to recover in the present
paper all the D1 detections of Mazure et al. (2007). Assuming the success
rates computed in the present paper via the Millennium simulations, we
expect to detect 2.7$\pm$1.4 times more clusters in the Deep D1 than
in the W1 data (uncertainty from Poisson estimates). Experimentally, we
detect 23 clusters in the W1 data
out of the 44 detected in the D1 data by Mazure et al.  (2007) with
exactly the same method. The ratio is 1.9, in agreement with the
expectations.

\begin{figure}[hbt]
\centering
\mbox{\psfig{figure=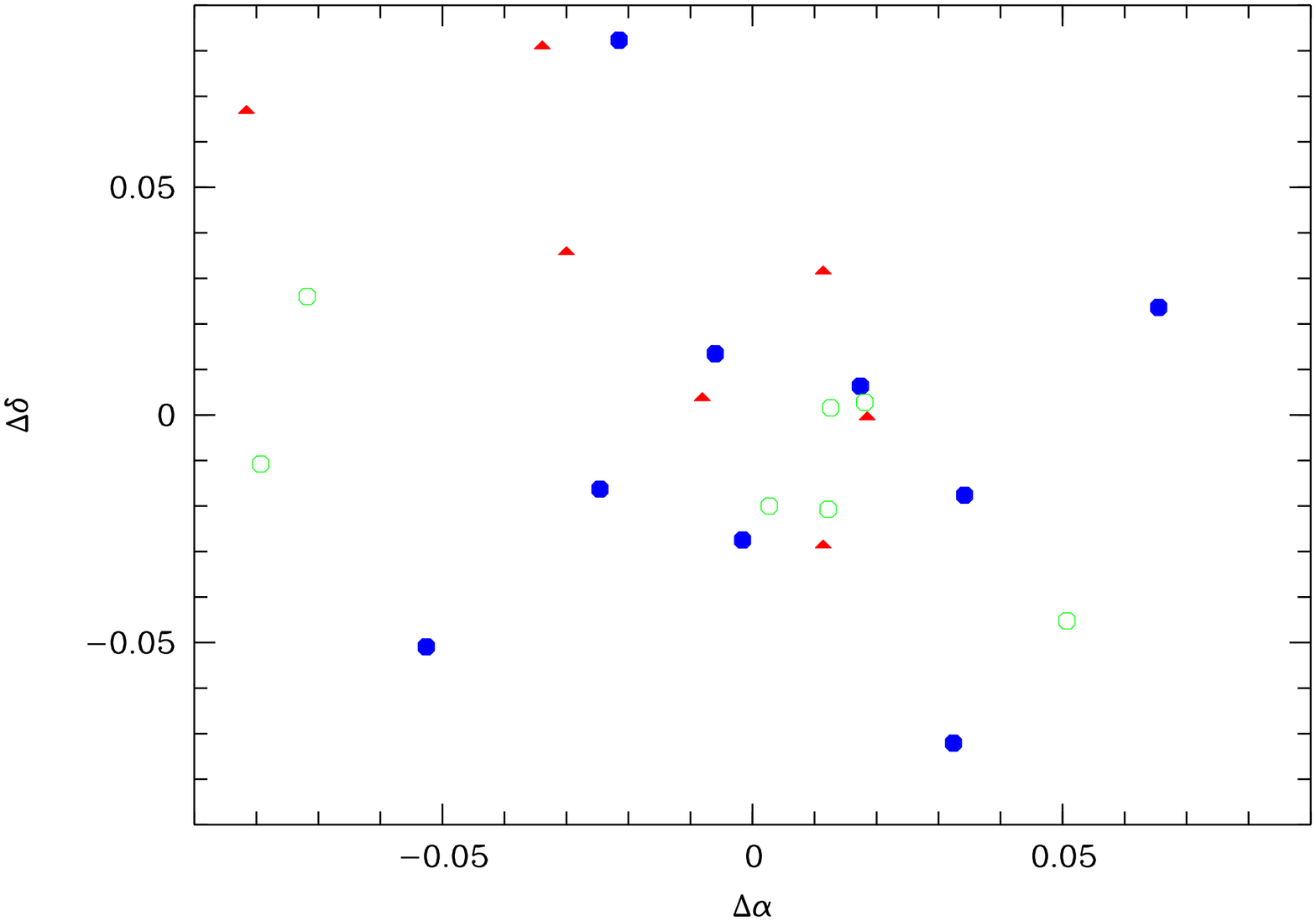,width=8.cm,angle=270}}
\mbox{\psfig{figure=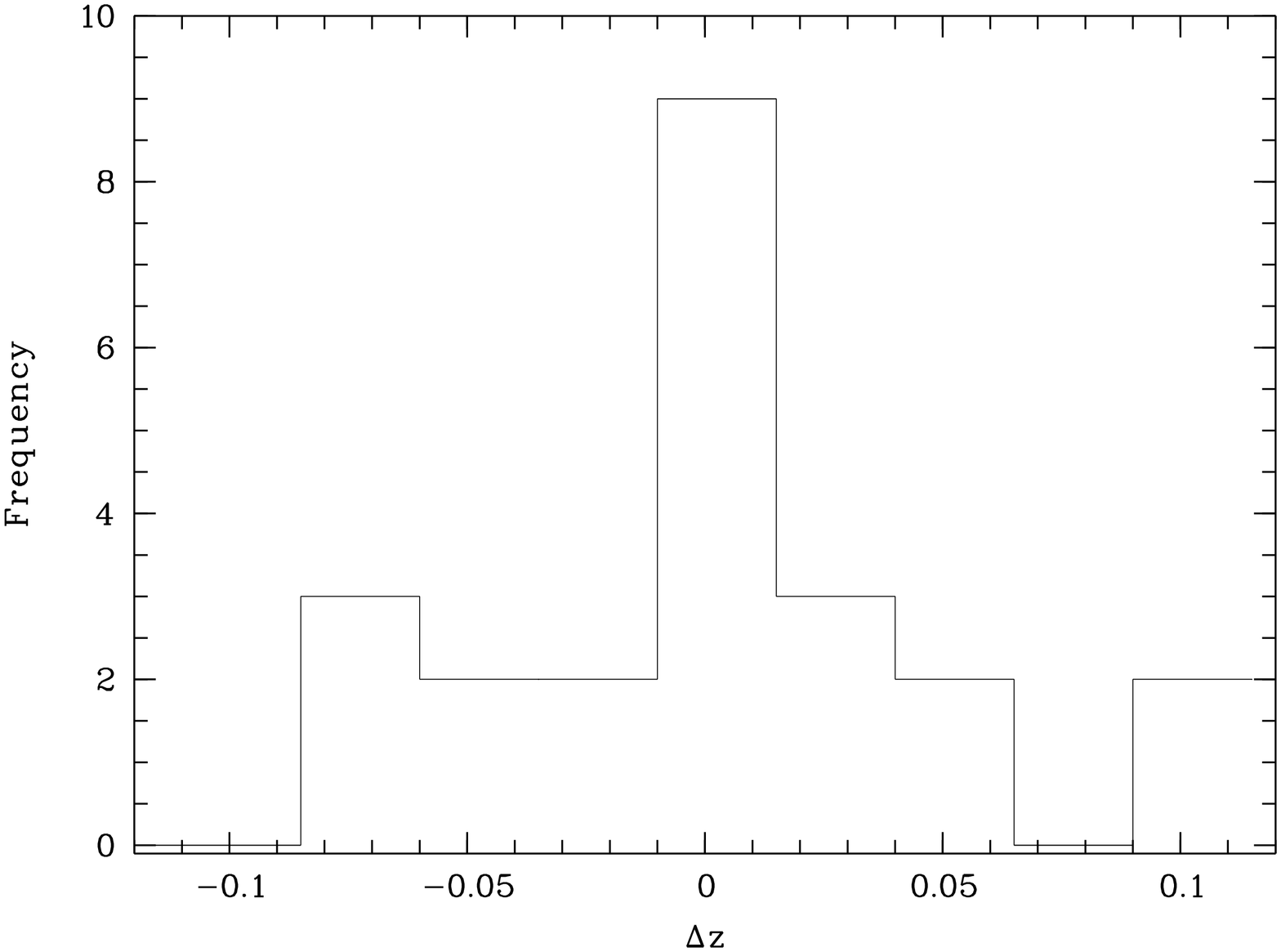,width=8.cm,angle=270}}
\caption[]{Upper figure: center shifts in deg (blue filled circles:
  z$\leq$0.5, green open circles: z=[0.5,0.8], red triangles:
  z$\geq$0.8). Lower figure: histogram of redshift difference between
  our D1 and W1 $detections$ (S/N$\geq$2).}
\label{fig:intcheck}
\end{figure}

We show in Fig.~\ref{fig:intcheck} the position and redshift
differences between the present W1 and D1 $detections$ of Mazure et
al. (2007). The mean center shift is 0.01$\pm$0.04 deg (0.6 arcmin)
and the mean redshift difference is -0.002$\pm$0.05.  We also note
that there is no significant variation in the center precision as a
function of redshift. This demonstrates that we are limited by the
pixel size used in the galaxy density map to define a cluster center.

\subsection{External assessment of our cluster detections.}

As in Mazure et al. (2007), we compare our $detections$ with the XMM-LSS
X-ray clusters published for the W1 field by Pacaud et al. (2007),
limiting our comparison to the 15 z$\geq$0.1 clusters of Pacaud et
al. (2007) included in the portion of the W1 field we analysed.  Ten
of our $detections$ are identified with these clusters. The remaining
ones (but one) are all affected by masked areas in the optical data
and we can reasonably assume that this makes their detection
impossible by the present method.

We also compare our $detections$ with the lensing searches made in the
CFHTLS areas. Limousin et al. (private communication) discovered a galaxy cluster at
z$\sim$0.88 in the CFHTLS D3 field (SL2S2214-1730) based on a strong lensing
analysis. This cluster is also detected with our method (D4-7, S/N=2) at
z=0.90. A more complete list of group detections in the SL2S survey is given in
Limousin et al. (2009). Among the 13 group detections of this paper, a single
one (SL2S2140-0532 at z=0.444) is included in our surveyed area (others are
outside the area where photometric redshifts were computed) and we detect
it at z=0.45 with S/N=6.

We compared our $detections$ with the Gavazzi \& Soucail (2007)
cluster sample. If we limit our search to clusters with a photometric
redshift in their paper (8 clusters), we redetect 7 of these clusters
with the present method. The last one is partially located in a masked
region in our data and this probably prevents its detection.

We finally compared our $detections$ with the matched filter detections of Olsen
et al. (2008). Still limiting the comparison to clusters included in our
surveyed area and at z$\geq$0.1, we detect 14 of the 16 Olsen et al. (2008)
spectroscopically confirmed clusters.

\begin{figure}[hbt]
\centering
\mbox{\psfig{figure=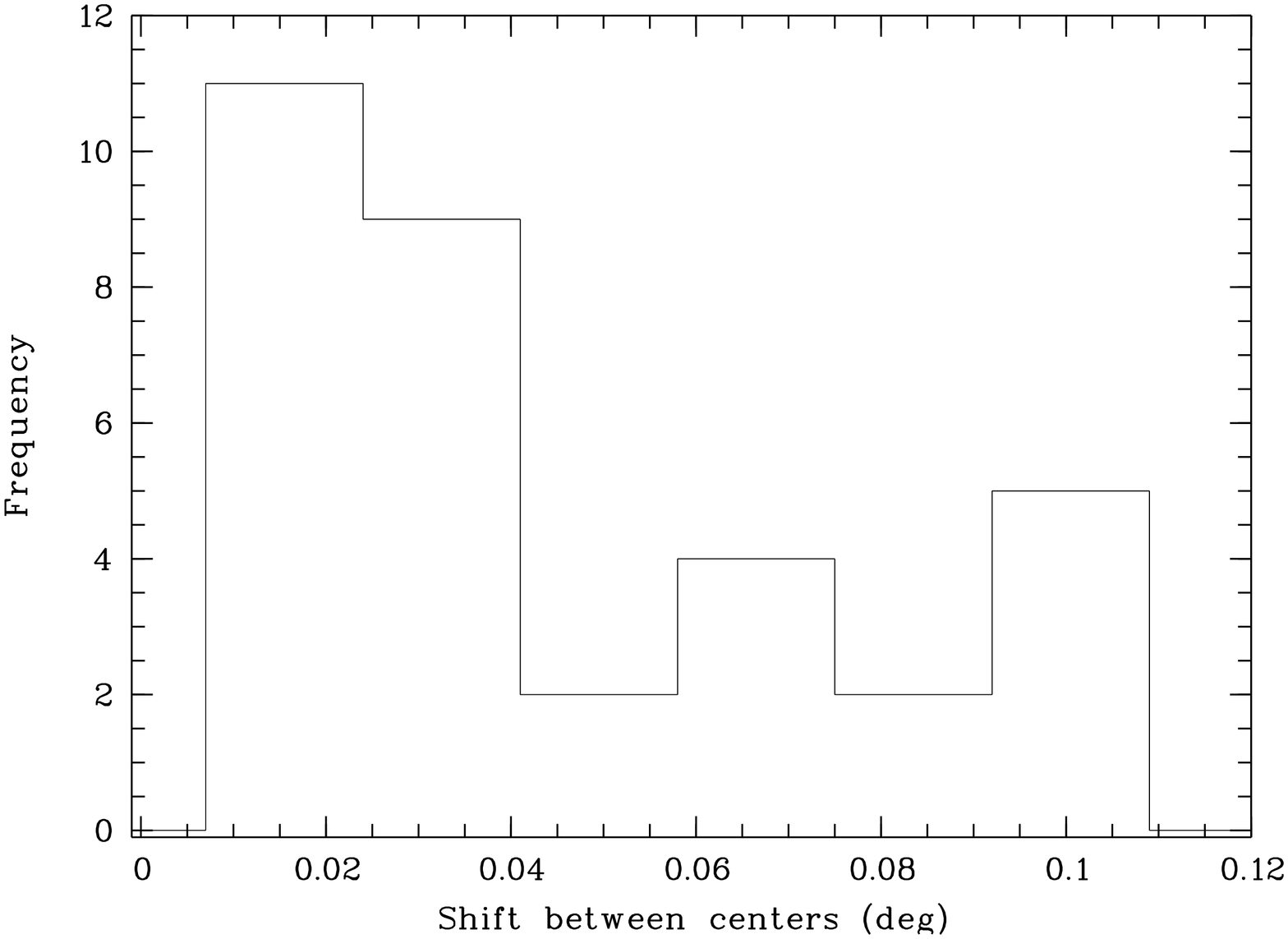,width=8.cm,angle=270}}
\mbox{\psfig{figure=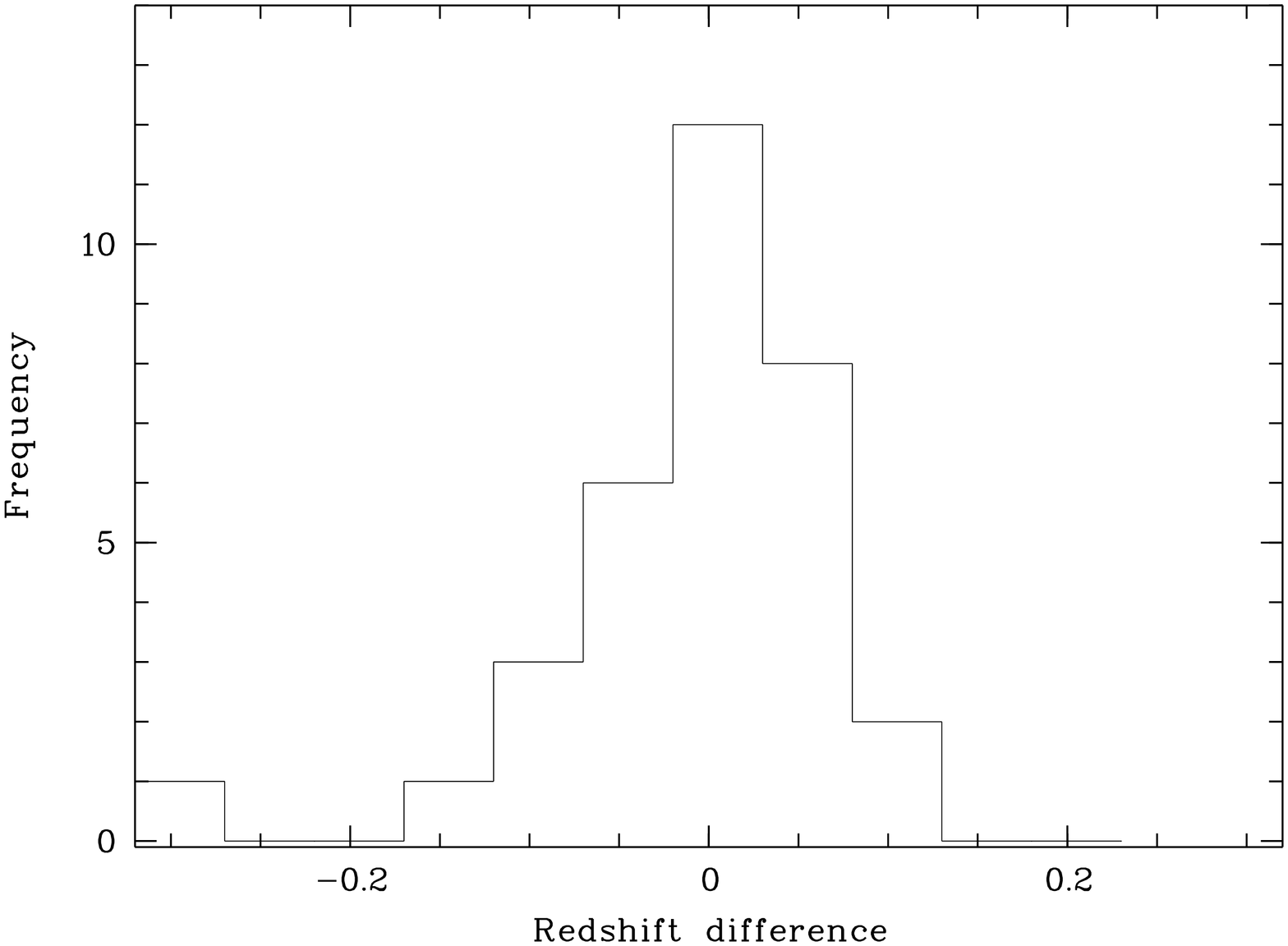,width=8.cm,angle=270}}
\caption[]{Histograms of the center shifts in degrees (upper figure)
  and redshift differences (lower figure) between our $detections$
  (S/N$\geq$2) and the literature clusters quoted in the text.}
\label{fig:extcheck}
\end{figure}

These high recovery rates therefore put our detection method on a firm
ground.  We give in Fig.~\ref{fig:extcheck} the histograms of the
center and redshift differences between our $detections$ and the
clusters previously quoted in the literature. The mean center shift is
0.045$\pm$0.03 deg (2.7 arcmin) and the mean redshift difference is
-0.01$\pm$0.08.

\clearpage

\section{Discussion and Conclusions}

We have detected 1200 candidate clusters in the CFHTLS Deep and Wide
fields. Statistically, more than 80$\%$ are real structures at z$\leq$1. This
is confirmed by internal and external comparisons with literature catalogs.

\begin{table}
\caption{Number and density of $detections$ as a function of S/N.}
\begin{center}
\begin{tabular}{ccc}
\hline
SExtractor threshold & number of $detections$ & density of $detections$ \\
 &  &  deg$^{-2}$ \\
\hline
2 & 535 & 19.2 \\
3 & 262 & 9.4 \\
4 & 170 & 6.1 \\
5 & 98 & 3.4 \\
6 & 135 & 4.8 \\
\hline
\end{tabular}
\end{center}
\label{tab:summary}
\end{table}

Table~\ref{tab:summary} gives the number and density of $detections$ as a
function of S/N. Over an effective area of $\sim$28 deg$^2$ (see
Coupon et al. 2009) this means that we detect 19.2 candidate clusters
per deg$^2$  for mass $\geq$ 1.0 10$^{13}$M$_\odot$, 9.4
candidate clusters per deg$^2$  for mass $\geq$ 1.3
10$^{13}$M$_\odot$, 6.1 candidate clusters per deg$^2$  for mass
$\geq$ 3.3 10$^{13}$M$_\odot$, 3.4 candidate clusters per deg$^2$  for
mass $\geq$ 3.5 10$^{13}$M$_\odot$, and 4.8 candidate clusters
per deg$^2$  for mass $\geq$ 5.5 10$^{13}$M$_\odot$.  Given the
typical uncertainty on detection rates, the two last candidate cluster
densities are compatible. We note that these numbers are not corrected for
detection efficiency. These numbers are also fully compatible with
recent X-ray estimates (e.g. Pacaud et al. 2007) showing detections of 5.8 clusters
per deg$^2$ for masses greater than 4.1 10$^{13}$M$_\odot$ (number
scaled to our cosmology).

If we compare our results to published optically based cluster
catalogs, our survey represents a major step forward (see
Table~\ref{tab:summary2}). We have one of the deepest and largest
cluster catalogs in the CFHTLS area by a factor of $\sim$10 most of
the time. We also basically provide the only cluster detections at
z$\geq$1 using CFHTLS data. The comparison of our resuls to the well
known MaxBCG SDSS catalog (Koester et al. 2007) provides similar
numbers in terms of cluster spatial density. The present deep and wide
surveys provides more than 13000 and 10500 $detections$ per
Gpc$^3$. This is comparable to the 16735 clusters per Gpc$^3$ of
Koester et al. (2007), assuming the values of Table~\ref{tab:summary2}.

\begin{table*}
\caption{Comparison of our cluster detections with public CFHTLS optically based
  cluster catalogs and with the MaxBCG SDSS catalog.}
\begin{center}
\begin{tabular}{rrrrr}
\hline
Authors & Detection method & Covered area & Number of detections & Maximal redshift \\
 &  & deg$^2$ &  &  \\
\hline
Present paper Deep & Photometric redshifts & 2.5 & 171 & 1.5 \\
Present paper Wide & Photometric redshifts & 28 & 1029 & 1.2 \\
Olsen et al. (2007) & Matched Filter & 4 & 162 & 1.15 \\
Mazure et al. (2007) & Photometric redshifts & 1 & 44 & 1.5 \\
Cabanac et al. (2007) & Strong lensing & 28 & 40 & 1. \\
Limousin et al. (2009) & Strong lensing & 102 & 13 & 0.85 \\
Gavazzi $\&$ Soucail (2007) & Weak lensing & 4 & 14 & 0.55 \\
Berg\'e et al. (2008) & Weak lensing & 4 & 7 & 0.5 \\
Koester et al. (2007) & Max BCG & 7500 & 13823 & 0.3 \\
\hline
\end{tabular}
\end{center}
\label{tab:summary2}
\end{table*}

These results illustrate the power of optical deep and wide field
surveys to provide large samples of galaxy clusters. These
samples could be used for pure cosmological applications (e.g. based
on cluster counts, Romer et al.  2001) or more generally for the study of
structures within a broad mass range. In particular, it is remarkable that
our deep catalogs are among the first ones to provide numerous group
detections at redshifts greater than 1.

In this picture, the perspectives of our work are:

- To perform a more homogeneous comparison with matched filter detections
across the Wide fields, and this will be the subject of a future
paper. A by-product of this future work will be the center refinement of our
candidate clusters. 

- To define a more precise optically-based mass estimator, via
e.g. the galaxy luminosity functions. This requires however to have performed
the previous match.

- To assess more precisely the detection rate for very massive clusters.
The Millennium simulation as it is now does not provide
a large enough number of very massive structures and this produces
too large an uncertainty. This is prohibitive for any serious
cosmological application based on cluster counts, since the most
massive clusters are the most constraining for cosmology (e.g. Romer
et al. 2001) and detection rates for these massive clusters need to be
precisely evaluated. This aspect is
currently under work and will also be developed in future
papers.

- Finally, the spectroscopic confirmation of the z$\geq$1 candidate clusters 
we have detected would be crucial for cosmology.

The cluster list will be available via the Cencos database at
http://cencosw.oamp.fr/ in a near future.

\begin{acknowledgements}
The authors thank the referee for useful and constructive
comments.
Authors thank M. Limousin for useful discussions.
We acknowledge support from the French Programme National Cosmologie, CNRS.
\end{acknowledgements}

\begin{appendix}

\onecolumn
\begin{longtable}{rrrrrrrrrrrrrrr}
  \caption{Candidate clusters detected in the D2 CFHTLS field. The
    first column is the cluster Id, the second and third columns are
    the J2000 coordinates given in decimal degrees, the fourth column
    is the mean redshift, and the fifth column is the
    SExtractor detection level. }\\
  \hline\hline
  Id & $\alpha$ & $\delta$ & z & $\sigma_S$ \\
  \hline
\endfirsthead
\caption{continued.}\\
\hline
Id & $\alpha$ & $\delta$ & z & $\sigma_S$ \\
\hline
\endhead
D2- 1 & 149.6931 &   1.9613 & 1.20 & 3. \\
D2- 2 & 149.6971 &   2.2002 & 0.60 & 2. \\
D2- 3 & 149.7092 &   1.8994 & 0.35 & 2. \\
D2- 4 & 149.7174 &   1.8370 & 0.63 & 2. \\
D2- 5 & 149.7198 &   2.2413 & 1.00 & 4. \\
D2- 6 & 149.7242 &   1.8689 & 1.42 & 2. \\
D2- 7 & 149.8395 &   2.2761 & 0.40 & 4. \\
D2- 8 & 149.8398 &   1.8005 & 0.15 & 3. \\
\end{longtable}

\onecolumn
\begin{longtable}{rrrrrrrrrrrrrrr}
\caption{Same as Table~\ref{tab:candD2} for the D3 CFHTLS field.}\\
\hline\hline
Id & $\alpha$ & $\delta$ & z & $\sigma_S$ \\
\hline
\endfirsthead
\caption{continued.}\\
\hline
Id & $\alpha$ & $\delta$ & z & $\sigma_S$ \\
\hline
\endhead
D3- 1 & 214.1574 &  52.4575 & 0.95 & 2. \\
D3- 2 & 214.1697 &  52.4380 & 0.85 & 2. \\
D3- 3 & 214.1903 &  52.6105 & 0.60 & 3. \\
D3- 4 & 214.1948 &  52.8756 & 1.20 & 2. \\
D3- 5 & 214.2042 &  53.0803 & 0.95 & 5. \\
D3- 6 & 214.2207 &  53.0568 & 0.60 & 4. \\
D3- 7 & 214.2221 &  52.3024 & 0.40 & 4. \\
D3- 8 & 214.2871 &  52.2871 & 0.83 & 2. \\
\end{longtable}

\onecolumn
\begin{longtable}{rrrrrrrrrrrrrrr}
\caption{\label{candD4}Same as Table~\ref{tab:candD2} for the D4 CFHTLS field.}\\
\hline\hline
Id & $\alpha$ & $\delta$ & z & $\sigma_S$ \\
\hline
\endfirsthead
\caption{continued.}\\
\hline
Id & $\alpha$ & $\delta$ & z & $\sigma_S$ \\
\hline
\endhead
D4- 1 & 333.4241 & -18.0281 & 0.50 & 3. \\
D4- 2 & 333.4323 & -17.8874 & 1.15 & 2. \\
D4- 3 & 333.4707 & -17.3199 & 0.65 & 2. \\
D4- 4 & 333.4883 & -17.8538 & 1.27 & 2. \\
D4- 5 & 333.4921 & -18.1523 & 0.60 & 3. \\
D4- 6 & 333.5163 & -17.4678 & 0.90 & 2. \\
D4- 7 & 333.5186 & -17.3119 & 0.15 & 2. \\
D4- 8 & 333.5498 & -17.6376 & 1.40 & 2. \\
\end{longtable}

\onecolumn
\begin{longtable}{rrrrrrrrrrrrrrr}
\caption{\label{tab_a1}Same as Table~\ref{tab:candD2} for the W1 CFHTLS field.}\\
\hline\hline
Id & $\alpha$ & $\delta$ & z & $\sigma_S$ \\
\hline
\endfirsthead
\caption{continued.}\\
\hline
Id & $\alpha$ & $\delta$ & z & $\sigma_S$ \\
\hline
\endhead
W1-  1 &  33.1280 &  -5.5454 & 0.32 & 6. \\
W1-  2 &  33.1363 &  -3.9767 & 0.57 & 6. \\
W1-  3 &  33.1392 &  -4.3235 & 0.75 & 6. \\
W1-  4 &  33.1402 &  -3.9087 & 1.07 & 6. \\
W1-  5 &  33.1512 &  -3.9950 & 0.90 & 6. \\
W1-  6 &  33.1601 &  -4.1575 & 0.60 & 6. \\
W1-  7 &  33.1720 &  -4.4264 & 1.00 & 6. \\
W1-  8 &  33.1771 &  -4.9606 & 0.55 & 6. \\
\end{longtable}

\onecolumn
\begin{longtable}{rrrrrrrrrrrrrrr}
\caption{\label{candW3}Same as Table~\ref{tab:candD2} for the W3 CFHTLS field.}\\
\hline\hline
Id & $\alpha$ & $\delta$ & z & $\sigma_S$ \\
\hline
\endfirsthead
\caption{continued.}\\
\hline
Id & $\alpha$ & $\delta$ & z & $\sigma_S$ \\
\hline
\endhead
W3-  1 & 209.2892 &  52.6450 & 1.05 & 4. \\
W3-  2 & 209.2931 &  52.9628 & 0.82 & 4. \\
W3-  3 & 209.3033 &  52.2869 & 0.15 & 4. \\
W3-  4 & 209.3059 &  52.7142 & 0.53 & 4. \\
W3-  5 & 209.3703 &  52.5734 & 0.15 & 6. \\
W3-  6 & 209.3849 &  52.6942 & 0.95 & 4. \\
W3-  7 & 209.4233 &  53.0499 & 0.55 & 5. \\
W3-  8 & 209.5383 &  53.0173 & 0.15 & 4. \\
\end{longtable}

\onecolumn
\begin{longtable}{rrrrrrrrrrrrrrr}
\caption{\label{candW4}Same as Table~\ref{tab:candD2} for the W4 CFHTLS field.}\\
\hline\hline
Id & $\alpha$ & $\delta$ & z & $\sigma_S$ \\
\hline
\endfirsthead
\caption{continued.}\\
\hline
Id & $\alpha$ & $\delta$ & z & $\sigma_S$ \\
\hline
\endhead
W4- 1 & 331.9663 &  -0.8770 & 0.85 & 2. \\
W4- 2 & 331.9718 &   0.6677 & 0.20 & 2. \\
W4- 3 & 331.9737 &   0.2738 & 0.45 & 3. \\
W4- 4 & 331.9771 &  -0.1417 & 0.94 & 2. \\
W4- 5 & 331.9846 &  -0.9280 & 1.10 & 2. \\
W4- 6 & 331.9849 &   0.7037 & 1.10 & 2. \\
W4- 7 & 331.9857 &   0.4758 & 0.80 & 4. \\
W4- 8 & 331.9902 &  -0.1315 & 0.75 & 2. \\
\end{longtable}

\end{appendix}

\end{document}